\documentclass[12pt]{article}
\usepackage{scicite}
\usepackage[flushleft]{threeparttable}
\usepackage{longtable}
\usepackage{times}
\usepackage[utf8]{inputenc}
\usepackage{amsfonts}       
\usepackage{nicefrac}       
\usepackage{microtype}      
\usepackage{amssymb,amsmath,amsthm,mathtools,bbm}
\usepackage{algorithm,algpseudocode}
\usepackage{enumitem}
\usepackage{subcaption}
\usepackage{graphicx}
\usepackage{blindtext}
\usepackage{capt-of}
\usepackage{booktabs}
\usepackage{url}
\usepackage{float}
\usepackage[margin=0.5in]{geometry}
\usepackage{array}
\usepackage{ragged2e}
\usepackage{tikz}
\usetikzlibrary{positioning, arrows.meta, quotes, calc}
\usepackage{xurl} 
\usepackage{hyperref}
\usepackage{seqsplit}


\usepackage{listings}
\definecolor{meaningfulcolor}{RGB}{235, 255, 235}
\definecolor{notmeaningfulcolor}{RGB}{255, 235, 235}

\usepackage[most]{tcolorbox}

\lstdefinestyle{promptstyle}{
  basicstyle=\ttfamily\footnotesize,
  columns=fullflexible,
  breaklines=true,
  breakatwhitespace=true,
  showstringspaces=false,
  keepspaces=true,
  tabsize=4
}

\newtcblisting{promptbox}[1]{
  listing engine=listings,
  listing only,
  breakable,
  colback=black!1,
  colframe=black!35,
  left=1.5mm, right=1.5mm, top=1mm, bottom=1mm,
  title={#1},
  fonttitle=\bfseries,
  listing options={style=promptstyle}
}

\usepackage{varwidth}
\usepackage{authblk}
\usepackage{adjustbox}
\usepackage{tabularx}
\usepackage{caption}
\usepackage{xcolor}
\usepackage{rotfloat}
\usepackage[left]{lineno}
\usepackage{makecell}
\usepackage{multirow}
\usepackage{csquotes}
\usepackage[normalem]{ulem}
\usepackage{comment}
\usepackage{placeins}
\usepackage{pdfpages}
\newtcolorbox{exampleq}[1]{
  title=#1,
  fonttitle=\bfseries,
  colback=white,
  colframe=black!40,
  boxrule=0.6pt,
  left=6pt,right=6pt,top=4pt,bottom=4pt
}
\tcbset{
  colback=gray!3,
  colframe=gray!45,
  boxrule=0.4pt,
  arc=2pt,
  left=6pt,right=6pt,top=6pt,bottom=6pt
}
\usepackage[algo2e]{algorithm2e}

\renewcommand{\arraystretch}{1.15}

\newcommand{\osbertlater}[1]{}

\newenvironment{sciabstract}{
\begin{quote} \bf}
{\end{quote}}

\title{Effective Personalized AI Tutors via LLM-Guided Reinforcement Learning}

\author{Angel Tsai-Hsuan Chung,$^{1,\ast}$
Botong Zhang,$^{2}$
Ling-Chieh Kung,$^{3}$\\
Hamsa Bastani,$^{1,\ast,\dagger}$ Osbert Bastani$^{2,\ast,\dagger}$\\
\normalsize{$^{1}$Department of Operations, Information, and Decisions, University of Pennsylvania,}\\
\normalsize{Philadelphia, USA}\\
\normalsize{$^{2}$Department of Computer and Information Science, University of Pennsylvania,}\\
\normalsize{Philadelphia, USA}\\
\normalsize{$^{3}$Department of Information Management, National Taiwan University}\\
\normalsize{Taipei, Taiwan}\\
\normalsize{$^\ast$To whom correspondence should be addressed (emails: angelchg@wharton.upenn.edu, hamsab@wharton.upenn.edu, obastani@seas.upenn.edu); \quad $\dagger$Equal last author}
}

\date{}

\begin{document} 

\baselineskip24pt

\maketitle

\begin{sciabstract}
Generative AI (GenAI) is rapidly reshaping education by unlocking the potential for personalized tutoring. Yet, emerging platforms largely focus on GenAI chatbot tutors that reactively answer student questions.
We hypothesize that the efficacy of GenAI chatbot tutors can be substantially improved by proactively guiding student learning. To test this, we design a novel tutoring platform that tightly integrates a carefully-designed GenAI chatbot with a reinforcement learning algorithm for sequencing practice problems. Critically, this algorithm  leverages rich signals from student-chatbot interactions to adaptively select practice problems of an appropriate difficulty level. In partnership with the Taipei City Government and American Institute in Taiwan, we deployed our tutoring platform in conjunction with a five-month course to teach Python to students across ten high schools. We randomized students between a fixed practice problem sequence and our adaptive sequencing algorithm. We find that adaptive sequencing increased unassisted final exam performance by 0.15 standard deviations (equivalent to 6-9 months of schooling by some estimates); mediation analysis suggests that gains were driven by increased engagement. Our work provides large-scale field evidence that student-chatbot interactions provide valuable signals for proactively optimizing and personalizing student learning.
\end{sciabstract}

\section{Introduction}

With the rapid recent progress in Generative Artificial Intelligence (GenAI), there has been substantial interest in leveraging it to improve education. A major focus has been on designing personalized tutors based on Large Language Models (LLMs)~\cite{extance2023chatgpt, bastani2025generative, wang2025effect,sun2024would,kasneci2023chatgpt,johnson2024using}, which gives students on-demand access to a GenAI chatbot that answers student questions and attempts to resolve their misunderstandings. Developing effective tutoring systems has become increasingly urgent as technological change is accelerating the need for worker reskilling~\cite{eloundou2024gpts,morandini2023impact}.

A key lesson from decades of research---from Computer-Assisted Instruction~\cite{cristia2017technology} to MOOCs~\cite{papadakis2023moocs} to initiatives such as One Laptop per Child~\cite{lamb2018institutional}---is that access to educational technology alone is insufficient to improve outcomes; successful deployments require careful design and implementation that aims to integrate with existing pedagogical context rather than replace it. GenAI chatbot tutors are no different---a recent meta-analysis synthesizing 51 experimental studies finds that while ChatGPT has positive impacts on learning on average, there is substantial heterogeneity in outcomes~\cite{wang2025effect}. Recent work has highlighted the importance of good design on the efficacy of GenAI chatbot tutors, studying the impact of prompt design~\cite{bastani2025generative}, optimizing chatbot features~\cite{liu2024teaching, wang2024tutor, shetye2024evaluation, kumar2025math, kestin2025ai, nam2025efficient}, or improving LLM accuracy on educational tasks~\cite{trinh2024solving, he2024evaluating}.

However, these approaches all focus on the problem of improving the chatbot, and stay within the reactive paradigm of responding to student queries (e.g., debugging code upon request)~\cite{sidorkin2025leapfrogging}. Yet, effectively promoting learning requires substantially more than simply conversing with the student. Evidence suggests that learning requires students to engage in productive struggle~\cite{warshauer2015productive, bastani2025incentive}, spending time working on challenging problems that help develop critical thinking and problem solving skills. The challenge with GenAI chatbot tutors is that students often lack the skills needed to effectively self-regulate their learning~\cite{bjork2013self, dunlosky2007metacomprehension, poulidis2025self}, making it difficult for them to formulate queries that significantly improve their learning outcomes.

In contrast, effective techniques such as mastery learning~\cite{bloom1968learning,kulik1990effectiveness}, desirable difficulty~\cite{bjork2011making}, zone of proximal development~\cite{vygotsky1978mind}, and productive struggle~\cite{kapur2008productive, warshauer2015productive} focus on engaging the student by providing them with challenging practice problems that are suitable for their current knowledge and skill. While these approaches traditionally require extensive instructor oversight~\cite{kulik1979meta,guskey1986defining,slavin1987mastery}, there has been substantial interest in scaling these techniques by designing algorithms that automate them~\cite{clement2015multiarmed,rafferty2016faster,shen2018improving,segal2018combining}. A fundamental challenge facing these algorithms is the need to estimate the student's current knowledge state, which captures how well they have learned the underlying concept. Accurate and efficient estimation of a rich knowledge state is key to rapidly adapting difficulty so the student does not waste valuable time solving problems they have already mastered. However, existing algorithms face a critical information bottleneck---they only have access to coarse proxies for the knowledge state, such as whether the student's solutions are correct or how much time they spent solving each problem; as a consequence, they can only effectively estimate a highly simplified approximation of the student's true knowledge state. For instance, the popular Bayesian knowledge tracing (BKT) algorithm~\cite{corbett1994knowledge,vsaric2024twenty} for personalized learning relies on binary signals (correct or incorrect solutions) to estimate a binary knowledge state (learned or unlearned), which is a very coarse approximation of student understanding.




We argue that these shortcomings can be addressed by tightly integrating adaptive problem sequencing into GenAI chatbot tutors, through two key strategies: (1) using LLMs to extract richer semantic signals from student solution attempts, and (2) extracting signals from student-chatbot interactions. To test our hypothesis, we propose a novel reinforcement learning algorithm that leverages signals from student-platform interactions---using LLMs both to measure quality of student-chatbot conversations and to analyze student solution attempts---to accurately estimate a continuous measure of the student's current knowledge state. To facilitate this rich feedback and knowledge state representation, our algorithm leverages a novel state estimation procedure based on particle filtering~\cite{thrun1999monte}, as well as a novel planning procedure based on model predictive control (MPC)~\cite{garcia1989model}. Our algorithm rapidly estimates a rich approximation of the student's knowledge state, and performs granular personalization of practice problem difficulty levels based on this estimate.
Furthermore, we developed a tutoring platform that tightly integrates this algorithm with a carefully-designed GenAI chatbot tutor.


In collaboration with the Taipei City Government, the American Institute in Taiwan, and the American Innovation Center (which operates under U.S. Department of State oversight), we deployed our tutoring platform through a five-month-long ``AI for Python Learning'' course. This course augmented standard lecture-based instruction with practice on our tutoring platform. We conducted a randomized controlled trial (RCT) across 10 participating high schools in Taipei, and we evaluate the impact of our system on semester-long learning outcomes for 770 students. All students were provided with access to the same course material and GenAI chatbot tutor, but we randomized students between our personalized problem sequencing algorithm versus a fixed problem sequence based on standard practice. To the best of our knowledge, our study provides a substantially larger-scale and longer-horizon assessment of the efficacy of GenAI tutoring compared to existing work.\footnote{The meta-analysis~\cite{wang2025effect} synthesizes 51 (quasi-)experimental studies published between November 2022 and February 2025 (72 coded effect sizes). For the learning-performance outcome specifically ($n=44$ independent studies), the duration distribution is heavily short-horizon that only $n=13$ studies lasting $>$8 weeks (i.e., 31/44 $\approx$ 70\% are $\le$8 weeks). Reported samples are typically small: median per-arm sample sizes are $\approx$35--37 students, and $\sim$89\% of comparisons involve fewer than 100 students per arm. In addition, we surveyed existing experimental evidence for the efficacy of adaptive problem sequencing, and find that most existing experiments conflate multiple interventions or compare to weak baselines; see Supplement~\ref{sup:literature} for a detailed discussion.}

Our results show that personalized sequencing improves outcomes compared to fixed sequencing by 0.15 standard deviations on an in-person written exam completed without AI assistance, which translates into as much as 6-9 months of additional schooling according to some estimates~\cite{evans2019equivalent, ritchie2018much}. Notably, these gains were achieved without increasing instruction time or teacher workload. Furthermore, our mediation analysis suggests that these gains are mediated by increased student engagement, addressing a long-standing challenge in educational technology of sustaining engagement beyond the ``novelty effect," where interest typically rapidly declines following initial excitement~\cite{tsay2020overcoming,reich2014mooc,eysenbach2005law}. Our work provides rigorous field evidence that leveraging rich behavioral signals from student-platform interactions can drive automated personalization of student practice to substantially improve learning outcomes.


\section{GenAI Tutoring System}
\label{sec:AIplatform}

\begin{figure}
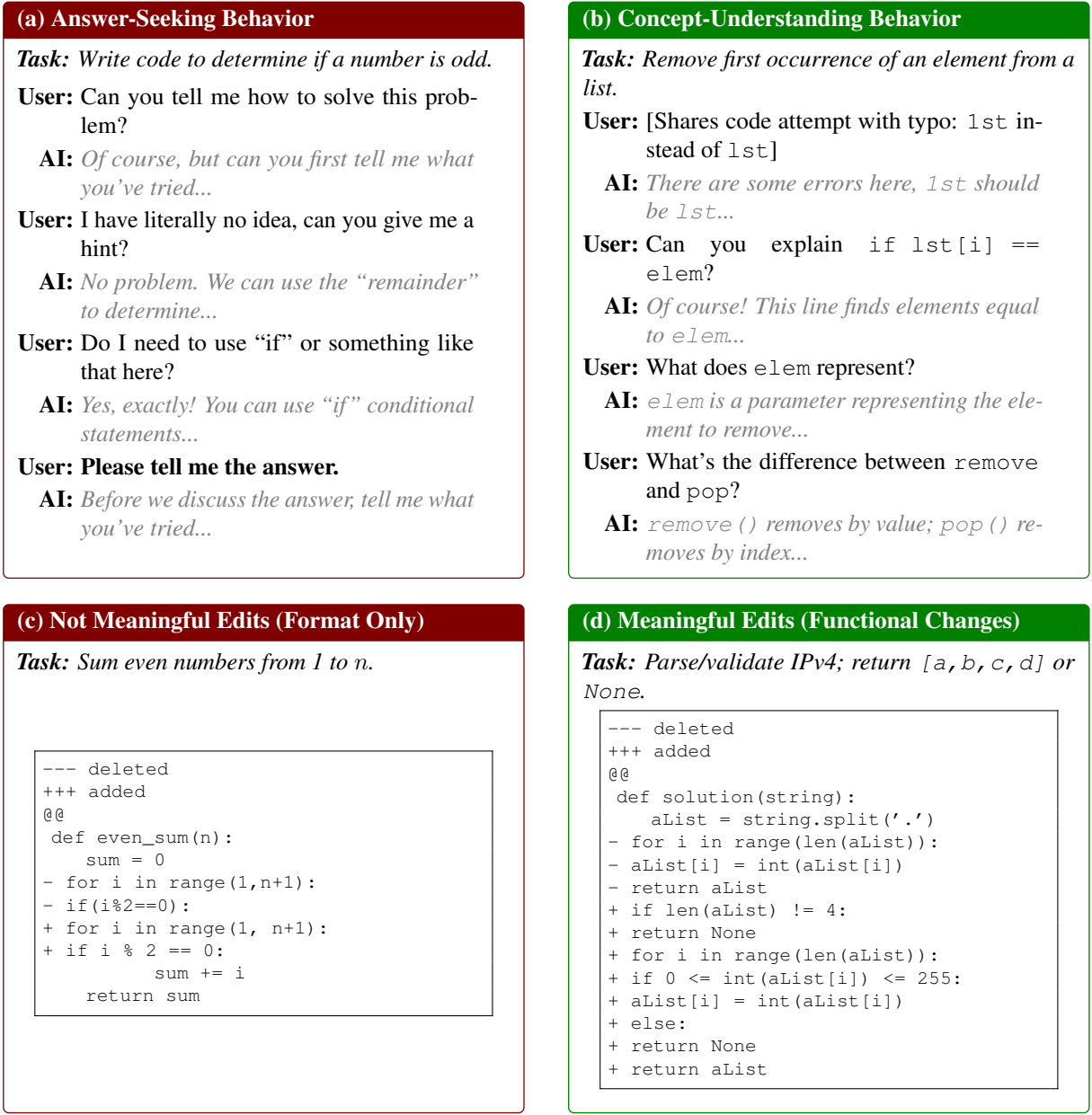

\centering
\begin{minipage}[t]{0.48\textwidth}
\begin{tcolorbox}[colback=white, colframe=red!50!black, title=\textbf{(a) Answer-Seeking Behavior}, fonttitle=\bfseries\footnotesize, boxrule=0.5pt, left=3pt, right=3pt, top=2pt, bottom=2pt, height=8.5cm, valign=top]
\footnotesize
\textit{\textbf{Task:} Write code to determine if a number is odd.}
\vspace{0.3em}

\begin{tabular}{@{}r@{\hspace{0.3em}}p{0.80\textwidth}@{}}
\textbf{User:} & Can you tell me how to solve this problem? \\
\textbf{AI:} & \textcolor{gray}{\textit{Of course, but can you first tell me what you've tried...}} \\
\textbf{User:} & I have literally no idea, can you give me a hint? \\
\textbf{AI:} & \textcolor{gray}{\textit{No problem. We can use the ``remainder'' to determine...}} \\
\textbf{User:} & Do I need to use ``if'' or something like that here? \\
\textbf{AI:} & \textcolor{gray}{\textit{Yes, exactly! You can use ``if'' conditional statements...}} \\
\textbf{User:} & \textbf{Please tell me the answer.} \\
\textbf{AI:} & \textcolor{gray}{\textit{Before we discuss the answer, tell me what you've tried...}} \\
\end{tabular}
\end{tcolorbox}
\end{minipage}
\hfill
\begin{minipage}[t]{0.48\textwidth}
\begin{tcolorbox}[colback=white, colframe=green!50!black, title=\textbf{(b) Concept-Understanding Behavior}, fonttitle=\bfseries\footnotesize, boxrule=0.5pt, left=3pt, right=3pt, top=2pt, bottom=2pt, height=8.5cm, valign=top]
\footnotesize
\textit{\textbf{Task:} Remove first occurrence of an element from a list.}
\vspace{0.3em}

\begin{tabular}{@{}r@{\hspace{0.3em}}p{0.80\textwidth}@{}}
\textbf{User:} & [Shares code attempt with typo: \texttt{1st} instead of \texttt{lst}] \\
\textbf{AI:} & \textcolor{gray}{\textit{There are some errors here, \texttt{1st} should be \texttt{lst}...}} \\
\textbf{User:} & Can you explain \texttt{if lst[i] == elem}? \\
\textbf{AI:} & \textcolor{gray}{\textit{Of course! This line finds elements equal to \texttt{elem}...}} \\
\textbf{User:} & What does \texttt{elem} represent? \\
\textbf{AI:} & \textcolor{gray}{\textit{\texttt{elem} is a parameter representing the element to remove...}} \\
\textbf{User:} & What's the difference between \texttt{remove} and \texttt{pop}? \\
\textbf{AI:} & \textcolor{gray}{\textit{\texttt{remove()} removes by value; \texttt{pop()} removes by index...}} \\
\end{tabular}
\end{tcolorbox}
\end{minipage}

\vspace{0.8em}

\begin{minipage}[t]{0.48\textwidth}
\begin{tcolorbox}[colback=white, colframe=red!50!black, title=\textbf{(c) Not Meaningful Edits (Format Only)}, fonttitle=\bfseries\footnotesize, boxrule=0.5pt, left=3pt, right=3pt, top=2pt, bottom=2pt, height=7.5cm, valign=top]
\footnotesize
\textit{\textbf{Task:} Sum even numbers from 1 to $n$.}
\vspace{3em}

\centering
\begin{minipage}{0.9\textwidth}
\begin{lstlisting}[basicstyle=\ttfamily\scriptsize,breaklines=true,frame=single]
--- deleted 
+++ added
@@
 def even_sum(n):
     sum = 0
-    for i in range(1,n+1):
-        if(i%2==0):
+    for i in range(1, n+1):
+        if i % 2 == 0:
             sum += i
     return sum
\end{lstlisting}
\end{minipage}
\end{tcolorbox}
\end{minipage}
\hfill
\begin{minipage}[t]{0.48\textwidth}
\begin{tcolorbox}[colback=white, colframe=green!50!black, title=\textbf{(d) Meaningful Edits (Functional Changes)}, fonttitle=\bfseries\footnotesize, boxrule=0.5pt, left=3pt, right=3pt, top=2pt, bottom=2pt, height=7.5cm, valign=top]
\footnotesize
\textit{\textbf{Task:} Parse/validate IPv4; return \texttt{[a,b,c,d]} or \texttt{None}.}
\vspace{0.3em}

\centering
\begin{minipage}{0.9\textwidth}
\begin{lstlisting}[basicstyle=\ttfamily\scriptsize,breaklines=true,frame=single]
--- deleted 
+++ added
@@
 def solution(string):
     aList = string.split('.')
-    for i in range(len(aList)):
-        aList[i] = int(aList[i])
-    return aList
+    if len(aList) != 4:
+        return None
+    for i in range(len(aList)):
+        if 0 <= int(aList[i]) <= 255:
+            aList[i] = int(aList[i])
+        else:
+            return None
+    return aList
\end{lstlisting}
\end{minipage}
\end{tcolorbox}
\end{minipage}

\caption{\textbf{Examples of LLM-based features showing a student's learning progress}. The top row shows AI tutor chat interactions (translated to English): (a) answer-seeking behavior vs. (b) concept-understanding engagement. The bottom row shows student's code editing patterns: (c) format-only changes (in students’ code-edit histories, the most common non-meaningful edits include typos, formatting, or stylistic changes) vs. (d) meaningful functional changes. }
\label{fig:LLMFeatureExamples}
\end{figure}

In this section, we describe our GenAI tutoring system. Both lecture-based instruction and solving practice problems occur on our platform, organized into modules that the students must complete in order. Our GenAI tutoring system is responsible for (i) generating a bank of practice problems (this step happens before the start of the course), (ii) helping students solve a sequence of practice problems for each module, and (iii) selecting the sequence of practice problems to maintain engagement. A key novelty of our system is that it leverages LLM-based features---e.g., student-tutor interactions and code-edit traces---to estimate and track each student’s learning progress (see Fig.~\ref{fig:LLMFeatureExamples} for illustrative examples); we find that these traces serve as a more precise proxy for learning progress than conventional measures (see Supplement~\ref{sup:simresult}). In our experimental design, (i) and (ii) are provided across all conditions, whereas access to (iii) is randomized, with the goal of evaluating whether it is effective at improving learning outcomes. We describe these three key components of our GenAI tutoring system here, and provide additional details in Supplement~\ref{supp:system_details}.

\subsection{Problem Bank Generation}
\label{sec:rag}

Personalized problem sequencing requires a large bank of candidate problems, which requires significantly more effort to generate. To scale this process, our system leverages a ``self-validating'' LLM to help generate high-quality, auto-gradable practice problems that are aligned with the course materials (see details in Supplement~\ref{sup:rag}). This component takes a human-in-the-loop approach, generating candidate problems that are edited by a human expert before being added to our problem bank. We adopt a human-in-the-loop workflow where teaching assistants review and edit validated candidates for quality and correctness before adding them to the problem bank.

To ensure that the problems are aligned with the course material, our problem generator uses retrieval-augmented generation (RAG)~\cite{lewis2020retrieval} from instructional materials, including lecture slides and past practice problems.
Given this knowledge base, we prompt an LLM (specifically, GPT-4o) to generate candidate practice problems. Prompts specify the course module, target concept, desired problem type (coding, debugging, or true/false) (see details in Supplement~\ref{sup:practiceQ}), and difficulty level. The LLM also generates the scaffolding required for autograding, including either the answer for true/false problems or test cases for coding and debugging problems. To ensure quality and correctness, each generated problem is validated through a Python script checking consistency among the question, solution, and test cases. When validation fails, error messages are passed to a separate LLM (specifically, Claude) for iterative revision until the problem passes.
Finally, a teaching assistant checks all validated problems for language clarity and alignment with course objectives before adding them to the problem bank. We provide additional details in Supplement~\ref{sup:practiceQ}.

\subsection{GenAI Chatbot Tutor}
\label{sec:platform_short}

The core of our GenAI tutor is a web interface that presents students with a sequence of practice problems along with a chatbot designed to help answer questions and resolve any confusion. Our interface shows the current practice problem to the student, and provides them with space to write and execute code and to chat with the GenAI tutor. Students can submit solutions and receive immediate feedback from our autograder; they can submit as many times as desired without penalty. They must correctly solve one problem before continuing to the next. The interface also disables copy/paste functionality to ensure students remain on our platform.


Our platform includes an LLM-based chatbot that is available throughout practice, enabling students to ask questions, request hints, and interpret error messages. The tutor is prompted to avoid providing answers unless students first demonstrate substantial effort (see Fig.~\ref{fig:LLMFeatureExamples}(a)-(b) for examples, and Supplement~\ref{sup:aitutor} for details). Notably, our platform records time-stamped logs of learner activity, including student and chatbot messages as well as all code edits and submissions (see Fig.~\ref{fig:LLMFeatureExamples}(c)-(d) for examples). These behavioral traces are critical for us to infer the student's knowledge state; in addition, we also use them in our empirical analysis to help understand the mechanisms underlying our improvements in learning outcomes. We provide additional details on interface, including screenshots, in Supplement~\ref{supp:system_details}.

\subsection{Personalized Problem Sequencing}
\label{sec:pomdp_short}


A key novelty of our system is how it selects the sequence of problems for students to solve. A solution requires two components: (i) a method for estimating the student's current \emph{knowledge state}---which summarizes their mastery of the skills in the current module---based on observations of their performance (e.g., the rate of correct solutions or the time taken to solve problems), and (ii) a policy for selecting the next optimal problem to give the student, based on the estimated knowledge state. Existing algorithms rely on very limited feedback for knowledge state estimation, which in turn limits the complexity of the knowledge state representation. We build on Bayesian knowledge tracing~\cite{corbett1994knowledge}, which formulates the optimal problem sequencing problem as a Partially Observed Markov Decision Process (POMDP)~\cite{kaelbling1998planning}, a general framework for encoding sequential decision-making problems where the true state (in our case, the student's knowledge state) is unobserved and must be estimated based on observations. However, their POMDP is simple---their estimation algorithm only considers binary observations of whether the student successfully solved a problem, so they correspondingly encode the knowledge state as a binary indicator of mastery of the current topic. In their setting, simplicity is necessary due to limited feedback from student interactions. Furthermore, their simple knowledge state also makes selecting the optimal problem straightforward.


In our setting, binary feedback is untenable since students rarely obtain the correct solution on their first try (e.g., due to syntax errors rather than misunderstandings); thus, the feedback would always be negative, and Bayesian knowledge tracing would conclude that the student never achieves mastery. Instead, we design a POMDP that leverages the substantially richer feedback provided by our system. We consider a knowledge state that encodes a continuous measure of mastery; it includes three parameters (i) the student's initial performance, (ii) the student's mastery ceiling (the maximum performance they achieve if they hypothetically solve all of the problems), and (iii) their learning speed (how much they improve after solving a problem). Based on its current estimate of the knowledge state, our algorithm (described below) must select the difficulty level of the student's next practice problem (out of four). The knowledge state evolves according to a deterministic transition function; at a high level, the student's mastery increases depending on the student's learning speed, the number of attempts they needed to solve the problem, and the selected problem difficulty; hyperparameters in the transition function were calibrated based on data collected from a pilot study, which we describe in Section~\ref{sec:field} and Supplement~\ref{sup:pilot}. Finally, our algorithm receives stochastic feedback based on the student's interaction with our platform---most notably, an LLM's estimate of the number of serious attempts required to solve the problem and the quality of the student-chatbot interactions---and updates its estimate of the knowledge state accordingly.

Our algorithm uses a particle filter for belief state estimation~\cite{thrun1999monte,arulampalam2002tutorial}; it maintains a discrete probability distribution over a beam of plausible knowledge states, and performs posterior updates to this distribution whenever it receives feedback. The belief state estimate is initialized based on the prediction from a random forest model trained on historical data from our pilot study; this model uses features from both a pre-study survey as well as performance in earlier modules when available. To select an optimal decision, our algorithm uses model predictive control (MPC)~\cite{garcia1989model}; at each step, it samples a large number of possible decision sequences up to a finite number of steps, simulates the outcome for each decision sequence, selects the sequence that achieves the highest performance, and takes the first decision in that sequence.

Finally, to speed up convergence, our algorithm uses model-based reinforcement learning, where it trains a machine learning model on historical data to help predict the initial belief state for a given student-module pair. At a high level, we derive features from historical data on each student's interactions with our platform, including features based on using an LLM to analyze student-chatbot conversations. This strategy enables our belief state estimates to converge more rapidly, enabling us to achieve strong performance earlier in the problem sequence.

We provide additional details on our algorithm in Supplement~\ref{sup:rl}, as well as simulation experiments supporting the importance of LLM-derived behavioral information from student-platform interactions.


\section{Field Experiment}
\label{sec:field}

In partnership with the Taipei City Government, the American Institute in Taiwan, and the American Innovation Center, we developed and ran an ``AI for Python Learning'' Program based on our platform. The goal of this program was to teach high school students introductory data science skills in parallel with Python programming skills, in alignment with the government's broader digital literacy and bilingual education goals.
The instructional content and materials were based on a ten‑module online course developed by professors at National Taiwan University for Coursera, where it has enrolled more than 44,000 learners and received a 4.9/5 average rating, demonstrating its high quality. The program included a formal government-endorsed certification (valid for college applications), providing a strong incentive for students to perform well. To earn the certificate, each student had to (i) complete the required homework for each of the ten modules on our platform (see program schedule in Supplement~\ref{sup:programschedule}), and (ii) pass a final, in-person, written certification examination. 

In Summer 2024, we ran a pilot study on an early version of the platform, which we detail in Supplement~\ref{sup:pilot}. Then, in Fall 2024, we worked with our government partners to recruit schools and students for the main certification program; our government partners helped recruit and obtain informed consent from recruited participants (see Supplement~\ref{sup:briefing} for details). Both the pilot and the main study were IRB-approved at the University of Pennsylvania; the main study was pre-registered (see Supplement~\ref{sup:preregist} for details).

A total of 1,047 students from 10 high schools in Taipei (8 public, 2 private) participated in the program for the main study. We report results on a sample of 770 students, excluding students that did not meet our pre-registered criteria (e.g., dropped out, flagged for cheating)\footnote{We do not find significant differences in attrition across both arms; see Supplement~\ref{sup:attrition}.} or that are not directly comparable to the standard program (see Supplement~\ref{sup:exclusion} for details). In January 2025, the government officially launched the program, which ran through the end of June 2025, for a total learning period of approximately five months (see Supplement~\ref{sup:programschedule} for details). In June 2025, we administered the final in-person certification examination and hired external graders to score all the exams. We randomized all students at the individual level into one of two groups with equal probability:
\begin{enumerate}
\item \textbf{Treatment (Personalized problem sequence):} Students solve practice problems selected adaptively by our POMDP-based personalized problem sequencing algorithm.
\item \textbf{Control (Fixed problem sequence):} Students solve a fixed sequence of problems that progresses from easy to hard within each module; this strategy mirrors common practice.
\end{enumerate}
In particular, the learning platform, including instructional content and tutor chatbot are the same in both groups, and the only difference is in terms of the problem sequence. To ensure comparable exposure by difficulty, all students---regardless of the condition---are required to complete at least two ``very hard,'' two ``hard,'' and two ``medium'' questions in each module.\footnote{Each module contains 40 auto-graded practice problems (coding, debugging, and true/false), generated via the LLM-based pipeline described in Section~\ref{sec:rag}. Homework assignments typically require 15 questions per module.}

We conducted baseline surveys at the beginning of the program to collect details on students' demographics, parents' education, motivation, and pre-existing Python ability (see Supplement~\ref{sup:survey}). Table~\ref{tab:balance} in Supplement~\ref{sup:balancedTable} shows that these student characteristics are well balanced between experimental conditions.

Our primary outcome is student performance on the final certification exam administered in-person at participating schools with no digital devices allowed; this exam assesses conceptual understanding and problem-solving skills in Python (see Supplement~\ref{sup:exam} for details). Finally, in Summer 2025, trained graders scored all exams using a common rubric. We provide additional details on our experimental design in Supplement~\ref{sup:expprotocol}. 

\section{Evaluation}
\label{sec:eval}

\subsection{Main Results}
\label{sec:main}

\begin{figure}[!t]
\centering
\captionsetup[subfigure]{labelformat=parens}

\begin{tabular}{@{}p{0.56\textwidth}@{\hspace{0.04\textwidth}}p{0.40\textwidth}@{}}

\begin{subfigure}[t]{\linewidth}
\centering
\vspace{0pt}
\includegraphics[width=\linewidth,trim=0 0mm 0 0,clip]{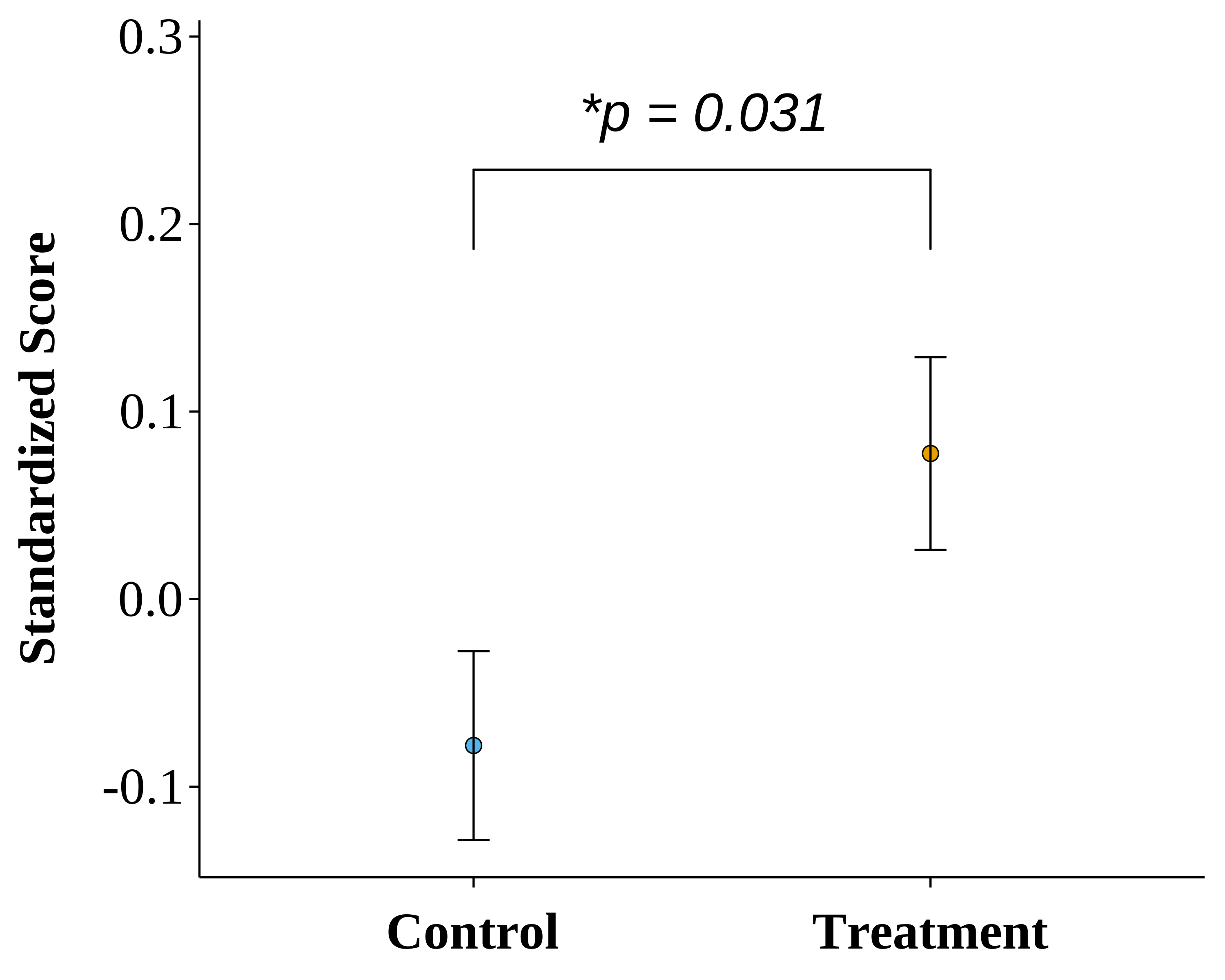}
\caption{Average standardized score by treatment condition.}
\end{subfigure}
&
\begin{subfigure}[t]{\linewidth}
\centering
\vspace{0pt}
\vspace{1.5mm} 
\footnotesize
\setlength{\tabcolsep}{4pt}
\renewcommand{\arraystretch}{1.15}

\begin{threeparttable}
\begin{tabular}{lcc}
\toprule
& \multicolumn{2}{c}{\textit{Dependent variable:}} \\
\cmidrule(lr){2-3}
& \multicolumn{2}{c}
{Standardized exam score} \\
& \multicolumn{1}{c}{(1)} & \multicolumn{1}{c}{(2)}\\
\midrule
Treatment & $0.156^{*}$ & $0.150^{*}$ \\
               & (0.072)           & (0.067)     \\
\midrule
Controls       & No                & Yes         \\
School fixed effect       & No                & Yes         \\
Grade-level fixed effect       & No                & Yes         \\
\midrule
Observations (N)            & 770               & 716         \\
Adjusted $R^{2}$        & 0.005             & 0.226       \\
\midrule
\bottomrule
\end{tabular}
\begin{tablenotes}[flushleft]
\footnotesize
\item \emph{Notes:} Standard errors in parentheses. $^{*}p<0.05$, $^{**}p<0.01$, $^{***}p<0.001$.
\end{tablenotes}
\end{threeparttable}
\caption{OLS result summary.}
\end{subfigure}
\end{tabular}

\caption{\textbf{Effect of personalized problem sequencing on exam performance.} (a) Dots show average standardized score by group with standard error. The bracketed annotation reports the two-sided t-test comparing average standardized scores between the treatment and control groups ($p = 0.031$), indicating a statistically significant difference in mean performance. (b) OLS estimates of the effect of personalized problem sequencing on standardized exam scores. Column (1) presents a baseline specification without controls, and Column (2) includes baseline controls and school and grade-level fixed effects. The treatment coefficient is positive and statistically significant across specifications.}
\label{fig:main}
\end{figure}

We evaluate the impact of our personalized problem sequencing algorithm on certification exam performance using an Intention-to-Treat (ITT) analysis with ordinary least squares regression (OLS); we provide details on this regression in Supplement~\ref{sup:main}. Our results show that the treatment group outperforms the control group by 0.156 standard deviations ($p<0.05$, see Fig.~\ref{fig:main}). This effect remains similar when we control for baseline covariates, school fixed effects, and grade-level fixed effects, showing gains of 0.150 standard deviations ($p<0.05$, see Fig.~\ref{fig:main}b, column 2). Based on some estimates from the literature~\cite{ritchie2018much,evans2019equivalent}, the average gain of 0.150 standard deviations is equivalent to approximately six to nine months of additional schooling;\footnote{Under the World Bank's ``Equivalent Years of Schooling'' (EYOS) framework~\cite{evans2019equivalent}, the pooled Skills Toward Employability and Productivity (STEP) benchmark implies approximately $0.15$--$0.21$ SD of learning per school year. However, the framework highlights substantial cross-system heterogeneity, with higher-performing settings (e.g., the U.S. and Finland) showing approximately $0.20$--$0.30$ SD annual learning gains per year. Using the faster rate as a more comparable benchmark for Taiwan implies that a 0.15 SD improvement in learning (our main finding) translates to approximately 6-9 months of schooling.} while these numbers may differ from one domain to another, they illustrate that the improvements we obtain are substantial. Our results demonstrate that integrating interventions such as a personalized problem sequencing can substantially improve the efficacy of GenAI tutors. We provide additional robustness checks in Supplement~\ref{sup:main}, including analyses using alternative standard error specifications (heteroskedasticity-robust and school-clustered) and the raw scores as the outcome; our findings are robust across specifications.

\subsection{Closing the Gap: Heterogeneity and Equity}
\label{sec:hte}

\begin{figure}[!t]
\centering
\includegraphics[width=0.85\columnwidth]{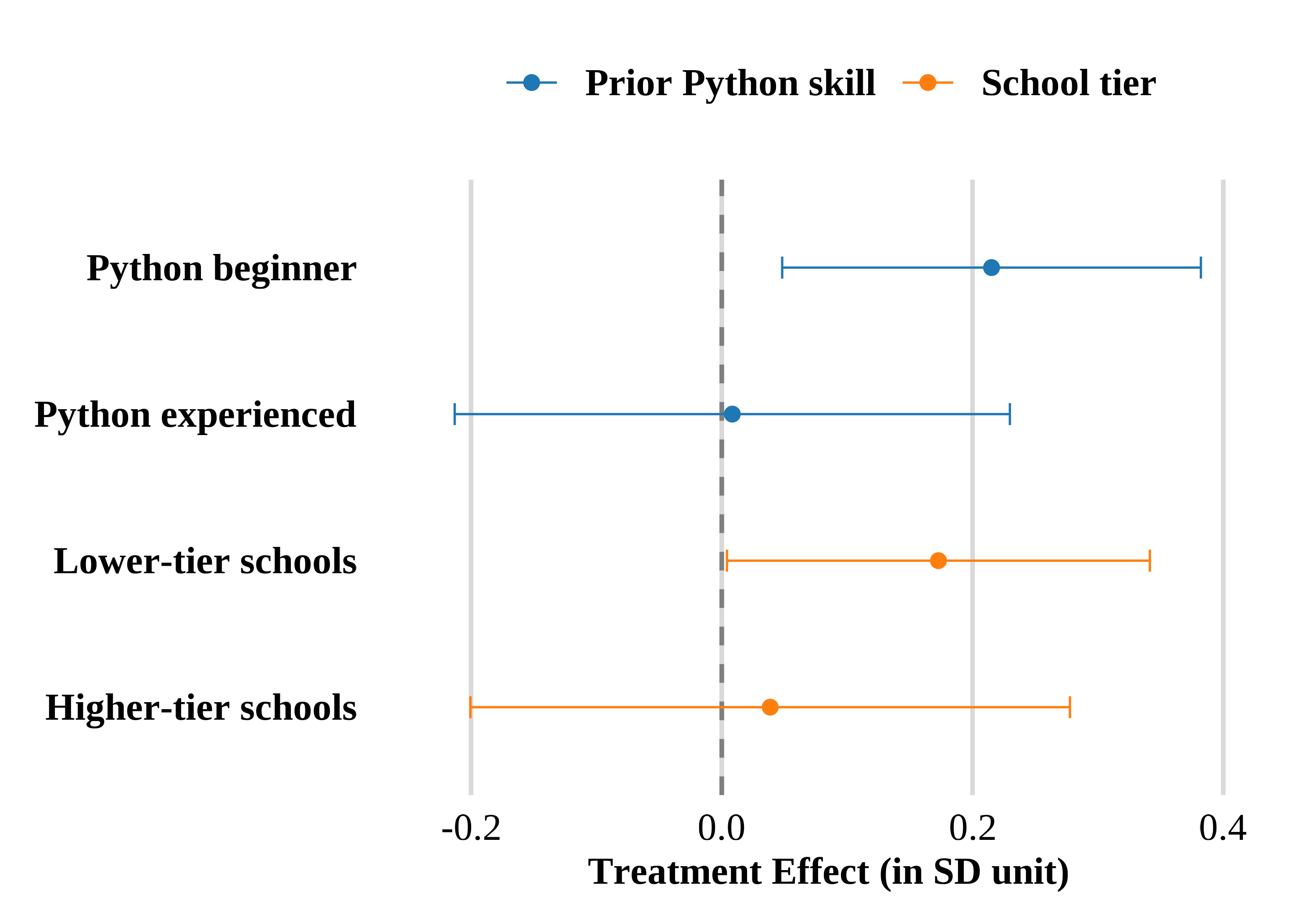} 
\caption{\textbf{Treatment effects by prior Python skill and school tier.}
The plot reports subgroup-specific treatment effects on the standardized outcome (in units of SD of exam performance across all students). Points denote estimated effects and horizontal bars denote 95\% confidence intervals; the dashed vertical line marks zero effect. Blue indicates subgroups defined by prior Python skill, and orange indicates subgroups defined by school tier.}
\label{fig:hteforest}
\end{figure}

Next, we explore treatment heterogeneity to understand whether the effects of personalized problem sequencing vary by student background. First, we classify students into two types based on their prior Python skill---specifically, beginners (53\%, consisting of students who self-identified as ``first-time learners'' or ``beginners'' in the baseline survey) and those with prior Python skill (39\%, consisting of students who self-identified as having ``intermediate'' or ``proficient'' coding skills in the baseline survey).\footnote{We exclude 6 students (1\%) who reported national competition-level Python skill, since they had already mastered the curriculum prior to the course and would reflect ceiling effects. $7\%$ of the students did not complete the baseline survey and are excluded from this analysis as well.} Our results (see Fig.~\ref{fig:hteforest}) indicate that beginners with no prior skill benefit more from personalization: among these students, personalization increases performance by 0.215 SD ($p=0.012$), whereas effects are negligible for students with prior skill (0.008 SD, $p=0.941$). This pattern is consistent with the literature that adaptive systems often deliver the largest gains for less-experienced learners~\cite{doroudi2019wheres}. Fixed problem sequences are typically calibrated for the median student, and lack the flexibility to accommodate heterogeneity in learning speeds.
Personalized problem sequencing addresses this issue by adapting difficulty based on the student's current performance.

We also examine the impact of our intervention for different school tiers---in Taiwan, high school admission is determined by a national standardized exam, the Comprehensive Assessment Program (CAP), so schools are clearly ranked by academic selectivity. Specifically, we classify schools as higher- or lower-tier based on entrance exam scores, and examine whether gains vary with school selectivity. As shown in Fig.~\ref{fig:hteforest}, estimated effects are positive for both school types, with larger gains in lower-tier schools (0.173 SD, $p=0.045$) and a smaller, statistically insignificant estimate in higher-tier schools (0.039 SD, $p=0.752$). For both prior Python skill and school tier, we also estimate interaction regressions with terms \textit{RL$\times$subgroup}, and find consistent results; see Supplement~\ref{sup:hte} for details.  

Overall, these results suggest that our personalized problem sequencing intervention can deliver equitable learning gains without widening educational inequality, and may even help bridge the gap between advantaged and disadvantaged students.

\subsection{Potential Mechanisms: Productive AI Use and Sustained Engagement}
\label{sec:mechanisms}

Finally, we examine the mechanisms underlying the performance gains from our personalized problem sequencing algorithm. First, we find that students in the treatment group interact with the tutor chatbot in more productive ways. Second, we find that our performance gains are driven by increased engagement (measured by time-on-task and persistence), but not by completing more problems or being pushed toward uniformly harder material. Together, these results suggest that personalized problem sequencing has the potential to improve student engagement, addressing the retention challenge in existing educational technologies~\cite{tsay2020overcoming,reich2014mooc,eysenbach2005law}.

\paragraph{More productive use of the tutor chatbot.} First, we analyze student interactions with our tutor chatbot. From the student-chatbot conversation history for each practice problem, we construct a chat quality score (on a scale of 1-10) by using LLM-as-a-judge~\cite{zheng2023judging} to distinguish productive conversations (e.g., explanation-seeking and iterative debugging) from unproductive ones (e.g., answer-seeking); see Figure~\ref{fig:LLMFeatureExamples}(a)-(b) for examples. Recall that all students have access to the same GenAI chatbot tutor with the same guardrails (e.g., prompted to
avoid providing answers unless students first demonstrate substantial effort, see Supplement~\ref{sup:aitutor}).

As shown in Fig.~\ref{fig:chatquality}, students in the treatment group exhibit significantly higher chat quality according to our metric. This pattern suggests that personalized problem sequencing encourages students to engage with the tutor chatbot in ways that support deeper understanding and active problem-solving, supporting our hypothesis that improving other aspects of GenAI tutors can be more productive than focusing on the tutor chatbot in isolation.

\begin{figure}[!t]
\centering
\begin{subfigure}[t]{0.48\textwidth}
\centering
\includegraphics[width=\textwidth]{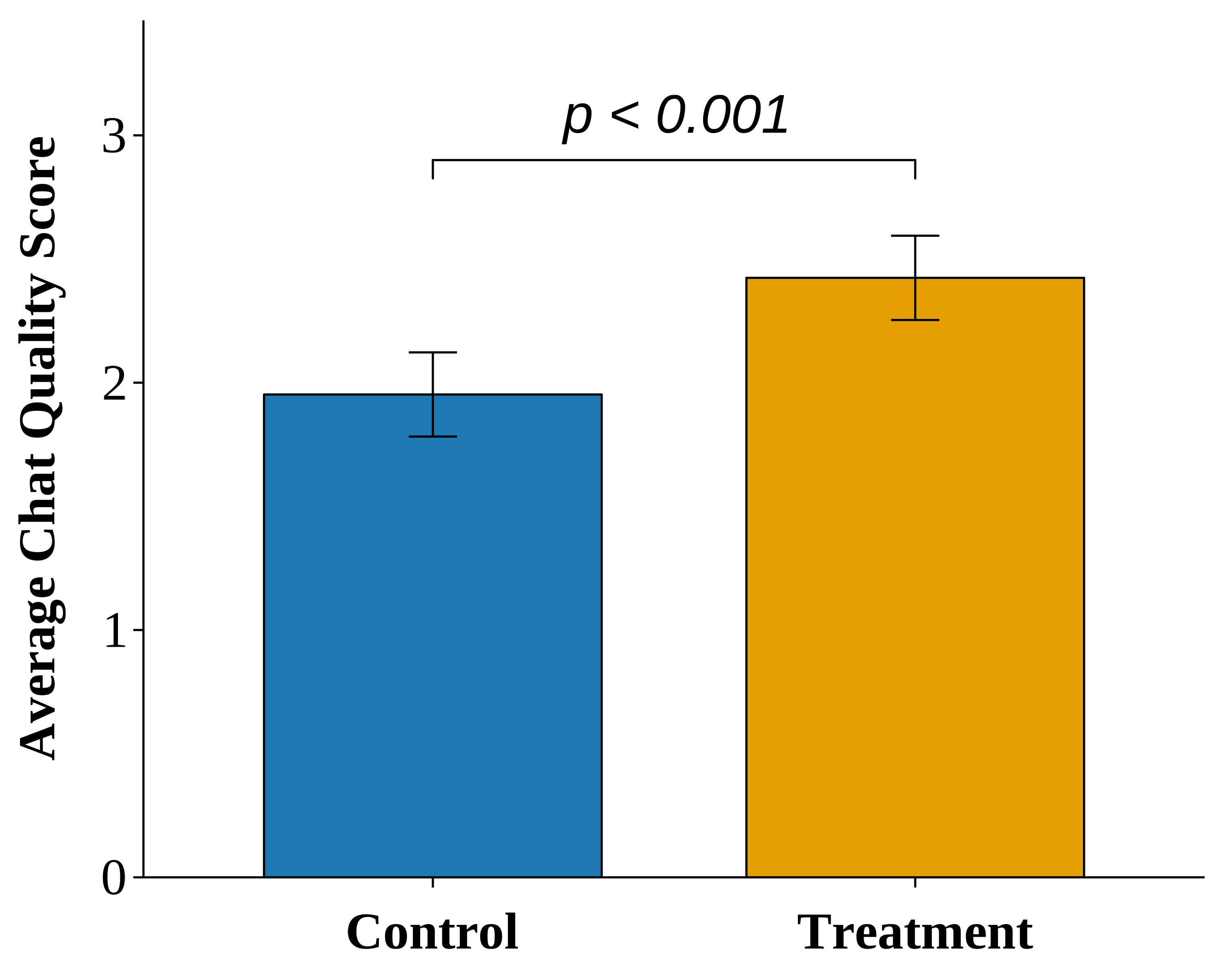}
\caption{Chat quality score}
\label{fig:chatquality}
\end{subfigure}
\hfill
\begin{subfigure}[t]{0.48\textwidth}
\centering
\includegraphics[width=\textwidth]{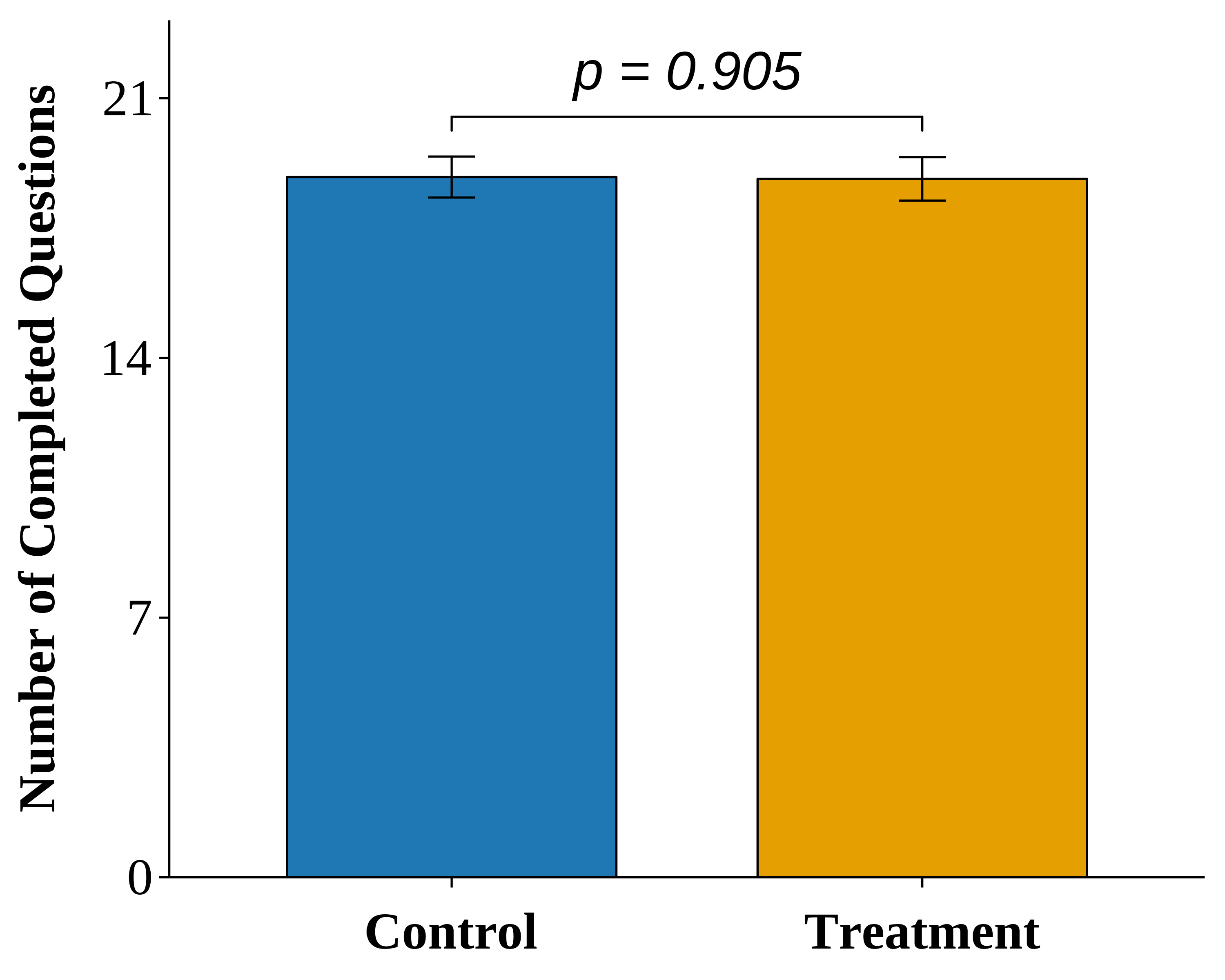}
\caption{Number of completed questions}
\label{fig:completedQ}
\end{subfigure}
\caption{\textbf{Engagement metrics by treatment assignment.} (a) Chat quality score comparing treatment and control groups, showing higher quality AI interactions in the treatment group. (b) Number of completed practice problems, showing similar completion rates across groups. Bars show group means with 95\% confidence intervals. Bracketed annotations report two-sided t-tests: chat quality shows a statistically significant difference ($p<0.001$), while completion rates do not differ significantly ($p=0.905$).}
\label{fig:engagement_metrics}
\end{figure}

\paragraph{Engagement mediates our performance improvement.} Next, we examine the mechanism by which personalized problem sequencing improves exam outcomes. Following a standard mediation analysis~\cite{imai2010general}, we study two engagement mediators: total time spent on the platform (\textit{totalTime}) and total number of attempts (\textit{totalAttempt}). We estimate the mediation models conditioning on the same rich set of observed controls as in our main analysis (e.g., baseline characteristics and fixed effects), which helps mitigate concerns that the mediator–outcome relationship is driven by observable differences in student background or program context. Our results show strong positive indirect effects through both measures (ACME: 0.185 SD via \textit{totalTime} and 0.149 SD via \textit{totalAttempt}, both $p<0.001$), while the direct effects are small and statistically insignificant (ADE: $-0.030$ SD, $p=0.690$; and 0.003 SD, $p=0.984$) (see Table~\ref{tab:med_results} in Supplement~\ref{sup:mediation}). Given a total effect of approximately 0.15 SD, these estimates imply that the treatment effect is almost fully mediated by increased engagement.\footnote{The estimated proportion mediated can exceed 1 in finite samples when the estimated direct and indirect components have opposite signs and/or due to sampling variability; we interpret these estimates as evidence that engagement accounts for essentially all of the total effect.}

Nevertheless, because engagement is not randomly assigned, mediator–outcome confounding may still arise from unobserved factors, so we do not interpret these mediation estimates as necessarily causal. To assess how strong such unobserved confounding would need to be to overturn our conclusions, we conduct sensitivity analyses that indicate the mediated effect would be eliminated only under moderate levels of unobserved confounding (see Supplement~\ref{sup:mediation} for details). These results support our hypothesis that personalized problem sequencing improves learning by sustaining student engagement throughout the semester. We provide detailed results and robustness checks, including an alternative path analysis specification using structural equation modeling (SEM), in Supplement~\ref{sup:mediation}.

\paragraph{Ruling out alternative explanations.}

Finally, we rule out two alternative explanations for our performance improvements. First, we show evidence that the gains are not driven by students simply completing more practice problems. We find that the average number of questions completed is nearly identical in the treatment and control groups (see Fig.~\ref{fig:completedQ}).\footnote{The majority of students complete no more than the required number of practice problems in the homework.} This result suggests that personalization benefits are derived from effort spent solving each problem instead of the number of problems solved---i.e., from increased practice quality instead of quantity.

Second, while the treatment group encounters harder questions on average, we find evidence that this increase is insufficient to explain our performance gains. We cannot perform either a regression or a mediation analysis on data from our main experiment, since practice problem difficulty is not assigned randomly; thus, we instead use data from our pilot study, where difficulty was assigned randomly.\footnote{Students used an earlier version of the platform that offered the same functionalities. Question difficulty was categorized into four discrete levels.} In a specification that does not allow for heterogeneity, average problem difficulty is not statistically significantly related to exam performance (see Table~\ref{tab:avgdif}, Column~1, where $\beta_{\text{Difficulty}}=0.417$, $p=0.283$). Furthermore, if we allow the effect to vary by prior coding background, then we find a significant positive interaction between difficulty and coding background (see Table~\ref{tab:avgdif}, Column~2, where $\beta_{\text{Difficulty}\times\text{PriorPython}}=2.542$, $p<0.05$). These results suggest that assigning harder questions disproportionately benefits students with prior Python skill, and does not help students without prior experience. In other words, problem sequencing must be personalized to provide benefits for all students.

\begin{table}[!t]
\centering 
\begin{threeparttable} 
\caption{\textbf{Regression summary: Effects of practice-question difficulty}}
\label{tab:avgdif}
\begin{tabular}{lcc} 
\toprule 
& \multicolumn{2}{c}{\textit{Dependent variable:}} \\
\cmidrule(lr){2-3}
& \multicolumn{2}{c}
{Standardized exam score} \\
& \multicolumn{1}{c}{(1)} & \multicolumn{1}{c}{(2)}\\
\midrule
Difficulty 
& 0.417 
& 0.271 \\
& (0.358) 
& (0.252) \\[2pt]

Prior Python 
& $0.909^{*}$
& -3.739 \\
& (0.375) 
& (1.598) \\[2pt]

Difficulty $\times$ Prior Python 
& 
& $2.542^{*}$ \\
& 
& (0.863) \\[2pt]
\midrule
Controls       & Yes                & Yes         \\
School fixed effect       & Yes                & Yes         \\

\midrule
Observations & 28 & 28 \\
Adjusted $R^{2}$ & 0.800 & 0.905 \\
\midrule
\bottomrule
\end{tabular}

\begin{tablenotes}[flushleft]
\footnotesize
\item \emph{Notes:} Standard errors in parentheses. $^{*}p<0.05$, $^{**}p<0.01$, $^{***}p<0.001$. All models include the same set of controls (AI tutor assignment, prior Python background, programming background, parent education, gender, and study majors) and school fixed effect (Supplement~\ref{sup:pilot}).
\end{tablenotes}
\end{threeparttable}
\end{table}


We provide details on these analyses in Supplement~\ref{sup:pilot}. Collectively, our results suggest that increased engagement rather than solving more practice problems or harder practice problems is the main driver behind our effect.

\section{Conclusion}

Our work demonstrates that the potential of GenAI tutors to improve educational outcomes lies not only in tutors based on chatbots, but by enabling students to engage productively with practice problems. To test this hypothesis, we have developed a GenAI tutoring system that combines a tutor chatbot with a reinforcement learning algorithm for personalized problem sequencing. Whereas prior algorithms are bottlenecked by coarse performance proxies, our algorithm translates complex behavioral traces including code edits and student-chatbot conversations into rich signals of student mastery. Based on this system, we have performed a randomized controlled trial involving 770 high school students, demonstrating that our system improved performance on an end-of-semester exam by 0.15 standard deviations---equivalent to as much as six to nine months of additional schooling by some estimates---without increasing instruction time or teacher workload. Importantly, our results suggest that these gains were driven by increasing student engagement with the practice problems. Maintaining student engagement has been a key challenge in education, especially for long-term educational technology programs. Our findings demonstrate how personalization---not of the tutor chatbot, but rather of the problem sequence---can improve student engagement in a way that benefits longer term outcomes.

More broadly, while our evaluation focused on introductory Python learning among high school students, introductory programming shares similar features with many other educational settings, including heterogeneity in baseline proficiency, steep learning curves, and the importance of productive struggle. Beyond traditional education settings, these features are also prominent in workplace reskilling, which is becoming an increasingly important problem due to the potential impact of GenAI on the labor market~\cite{eloundou2024gpts,morandini2023impact}. Our findings suggest that AI tutoring systems that tightly integrate task sequencing into chatbot tutors have the potential to enable effective and personalized learning/reskilling at scale.


\bibliography{scibib}

\bibliographystyle{Science}

\section*{End Notes}
\paragraph*{Acknowledgments}
We are grateful for the close partnership of the Department of Education of the Taipei City Government, the American Institute in Taiwan, American Innovation Center, and the local non-governmental organization Better Together for NextGen Taiwan. We acknowledge invaluable software development support from Bahati Tadjuidje Boniface, Jerry Chih-Yuan Huang, Zihan (Crescent) Xiong, Haosen Ge, and Hingis Chiao-Hsin Chang. We thank our research assistants---Allan Zhang, Cheng-Ying Wu, Harvani Sumawijaya, and Astrid (Pei-Chen) Hsu---as well as TA lead Hsiao-Chiao Lin, Chia-Chien Yang, and the many TAs and graders from the National Taiwan University. We especially thank the teachers at participating schools for their critical role in program implementation. We also thank Lizzy Pecora for administrative support, along with Eddy Yen-Ting Lin and Peter Jia-Wei Cui (Better Together NextGen Taiwan); Yu-Hsin Cho and Chiu-Fang Chiu of the Taipei City Government Department of Education---together with the Taipei City Government’s administrative staff, IT team, and government officials---for implementation and logistical support. Finally, we appreciate administrative support from Jerry Weng and the staff and government officials of the American Institute in Taiwan. 


\paragraph*{Funding:} This research was supported by generous funding from the Wharton AI \& Analytics Initiative, Wharton Mack Institute for Innovation Management, an Amazon Research Award (Fall 2023), the Taipei City Government, and the American Institute in Taiwan. 

\paragraph*{Author contributions:}
HB, OB, and AC contributed to the study design and methodology. HB, OB, AC, and BZ contributed to manuscript writing. AC and BZ analyzed the results under the supervision of HB and OB. AC and LK contributed to study implementation.

\paragraph*{Competing interests:}

There are no competing interests to declare.

\paragraph*{Ethics \& Inclusion Statement:} This research was approved by the University of Pennsylvania Institutional Review Board (Protocol \# 855951) and granted exempt status under Category 1. It was conducted in close partnership with the Taipei City Government and the American Institute in Taiwan, subject to a  formal Memorandum of Understanding (MOU) between the University of Pennsylvania and the Taipei City Government.  We collaborated closely with policymakers and frontline personnel, conducted training sessions, and aligned our approach with data privacy policy and existing government system and priorities. Roles and responsibilities were clearly defined in advance. 

\paragraph*{Materials \& Correspondence.}  Correspondence should be addressed to angelchg@wharton.upenn.edu, hamsab@wharton.upenn.edu, and obastani@seas.upenn.edu

\section*{Data and materials availability:}
To support further research, all data and code will be made available through a Github repository upon publication. Replication code and data is accessible here now: https://tinyurl.com/yx7vsrtc

\clearpage
\appendix

\begin{center}
\LARGE{\bf Supplementary Information}
\end{center}

\ref{supp:system_details} describes the personalized AI learning system. \ref{sup:rl} details the RL algorithm (POMDP formulation, belief updates, and MPC planning), along with supporting simulation results. Next,  \ref{sup:pilot} reports details of the pilot study, \ref{sup:expprotocol} reports the experimental protocol (pre-registration and implementation details), \ref{sup:databalanced} reports data processing and balance checks, and \ref{sup:eval} reports additional analyses supporting the main evaluation, including robustness checks, heterogeneity, mediation, and student feedback. We summarize related work in \ref{sup:literature} and provide course-related materials in \ref{sup:material}, including example practice problems and the certification exam.

\section{Personalized AI Learning System}
\label{supp:system_details}

\ref{sup:platform} introduces the user interface and function of our web platform; 
\ref{sup:rag} provides the RAG problem-generation pipeline; \ref{sup:aitutor} describes the embedded AI tutor.

\subsection{Web Platform}
\label{sup:platform}

Our web-based learning platform integrates lecture videos, coding practice problems, and AI tutoring into a single interface. It is integrated with the government's e-learning system (Taipei CooC-Cloud), which students access through government-authenticated accounts linked to their official academic records; this authentication mechanism mitigates spillover concerns as students cannot easily access other accounts, and furthermore, easily adheres to the Personal Data Protection Act (PDPA) by limiting researcher access to de-identified student information.

\begin{figure}[!t]
\centering
\includegraphics[width=0.75\columnwidth]{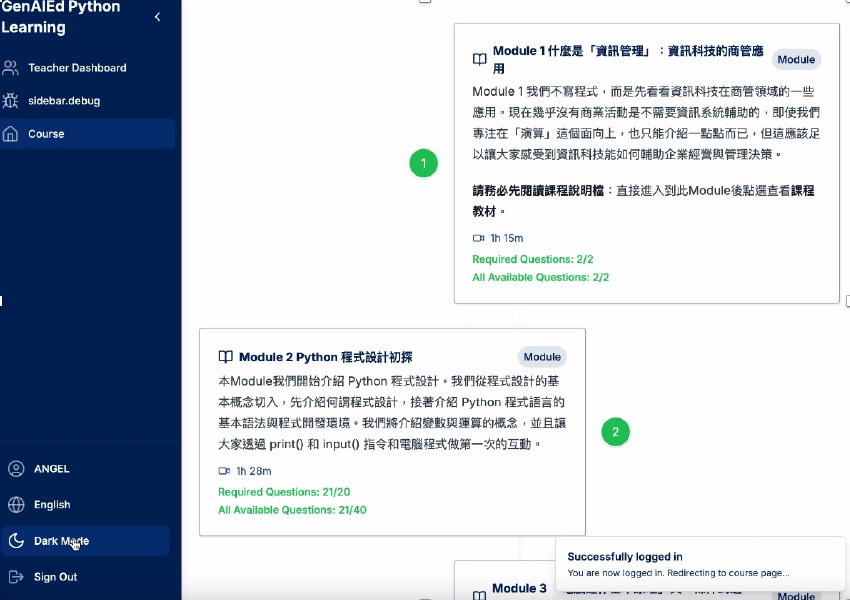}
\caption{\textbf{Menu of Modules on our AI Tutoring Platform}}
\label{fig:menumodule}
\end{figure}

After logging in, each student is taken to a course dashboard that lists all modules in the certificate curriculum (see Fig.~\ref{fig:menumodule}). Modules are released sequentially; a student can access module $m+1$ only after completing the required work in module $m$. This design enforces a common concept ordering while allowing within-module personalization of practice.

\begin{figure}[!t]
\centering
\includegraphics[width=0.75\columnwidth]{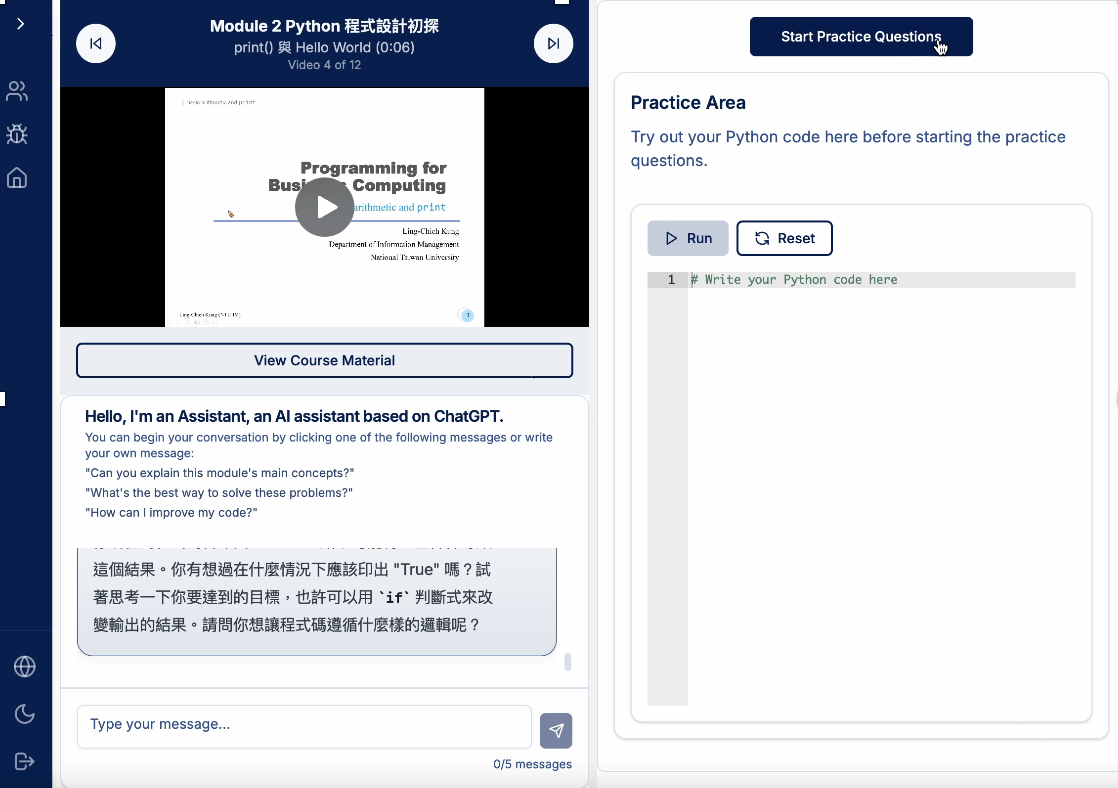}
\caption{\textbf{Learning Mode on our AI Tutoring Platform}}
\label{fig:instruction}
\end{figure}

Within a module, a student starts with a lecture video first. The main interface is divided into three functional areas. As Fig.~\ref{fig:instruction} shows, the upper left panel contains a video player that presents the instructor’s lecture for the module. In the upper right, students can work through
example problems that mirror those demonstrated in the video,
allowing learners to immediately practice the concepts they are seeing in the lecture without leaving the platform.

\begin{figure}[!t]
\centering
\includegraphics[width=0.6\columnwidth]{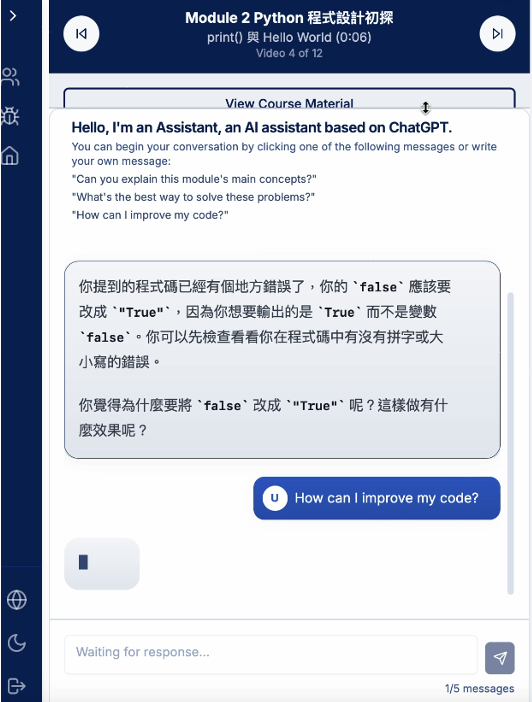}
\caption{\textbf{GenAI Chatbot Tutor on our AI Tutoring Platform.}}
\label{fig:aitutor}
\end{figure}

In the lower-left panel is an LLM-powered AI tutor that is available 24/7 (see Fig.~\ref{fig:aitutor}). Students can query the tutor with Python-related questions, such as asking for clarification of module concepts, hints on a problem, or explanations of error messages. The tutor is prompted to avoid providing direct answers, and to only provide hints after students have first demonstrated effort to solve the problem.

\begin{figure}[!t]
\centering
\includegraphics[width=0.75\columnwidth]{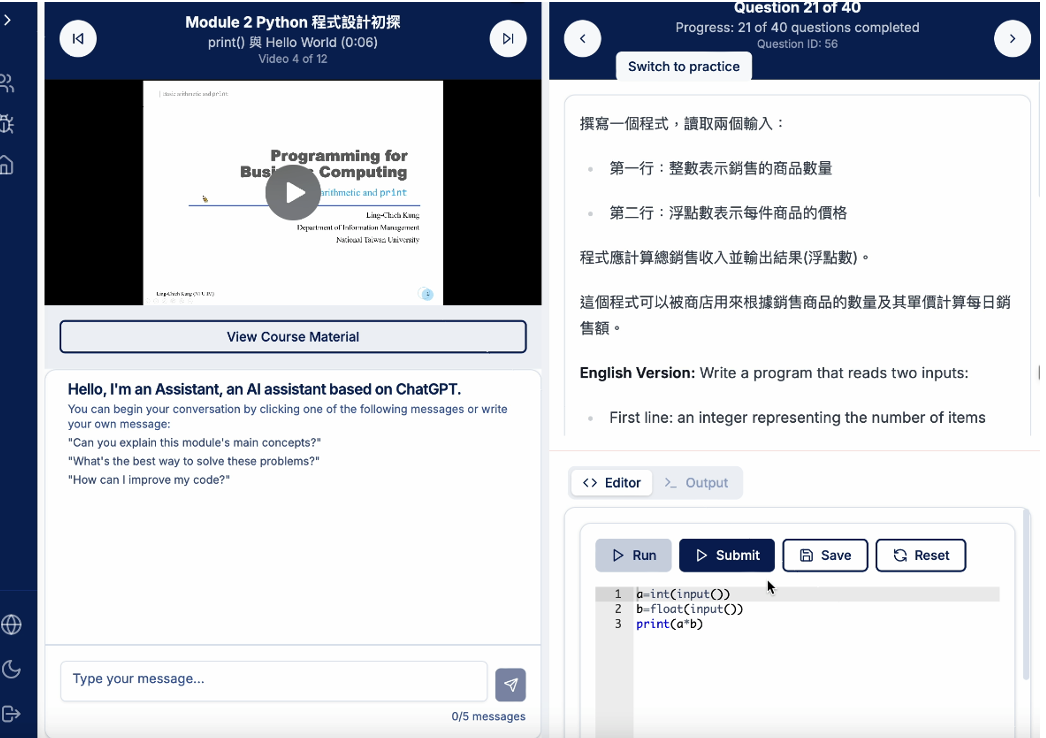}
\caption{\textbf{Practice Mode}}
\label{fig:practice}
\end{figure}

Once the student feels ready, they can click ``Start practicing" to enter the practice mode (see Fig.~\ref{fig:practice}). Practice problems appear one at a time and fall into three types (see examples in \ref{sup:practiceQ}): (i) coding questions, where students must write a Python
program to satisfy a set of hidden test cases; (ii) debugging questions, where students must identify and correct errors in provided code; and (iii) true/false questions assessing conceptual understanding. All work is completed directly in the browser-based editor; students do not need to open an external development environment. 

\begin{figure}[!t]
\centering
\begin{subfigure}[t]{0.48\textwidth}
\centering
\includegraphics[width=\textwidth]{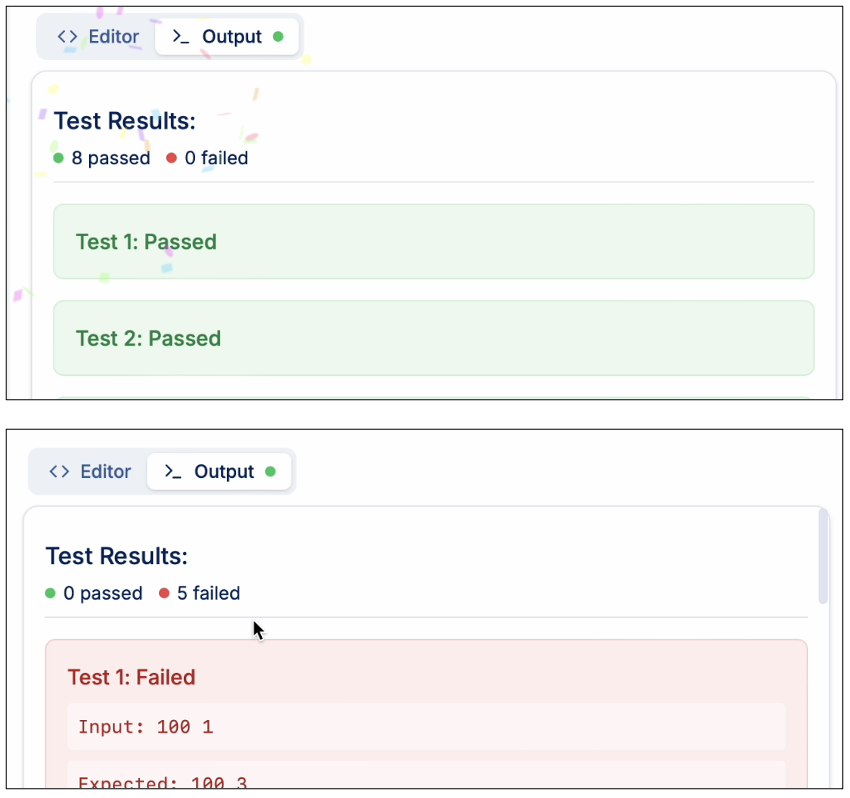}
\caption{Immediate practice feedback}
\label{fig:testfeedback}
\end{subfigure}
\hfill
\begin{subfigure}[t]{0.48\textwidth}
\centering
\includegraphics[width=\textwidth]{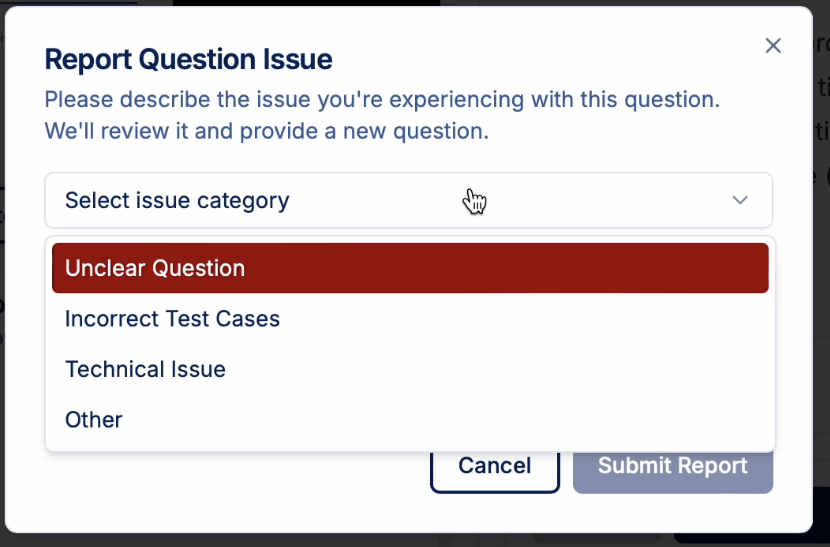}
\caption{Issue reporting}
\label{fig:reportissue}
\end{subfigure}
\caption{\textbf{Student practicing feedback and issue reporting interface.} (a) Students receive immediate feedback after they submit an answer. When tests fail, students can view detailed error information; (b) Students can use the report button to report any issues.}
\label{fig:testing_interface}
\end{figure}

For coding and debugging questions, the editor allows students to run code, save intermediate versions, and submit final answers. Submissions are executed on the server, and the platform returns immediate feedback on correctness; when a submission is incorrect, the interface reports failing test cases and/or runtime errors (see Fig.~\ref{fig:testfeedback}). Students can also report issues encountered during practice, to which the research team responded within 12 hours (see Fig.~\ref{fig:reportissue}).

Students can only move to the next question after answering the current question correctly. Each module contains a bank of 40 potential questions; homework assignments typically require each student to complete 15 questions. To ensure comparable exposure across experimental conditions, both treatment and control students were required to complete at least two ``very hard," two ``hard," and two ``medium" questions in each module. Aside from this constraint, for control students, the sequence follows a fixed ordering from easy to hard, and for treatment students, the RL policy selects the next question adaptively based on the inferred belief state.

The platform records a detailed, time-stamped log of all student activities, including code edits and saved versions, run and submission events, practice outcomes, navigation across modules and questions, and student's conversation history with the AI tutor. These data support both the construction of the learner model used by the RL policy and the
subsequent empirical analysis of learning behavior and outcomes.

\subsection{RAG System for Problem Generation}
\label{sup:rag}

\begin{figure}
\centering
\includegraphics[width=0.85\columnwidth]{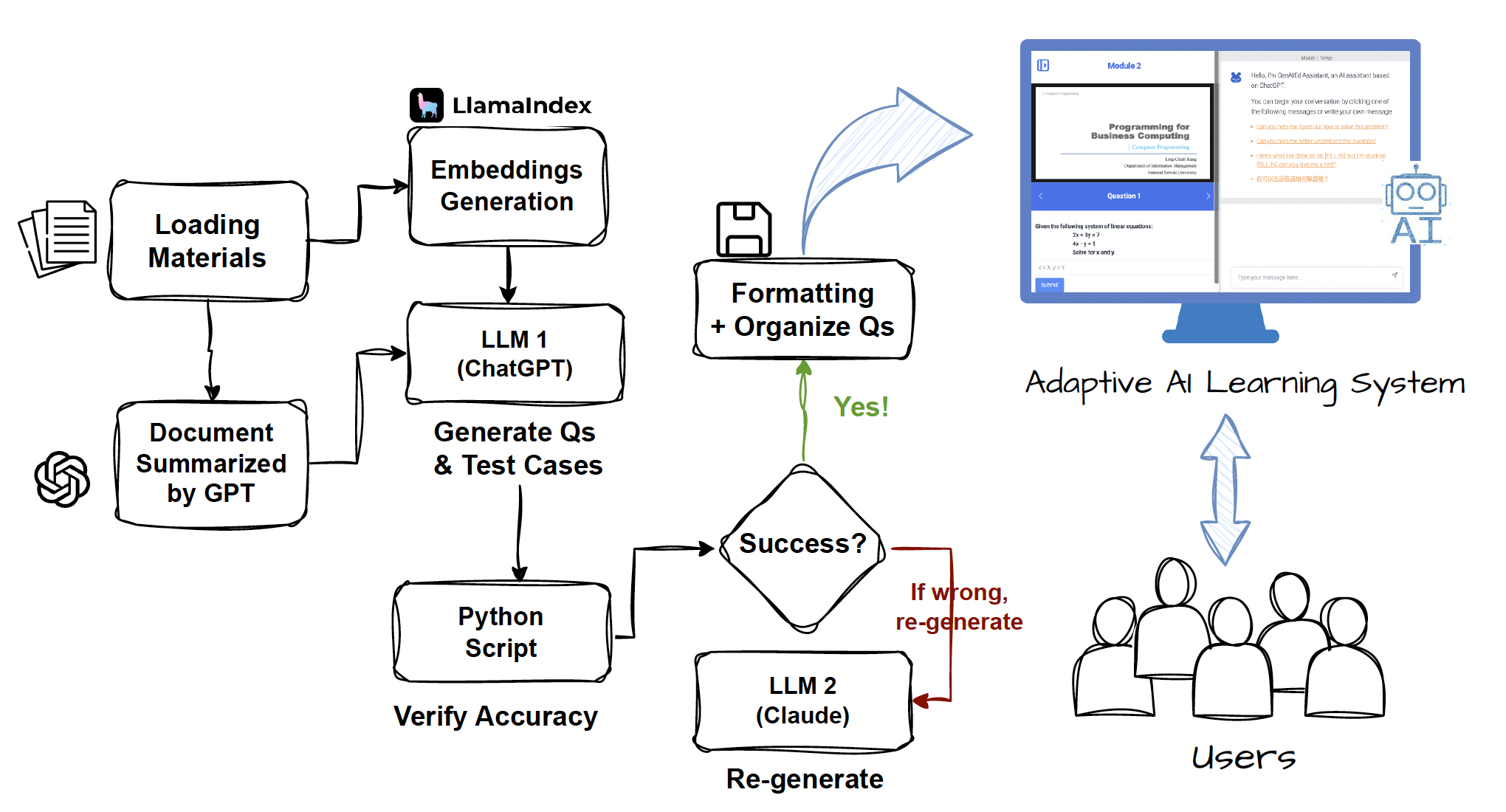}
\caption{\textbf{Human-AI RAG pipeline for practice problem generation.} Course materials and past exams are retrieved as reference contents for an LLM (ChatGPT) to generate programming questions, solutions, and test cases. Each generated output is validated using a Python script that checks functional correctness and consistency. When validation fails, a secondary LLM (Claude) iteratively revises the components until they pass. Teaching assistants then review finalized questions for clarity and quality before deployment.}
\label{fig:RAG}
\end{figure}

We generate Python programming questions based on instructional materials using the RAG-based LLM pipeline shown in Fig.~\ref{fig:RAG}. For each generation request, we prompt the LLM with (i) the main instructional content, (ii) optional additional content (e.g., prior modules), and (iii) a list of existing questions to discourage duplicates. The model is instructed to output a JSON object containing two coding questions (in English with a traditional Chinese translation), along with a template solution stub, a working sample solution, and a comprehensive assert-based test suite. To encourage variety in the generated questions (e.g., different scenarios and wording), we use sampling settings that allow some randomness  (temperature $=$ 1.0, top-$p=$ 0.7, top-$k=$ 200). 

The system prompt provides the educational content and high-level role instructions, while the user prompt specifies the required output format and constraints (difficulty levels, bilingual problem statements, and test-suite requirements); we show the system prompt in Figure~\ref{fig:ragsystemprompt} (we omit the user prompt due to its long length). We generate questions in batches of two.

\begin{figure}
\begin{promptbox}{System prompt}
You are a knowledgeable AI assistant designed to help generate questions based on educational content.
Main content:
{MAIN_DOCUMENT_TEXT}

Additional content:
{ADDITIONAL_DOCUMENT_TEXT}

Your goal is to create well-structured, meaningful, and educational coding questions based on the content provided.
You may combine the concepts taught in the additional contents with the main content for generating unique and insightful questions.
You should always ensure the correctness of your output and follow the question generation format carefully.
When answering, adhere strictly to Python programming conventions and ensure the output format is clean and structured.
\end{promptbox}
\caption{\textbf{System prompt for RAG practice problem generation pipeline.}}
\label{fig:ragsystemprompt}
\end{figure}

\subsection{GenAI Chatbot Tutor}
\label{sup:aitutor}

\paragraph{Tutor integration and shared context.}

Students interacted with our GenAI tutor (powered by OpenAI’s GPT-4o model) embedded in the learning platform during practice. The tutor was available in both the treatment (RL sequencing) and control (fixed sequencing) conditions, and the tutoring model and prompting pipeline were held constant across conditions.

At each practice step, the tutor received a user prompt formed by concatenating the student’s query with the current practice problem, relevant metadata (e.g., difficulty), and the sample solution code and/or answer key. The tutor was instructed in the system prompt (see Fig.~\ref{fig:baseprompt}) to keep responses brief and to explain concepts at a middle-school level. This design enabled the tutor to provide accurate, targeted hints while enforcing a strict ``no direct answers'' policy. 

\begin{figure}
\begin{tcolorbox}
\textbf{Base system prompt.}
\begin{quote}\ttfamily
You are a Python tutor. Your goal is to help students develop a deeper understanding of Python while keeping them focused on learning. Keep responses brief, concise, and to the point---exercise restraint.\\
Explain to me like I'm a middle school student and make sure to use the solution code provided in the context.
\end{quote}
\end{tcolorbox}
\caption{\textbf{Common base system instruction (used in all conditions)}}
\label{fig:baseprompt}
\end{figure}

\paragraph{Pedagogy randomization.}
The focus of this paper is on RL-based sequencing of practice problem difficulty. However, we also randomized the AI tutor’s system prompt in this study (see our pre-registered plan in \ref{sup:preregist}). 
All our pedagogy prompts shared the following instructions: (i) do not provide direct answers or full code; (ii) diagnose whether the student's difficulty is problem formulation versus Python syntax/logic and respond accordingly; (iii) keep responses concise; (iv) maintain role consistency; and (v) respond in Traditional Chinese when the student writes in Chinese. They only differed in pedagogical emphasis. We find no meaningful or statistically significant differences across these pedagogy conditions in exam performance or engagement outcomes; accordingly, we do not focus on pedagogy comparisons in this paper. We show that our results are robust to this additional randomization in \ref{sup:main}.

\section{Reinforcement Learning for Problem Sequencing}
\label{sup:rl}

In this section, we describe our reinforcement learning algorithm for adaptively sequencing student practice problems. At a high level, we use a model-based reinforcement learning algorithm, where we first predict the parameters $\theta$ of a Partially Observed Markov Decision Process (POMDP), and then use optimal control within this POMDP for problem sequencing. Each POMDP targets a single student-module pair---i.e., the sequence of problems that a specific student works on for a specific module. We train a machine learning model to predict the parameters of this POMDP based on historical data, using features such as student characteristics and their progression in previous modules.

Critically, LLMs perform several key roles in this algorithm: (1) they form the basis of the observation space of our POMDP, analyzing student code submissions to accurately measure progression, (2) they are used to construct several key features for the machine learning model, including analysis of student-tutor interactions. We provide simulation evidence that these features play a critical role in improving the performance of our algorithm, highlighting the benefits of tight integration of AI tutors and POMDP-based problem sequencing.

Due to the limited availability of pilot data, many of our design decisions had to be made on the basis of intuition. Thus, the primary validation of our reinforcement learning algorithm is our randomized controlled trial. Nevertheless, we provide accuracy and simulation-based results to help justify some of these decisions.

\ref{sup:pomdp} describes the POMDP design; \ref{sup:pilotestimation} details our model-based reinforcement learning approach; \ref{sup:simulation} outlines the simulation setup and baseline implementations; and \ref{sup:simresult} reports the simulation results, including empirical evidence for the value of using LLM-derived signals to estimate learning progression.

\subsection{Partially Observable Markov Decision Process (POMDP)}
\label{sup:pomdp}

We begin by providing a high-level overview of the POMDP, and then provide details on each component of the POMDP.

\subsubsection{Overview}

Modeling a student's learning progression is challenging because knowledge is \emph{latent} (i.e., unobserved); it evolves with practice and generates only noisy observations (e.g., how long they take to solve a problem). Partially observed Markov Decision Processes (POMDP) have been applied to solving this challenge. Following this line of work, we formulate the adaptive sequencing of our practice problems as a POMDP. This framework enables the system to select the next difficulty level while uncertain about the student's underlying state, refining its estimate via Bayesian filtering as new interaction data arrive.

\begin{figure}[!t]
\centering
\includegraphics[width=0.7\columnwidth]{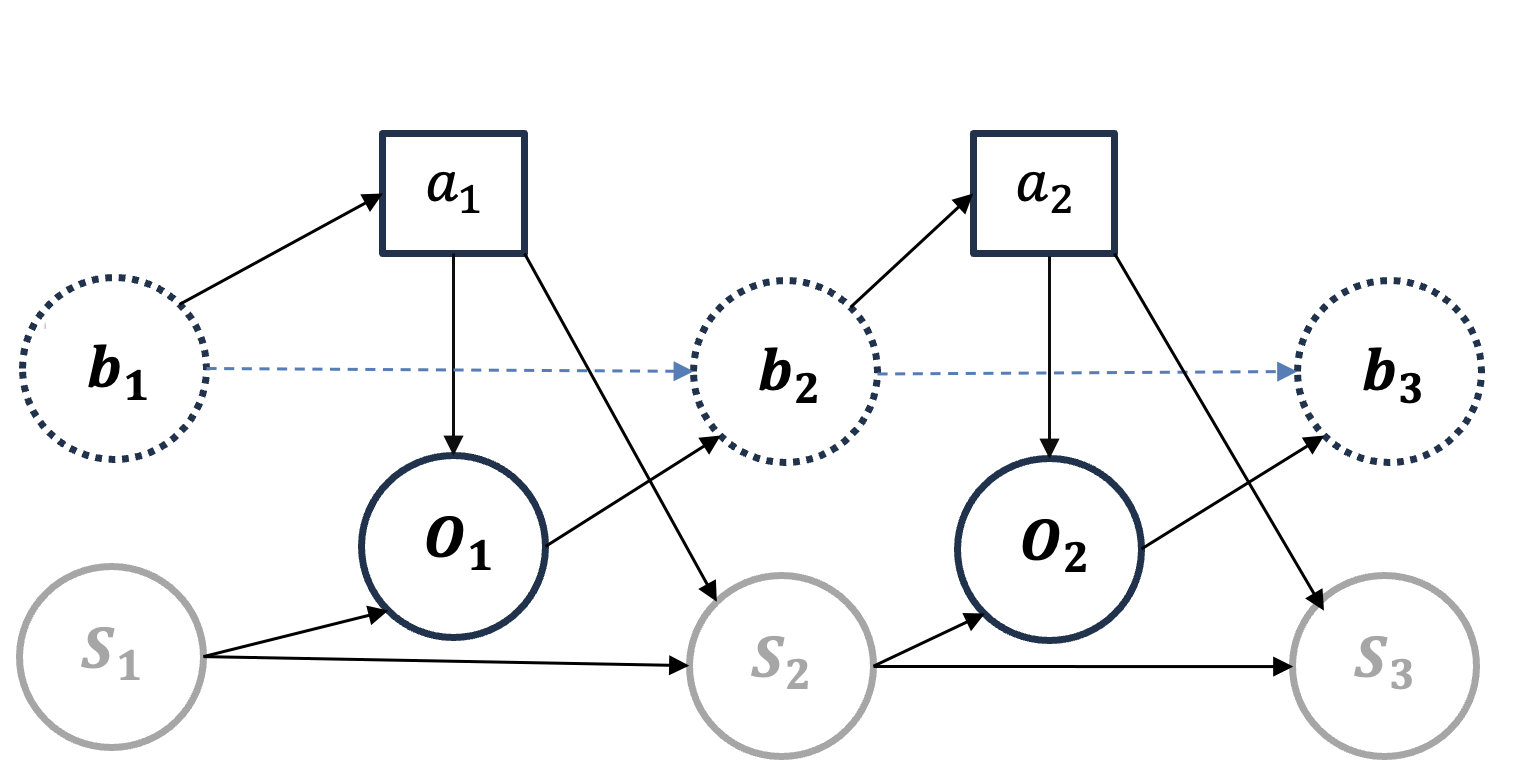}
\caption{\textbf{POMDP-Based Reinforcement Learning System}}
\label{fig:pomdp}
\end{figure}

Formally, our POMDP consists of: (i) a latent learner state $s_t$, (ii) an action $a_t$ selecting the difficulty, (iii) an observation $o_{t+1}$ summarizing the student's effort and performance, and (iv) a belief state $b_t$ representing a probability distribution over plausible learner models. At each decision epoch $t \in \{0, 1, \dots, T_{\max}-1\}$, the policy selects a \emph{difficulty level} for the next practice problem (in our setting, Easy/Medium/Hard/Very Hard) based on the current belief state $b_t$. After the student attempts the problem, the platform observes $o_{t+1}$ and computes the belief state $b_{t+1}$ for the next step. Fig.~\ref{fig:pomdp} illustrates this system.

\osbertlater{Can we give some numbers on all of this stuff?}
Our modeling choices diverge from standard approaches in the literature like Bayesian Knowledge Tracing (BKT). BKT uses a very simple POMDP that includes binary performance feedback (whether the student successfully solves the problem) to track a binary mastery indicator (whether the student has learned the concept). In our setting, however, binary feedback is uninformative: students iterate until they reach a correct solution, making eventual correctness universal; furthermore, correctness on the very first try is rare (due to syntax or other superficial errors). To obtain a signal that meaningfully tracks progress, we could use the number of attempts required to reach a solution. However, the raw attempt counts (i.e., how many times the student submitted the code to our server to check for correctness) can be inflated by superficial resubmissions, with variation across students on how frequently they make submissions (see our analysis in \ref{sup:simresult}).

To this end, we need to filter attempts to count \emph{meaningful} submissions. Na\"{i}ve metrics such as edit distance may not accurately capture whether one code submission has substantially semantic differences from another. Thus, a key feature of our system is to instead employ an LLM to identify meaningful attempts from the student's code edit trajectory. Specifically, we prompt an LLM to take the submission history as input and output whether the latest one is meaningful according to a set of guidelines. Given a sequence of student submissions, we provide this sequence to the LLM and ask it to count the number of meaningful code submissions; we show the prompt we use to do so in Figure~\ref{fig:meaningfulsubmissionsprompt}.

\begin{figure}[!t]
\begin{tcolorbox}[
    colback=gray!10, 
    colframe=gray!60, 
    title=\textbf{Prompt for counting meaningful attempts},
    fonttitle=\bfseries,
    arc=1mm
]
\ttfamily\small 
You are given a list of coding history for a user which contains various code submissions the question below.

\vspace{0.5em}
\{coding\_history\}

\vspace{0.5em}
\textbf{Question:}\\
\{question.content\}

\vspace{0.5em}
Count the number of meaningful code submissions in the user's history. 

For each code submission in the history, compare it with its previous submission and count it as 1 ONLY if it introduces a meaningful change in terms of logic, approach, or functionality.

The same code as the previous submission or like small typos, formatting, or minor variable name changes without changing logic, or provide empty or nonsense code should be counted as 0.

The final count must be an integer between 0 and \{len(coding\_history\_list)\}. Return only the integer value with no additional text.
\end{tcolorbox}
\caption{\textbf{Prompt used to ask an LLM to count the number of meaningful submissions.}}
\label{fig:meaningfulsubmissionsprompt}
\end{figure}

This design naturally leads to a discrete observation space $o_t\in\mathbb{N}$; we considered a number of different distributions, and found that Poisson distributions best matched our pilot data.
\osbertlater{Can we give details?}
To handle the increased complexity of the observation space, we also need to increase the complexity of the latent state; otherwise, the latent state would fail to accurately capture variation in observations. Since Poisson distributions have a single real-valued parameter, we correspondingly use a real-valued latent state $s_t\in\mathbb{R}$.

The transition function for our POMDP is hand-designed to match patterns observed in the historical data, including the following salient features: (i) students start from some high initial number of attempts, (ii) students gradually make fewer attempts as they solve problems, and (iii) students converge to some baseline number of attempts. In addition, our transitions assume that students learn more when given harder problems.

Finally, the main parameters of our POMDP are the student's initial state, learning speed, and asymptotic mastery. Given these parameters, the state transitions are deterministic; thus, our belief state is a distribution over these parameters. For state estimation, we use a modified version of a standard Bayesian particle filter for belief state estimation; roughly speaking, the particles come from historical data (specifically, from our pilot),
and we maintain a distribution over these particles. Then, belief state estimation amounts to performing posterior updates to this distribution. For control, we use a stochastic model predictive control strategy, where we estimate the value of each action using Monte Carlo rollouts and then select the action with the best estimated value.

\subsubsection{Latent States, Actions, Transitions, and Observations}
\label{sup:obs}

The design of our POMDP is guided by a key insight from our pilot data---namely, the distinction between ``fast'' and ``slow'' learners (here, we use the term ``learner'' to refer to a student-module pair). Even among students who ultimately performed well, we observed substantial heterogeneity in the speed of progression through difficulty levels. Slower progression did not imply lower capability, but instead reflected a different learning speed that required more practice opportunities before advancing. This pattern aligns with Mastery Learning~\cite{bloom1968learning} and Carroll's Model of School Learning~\cite{carroll1963model}, which emphasize that aptitude is closely related to the time required to learn rather than whether learning is possible.

Designing a POMDP with a real-valued latent state can lead to a complicated belief state estimation procedure. Thus, to simplify the design of our POMDP, we encapsulate the uncertainty over the latent state in a set of time-invariant parameters $\theta$, which are specific to the current learner (i.e., student-module pair). For a fixed $\theta$, the POMDP transitions are deterministic, depending only on the selected actions. With these two choices, the belief state can be expressed purely as a distribution over $\theta$, with the corresponding distribution $s_t$ over the latent state being derived by simulating the deterministic POMDP transitions for the chosen actions. The parameter vector
$\theta=(c_1,c_2,\tau)$ includes the following information:
\begin{itemize}
\item $c_1$ controls the learner's \emph{initial mastery}
\item $c_2$ is the learner's long-run \emph{mastery ceiling} (the asymptotic attempt rate under mastery)
\item $\tau$ controls the speed at which mastery improves (note that lower latent state corresponds to higher mastery).
\end{itemize}
These parameters are estimated from normalized historical data and lie within
the compact set
\begin{equation*}
c_1, c_2 \in [1.0, 5.0],
\qquad
\tau \in [0.1, 0.5].
\end{equation*}
Learning proceeds over a finite horizon of at maximum $T_{\max}=40$ steps.\footnote{40 is the maximum number of questions available for each module.} The latent effort variable 
$\mathrm{SE}_t$ 
denotes the learner's underlying expected effort required at time $t$ to solve a practice problem correctly (e.g., expected number of meaningful attempts on a reference task); importantly, lower values indicate higher proficiency. The full latent state is
\begin{equation*}
s_t = (\theta, t, \mathrm{SE}_t) = (c_1,c_2,\tau,t,\mathrm{SE}_t).
\end{equation*}
The initial effort level is specified as
\begin{equation*}
\mathrm{SE}_0 = c_1 + c_2.
\end{equation*}
At each step $t$, the platform selects a difficulty level
\begin{equation*}
a_t \in \mathcal{A}=\{0,1,2,3\},
\end{equation*}
with larger values indicating harder questions. Conditional on $(s_t,a_t)$, the latent effort evolves according to the transition function
\begin{equation}
\mathrm{SE}_{t+1}
= \bigl(\mathrm{SE}_t - c_2\bigr)\exp(-\Delta_t) + c_2,
\label{eq:transition}
\end{equation}
where
\begin{equation}
\Delta_t
= \tau (\eta+a_t\delta_{\text{knowledge}}),
\label{eq:delta}
\end{equation}
where $\eta=1.2$,  $\delta_{\text{knowledge}} = 1$.
Because 
$\mathrm{SE}_0>c_2$ and $\Delta_t>0$
, Eq.~\eqref{eq:transition} says the expected effort 
$\mathrm{SE}_t$
(e.g., expected number of attempts on a reference task) required to achieve a correct solution decreases monotonically toward $c_2$, with faster convergence for larger $\tau$ and higher assigned difficulty $a_t$.

After the learner interacts with the practice problem chosen at time $t$, the platform observes a noisy effort signal $o_{t+1}\in\mathbb{N}_0$. In our implementation, $o_{t+1}$ is the LLM-estimated number of \emph{meaningful} attempts required to reach a correct solution (i.e., substantively distinct solution attempts), which filters out superficial resubmissions and better reflects the student’s cognitive effort and learning progression. We model $o_{t+1}$ as a Poisson random variable, which is well-suited for nonnegative count data and captures the empirical pattern that effort becomes both larger on average and more variable for harder tasks:
\begin{equation*}
o_{t+1}\mid s_{t+1},a_t \sim \mathrm{Poisson}(\lambda_{t+1}),
\end{equation*}

with rate
\begin{equation}
\lambda_{t+1} = \max\{\text{SE}_{t+1}(1 + a_t)(1-\text{slip}), \epsilon\}
\label{eq:obs-rate}
\end{equation}
where $\text{slip}=0.1$ denotes the probability of an incorrect response despite having the required knowledge, and $\epsilon=10^{-6}$ is a small constant for numerical stability.

Thus, harder tasks (i.e., larger $a_t$) tend to induce larger observed effort, all else being equal. Under the Poisson model, larger $\lambda_{t+1}$ implies both a higher mean and a higher variance, which parsimoniously reflects increased dispersion in attempt counts on more difficult items.

\subsubsection{Belief State and Belief State Estimation}

The true state is unobserved because $\theta$ and 
$\mathrm{SE}_t$
are latent. Thus, we maintain a belief distribution over the learner's latent
parameters $\theta$ and propagate an associated derived effort trajectory. For computational feasibility, we use a standard particle filtering algorithm for belief state estimation. Concretely, we approximate the belief state over the hyperparameter $\theta$ based on a finite set of plausible hyperparameter choices (called \emph{particles}):
\begin{equation*}
\theta^{(j)} = \left(c_1^{(j)},c_2^{(j)},\tau^{(j)}\right),
\qquad j\in[n]=\{1,\dots,n\}.
\end{equation*}
These particles are constructed from historical data, as we describe in \ref{sup:pilotestimation}. The underlying effort 
$\mathrm{SE}_t^{(j)}$ 
is updated online for each particle via the transition function; that is, the belief space is approximated by propagating particles over $\theta$ according to the transition function. In more detail, we represent the belief state $b_t$ at time $t$ as a distribution over the time-invariant parameters $\theta$, with
\begin{equation}
b_t(\theta) \approx \hat{b}_t
= \sum_{j=1}^n p_t^{(j)}  \delta_{\theta^{(j)}}(\theta),
\qquad
p_t^{(j)} \ge 0,
\qquad
\sum_{j=1}^n p_t^{(j)} = 1,
\label{eq:belief-theta}
\end{equation}
where $\delta_{\theta^{(j)}}$ is the Dirac measure at $\theta^{(j)}$. The corresponding distribution over the latent state is
\begin{equation*}
s_t^{(j)}=\left(\theta^{(j)},t,\mathrm{SE}_t^{(j)}\right).
\end{equation*}
It remains to describe how our belief state is initialized and updated. First, the initial belief state $\hat{b}_0$ is based on predictions from historical data, as described in \ref{sup:pilotestimation}. Next, given belief $\hat{b}_t$, action $a_t$, and observation $o_{t+1}$, we update particle weights via three steps:
\begin{itemize}
\item \textbf{Step 1: Predictive propagation of derived effort.} For each particle $j$ with state
\begin{equation}
s_t^{(j)}=\left(\theta^{(j)},t,\mathrm{SE}_t^{(j)}\right),
\end{equation}
we propagate its derived effort using the transition function:
\begin{equation}
\mathrm{SE}_{t+1}^{(j)}
= \left(\mathrm{SE}_t^{(j)} - c_2^{(j)}\right)
\exp\left(-\Delta_t^{(j)}\right)
+ c_2^{(j)},
\end{equation}
where
\begin{equation}
\Delta_t^{(j)}
= \tau^{(j)} \left(\eta + a_t\delta_{\text{knowledge}}\right).
\end{equation}
The particle $\theta^{(j)}$ itself remains unchanged, while 
$\mathrm{SE}_t^{(j)}$
evolves deterministically given $a_t$.
\item \textbf{Step 2: Likelihood weighting using the Poisson observation model.} For each particle, the corresponding Poisson rate is
\begin{equation}
\lambda_{t+1}^{(j)}
= \max\{\mathrm{SE}_{t+1}^{(j)}
  (1 + a_t)(1-\text{slip}), \epsilon\},
\end{equation}
so the likelihood of observing $o_{t+1}$ is
\begin{equation}
\ell^{(j)} 
= P(o_{t+1} \biggm\vert \theta^{(j)}, \mathrm{SE}_{t+1}^{(j)}, a_t)
= \frac{\left(\lambda_{t+1}^{(j)}\right)^{o_{t+1}}
\exp\left(-\lambda_{t+1}^{(j)}\right)}
{o_{t+1}!}.
\end{equation}
We compute unnormalized posterior weights
\begin{equation}
\tilde{p}_{t+1}^{(j)}
= p_t^{(j)}  \ell_j,
\qquad j=1,\dots,n,
\end{equation}
and normalize:
\begin{equation}
p_{t+1}^{(j)}
= \frac{\tilde{p}_{t+1}^{(j)}}{
\sum_{j'=1}^n \tilde{p}_{t+1}^{(j')}
},
\qquad j=1,\dots,n.
\end{equation}
\item \textbf{Step 3: Posterior belief update.} The updated belief over $\theta$ is again represented as
\begin{equation}
b_{t+1}(\theta) \approx \hat{b}_{t+1}(\theta)
= \sum_{j=1}^n p_{t+1}^{(j)} \delta_{\theta^{(j)}}(\theta),
\end{equation}
and posterior expectations of any function $h(\theta)$ are approximated by
\begin{equation}
\mathbb{E}_{b_{t+1}}[h(\theta)]
\approx \sum_{j=1}^n p_{t+1}^{(j)} h\bigl(\theta^{(j)}\bigr).
\end{equation}
\end{itemize}

\subsubsection{Reward Function}


Our ultimate objective is to help learners achieve mastery of the material. While this goal can be captured by assessing whether the latent state converges to its asymptotic value, such an objective misses many nuances about whether the learner was engaged and whether they found the practice problems to be challenging yet rewarding to solve. These nuances are all critical for sustaining learning.

Thus, we instead design a reward function that captures the idea that students should solve problems at a difficulty level appropriate for their current skill level, which has been extensively studied as the \emph{zone of proximal development (ZPD)}~\cite{vygotsky1978mind}. There is substantial evidence that keeping students in the ZPD maintains engagement while improving learning. Concretely, our reward function specifies the best difficulty level for matching the learner's current skill level, thereby increasing difficulty at a pace that reflects their personal learning speed $\tau$. An additional advantage of this approach is that it provides a shaped reward function that guides our control algorithm described in \ref{sup:approx-planning}.


A challenge with this approach is that it requires substantial human expertise to specify the rate of progression. The pacing encoded by our reward function is based on expert pedagogical knowledge and was calibrated using pilot study trajectories, where we observed the variation in how quickly students progressed even among those who performed well. Our pacing is calibrated to accommodate both fast and slow learners, and to avoid penalizing students who require more attempts in the short run but are on track to achieve mastery.

Specifically, let 
$\mathrm{SE}_t$
be a measure of required effort to succeed where \emph{lower is better}. We define normalized skill 
$\mathrm{SE}_t$ as: 
\begin{equation}
\text{NSE}_t = \frac{\text{SE}_t - c_2}{c_1}
\end{equation}
so lower 
$\mathrm{NSE}_t$
corresponds to fewer attempts/time and thus higher proficiency. Then, we compare 
$\mathrm{NSE}_t$
to calibrated thresholds
\begin{equation*}
(\text{threshold}_1,\text{threshold}_2,\text{threshold}_3)=(0.2,0.4,0.6),
\end{equation*}
where the thresholds are hyperparameters that we manually selected based on the pilot data. Then, we define the desired (appropriate) difficulty level as
\begin{equation}
a_{\mathrm{exp}}
= \sum_{i=1}^3
\mathbbm{1}\!\left(
\mathrm{NSE}_t \le \text{threshold}_i\right),
\label{eq:aexp}
\end{equation}
so lower 
$\mathrm{NSE}$ corresponds to higher desired difficulty (from 0 to 3). Finally, the reward for choosing action $a_t$ at time $t$ is
\begin{equation}
r_t
=
\begin{cases}
1
& \text{if } a_t = a_{\mathrm{exp}}, \\
\max\left\{0, 1 - 0.25 \cdot |a_t - a_{\mathrm{exp}}|\right\}
& \text{otherwise}.
\end{cases}
\label{eq:reward}
\end{equation}
This function assigns a full reward of 1 when the chosen difficulty matches the desired level,
and penalizes deviations linearly in $\lvert a_t-a_{\mathrm{exp}}\rvert$. The corresponding objective for a control policy $\pi$ is to
maximize the expected cumulative reward:
\begin{equation}
J(\pi)
= \mathbb{E}_\pi \left[ \sum_{t=0}^{T_{\max}-1} r_t \right]
\end{equation}
where the expectation is taken with respect to the Markov chain induced by taking actions according to $\pi$.

\subsubsection{Model Predictive Control}
\label{sup:approx-planning}

Next, we describe our control algorithm for selecting actions (i.e., practice problem difficulty levels) to maximize our POMDP objective.
Computing the exact optimal solution is computationally intractable due to the large state space~\cite{papadimitriou1987complexity}. Thus, we select actions online using model predictive control (also called receding-horizon control) in conjunction with Monte-Carlo rollouts~\cite{tesauro1996online}. Recall that at step $t$, the belief state $b_t$ is represented by a distribution over particles, which we denote
$\{(s_t^{(j)},p_t^{(j)})\}_{j=1}^n$. To select the next action, we evaluate each candidate action $a\in\mathcal A$ by estimating its expected
cumulative shaped reward over a lookahead horizon $H$. Specifically, for each
action $a$, we sample $M$ Monte-Carlo rollouts. Specifically, for each rollout $m\in[M]$ (we use $M=50$), we first
sample an initial latent state index
$j\sim\text{Categorical}(p_t^{(1)},\dots,p_t^{(n)})$ and set the simulation
state $s_{t,0}^{(m)} = s_t^{(j)}$. Then, we simulate the POMDP over $H=10$ steps by (i) taking action $a$ at the first step and following a
handcrafted policy $\tilde\pi$ thereafter; (ii) propagating the latent dynamics via \eqref{eq:transition}--\eqref{eq:delta}; (iii) sampling observations from the Poisson model \eqref{eq:obs-rate}; and (iv) accumulating the shaped rewards defined in \eqref{eq:reward}. For $\tilde\pi$, we randomly sample a state from the current belief state and then act optimally according to that state. This strategy yields the following Monte-Carlo value estimate:
\begin{equation}
\widehat Q_t(b_t,a)
= \frac{1}{M}\sum_{m=1}^M \sum_{h=0}^{H-1} \gamma^hr_{t+h}^{(m)},
\end{equation}
where $\gamma$ is the discount factor (we use $\gamma=1$ since our horizon is short). We select the action maximizing this value:
\begin{equation}
a_t^* \in \arg\max_{a\in\mathcal A} \widehat Q_t(b_t,a),
\end{equation}
execute $a_t^*$ on the platform, and update the belief given the realized observation $o_{t+1}$ (\ref{sup:obs}). We re-plan at every time step using the current belief state.


\subsection{Model-Based Reinforcement Learning}
\label{sup:pilotestimation}

A key remaining question is what to use as the initial belief state over the POMDP parameters $(c_1,c_2,\tau)$. One strategy is to use a fixed choice across all learners, relying on the POMDP to update them correctly over time. However, the time horizon of our POMDP is relatively short, leaving little time to obtain accurate estimates for choosing high-reward actions. Thus, we use a model-based reinforcement learning approach, where we predict the parameters from historical data, and form an initial belief state based on these predictions.

Specifically, we train a machine learning model to predict the POMDP parameters based on features derived from behavioral logs from earlier modules, including time-stamped code edits, coding history, and AI-tutor interactions.
For our machine learning model, we use a gradient-boosted model (GBM), which achieved the best predictive performance among the models we evaluated. Specifically, we evaluated the performance of different models on data from our pilot; results in \ref{sup:simresult} show that GBMs substantially outperform other model families. In our deployment we used our data from our pilot to train the GBM in Modules~1--3 since data from our deployment was not yet available. Starting in Module~4, we instead train the GBM using data from earlier modules.

The outcome predicted by the GBM is whether the student is a slow or fast learner---i.e., the value of $\tau\in\{\tau_{\text{high}},\tau_{\text{low}}\}$. We found that this was the most important parameter to achieving good performance; furthermore, it was much more difficult to get good accuracy at predicting $c_1$ and $c_2$. We estimate the prediction targets for historical data by curve fitting. Specifically, for each given sequence of observations, we compute the \emph{maximum a posteriori} (MAP) estimate of the underlying sequence of states 
$\text{SE}_0,...,\text{SE}_T$
by maximizing the Poisson log-likelihood over $c_1,c_2,\tau$ using the \texttt{curve\_fit} routine in \texttt{scipy}.

Our features include (i) difficulty-level summaries across attempted items (mean, standard deviation, and slope of a least-squares linear fit over time), (ii) time-to-solve summaries in minutes (mean, standard deviation, and the corresponding linear-fit slope), and (iii) difficulty-adjusted time-to-solve (mean and linear-fit slope), where time is adjusted by assigned difficulty to separate effort from practice problem’s difficulty level. We also include LLM-based measures constructed from submission trajectories and AI-tutor conversations, including (iv) meaningful attempts (mean, standard deviation, and linear-fit slope), (v) the average growth rate of attempts required to solve a question correctly, and (vi) a chat-quality score capturing the quality of student--chatbot interactions (using the first observed value to reflect initial help-seeking behavior).

Finally, we use the prediction probability $p$ for $\tau=\tau_{\text{high}}$ from the GBM, along with historical data, to construct the initial belief state $b_0$. Specifically, we aggregate our MAP estimates from the historical data (our pilot data for Modules~1--3 , and data from earlier modules for Module~4 and onwards) to form $\mathcal{D}=\{(c_{1,i},c_{2,i},\tau_i)\}_{i=1}^n$; these estimates form the particles in our particle filter. Na\"{i}vely, we could use the uniform distribution over these particles as $b_0$; however, this strategy is not personalized to the current learner. Instead, we use the predicted probability $p$ of $\tau=\tau_{\text{high}}$ from the GBM to form the prior over $\tau$ (i.e., $P(\tau=\tau_{\text{high}})=p$ and $P(\tau=\tau_{\text{low}}=1-p$). For $c_1,c_2$, we construct the datasets
\begin{align*}
\mathcal{D}_{\text{high}}&=\{(c_{1,i},c_{2,i},\tau_i)\in\mathcal{D}\mid\tau_i=\tau_{\text{high}}\} \\
\mathcal{D}_{\text{low}}&=\{(c_{1,i},c_{2,i},\tau_i)\in\mathcal{D}\mid\tau_i=\tau_{\text{low}}\}.
\end{align*}
Then, we form the mixture distribution
\begin{align*}
\mathcal{D}_p&=p\cdot\text{Uniform}(\mathcal{D}_{\text{high}})+(1-p)\cdot\text{Uniform}(\mathcal{D}_{\text{low}}) \\
&=p\sum_{\theta\in\mathcal{D}_{\text{high}}}\frac{\delta_{\theta}}{n_{\text{high}}}+(1-p)\sum_{\theta\in\mathcal{D}_{\text{low}}}\frac{\delta_{\theta}}{n_{\text{low}}}
\end{align*}
where $n_{\text{high}}=|\mathcal{D}_{\text{high}}|$ and $n_{\text{low}}=|\mathcal{D}_{\text{low}}|$. In other words, we use a uniform distribution over $\mathcal{D}_{\text{high}}$ with probability $p$, and a uniform distribution over $\mathcal{D}_{\text{low}}$ with probability $1-p$. We use $b_0=\mathcal{D}_p$ as our initial belief state.

\subsection{Simulation Experiment}
\label{sup:simulation}

Next, we describe the setup of our simulation experiment used to evaluate our approach. This experiment is performed entirely on our pilot data (described in \ref{sup:pilot}), where problem sequences are randomized.\footnote{We cannot use data from our main study since problem sequences are either fixed (for the control group) or adaptive (for the treatment group), so there is no random variation that can be leveraged for evaluation purposes.} Our results (described in \ref{sup:simresult}) demonstrate the value of both our reinforcement learning algorithm as well as our use of LLM-derived features and observations.

\subsubsection{Experimental Setup}

\paragraph{Dataset.}

Our pilot dataset consists of 43 students with module records from 3 to 15, to simulate the algorithms. We restrict to 37 students who had completed at least 7 of the 15 modules to ensure reliability of the simulation.
We randomly split them into a training set of 27 students for fitting the prediction model and a test set of 10 students for evaluating both prediction accuracy and POMDP decision-making performance. For each module, we simulate 15 steps, as homework assignments typically consist of 15 questions per module (see Section~\ref{sec:field}).

\begin{algorithm}[H]
\caption{Simulation Evaluation of our POMDP-Based Decision-Making Policy}
\label{alg:pomdp}
\KwIn{Student ID, module IDs, ground truth $(c_1, c_2, \tau)$, initial belief probabilities $\{p_0^{(j)}\}_{j=1}^n$}
\KwOut{Sequences $\{a_{t}\}, \{o_{t}\}, \{r_{t}\}, \{b_{t}\}$}

\textbf{Initialize:}\\
$\{(c_1^{(j)}, c_2^{(j)}, \tau^{(j)})\}_{j=1}^n \leftarrow$ particles from historical data \\
$b_{0} \leftarrow$ belief state with probabilities $\{p_0^{(j)}\}_{j=1}^n$ over $n$ states \\
Initialize each state $s_0^{(j)}$ with $SE_0^{(j)} = c_1^{(j)} + c_2^{(j)}$ \\
\textbf{True State:} $SE^{\text{true}}_{0} \leftarrow c_1 + c_2$ \\

\For{$t = 1,2,\ldots,T$}{

\textbf{Action Selection:}\\
\For{each hypothesis state $s_{t-1}^{(j)}$ in belief $b_{t-1}$}{
$a^{(j)*} \leftarrow$ optimal action selected by our policy 
for state $s_{t-1}^{(j)}$
}
$a_{t} \leftarrow \arg\max_{a \in \{0,1,2,3\}} \sum_{j: a^{(j)*} = a} p_{t-1}^{(j)}$

\textbf{Environment Step:}\\
Execute $a_{t}$; observe $o_{t}$, reward $r_{t}$

\textbf{True State Update:}\\
$\Delta^{\text{true}} \leftarrow \tau \cdot (1.2 + a_{t}\delta_{knowledge})$ \\
$SE^{\text{true}}_{t} \leftarrow (SE^{\text{true}}_{t-1} - c_2) \cdot \exp(-\Delta^{\text{true}}) + c_2$

\textbf{Belief Update:}\\
\For{each hypothesis state $s_{t-1}^{(j)}$ in belief $b^{(t-1)}$}{
\textbf{State Transition:}\\
$\Delta^{(j)} \leftarrow \tau^{(j)} \cdot (1.2 + a_{t}\delta_{knowledge})$ \\
$SE^{(j)}_{t} \leftarrow (SE^{(j)}_{t-1} - c_2^{(j)}) \cdot \exp(-\Delta^{(j)}) + c_2^{(j)}$ \\
$s_{t-1}^{(j)} \leftarrow$ update state with new $SE^{(j)}_{t}$

\textbf{Observation Likelihood:}\\
$\lambda^{(j)} \leftarrow \max(SE^{(j)}_{t} \cdot (1 + a_{t})(1-\text{slip}), \epsilon)$ \\
$\ell^{(j)} \leftarrow P(o_{t} \mid s^{(j)}_{t}) = \frac{{(\lambda^{j})}^{o_{t}} \exp(-\lambda^{(j)})}{o_{t}!}$
}
\textbf{Belief Normalization:}\\
$p^{(j)}_{t} \leftarrow \frac{p^{(j)}_{t-1} \cdot \ell^{(j)}}{\sum_{i=1}^n p^{(i)}_{t-1} \cdot \ell^{(i)}}$ for all $j$ \\
$b_{t} \leftarrow$ updated belief with states $\{s_t^{(j)}\}$ and probabilities $\{p^{(j)}_{t}\}$

\If{maximum steps reached or all modules are completed}{
\textbf{break}
}
}

\Return All sequences
\end{algorithm}

\paragraph{Methodology.}

First, we describe our simulation environment. The main information required to construct our simulation environment is an initial state distribution. To this end, we use the method described in \ref{sup:pilotestimation} to estimate the initial state of each student in our dataset. Then, we use the uniform distribution over the training (resp., test) students as our initial state distribution for our training (resp., test) environment. Then, for each approach (our algorithm or one of our baselines, described below), we perform training on the training environment, and evaluate performance on the test environment. To do so, each algorithm is given simulation access to our training and test environments, which we perform as follows (see Algorithm~\ref{alg:pomdp} for a summary). First, we randomly sample an initial state, and initialize our POMDP with that initial state as the true state. Then, at each step, (1) the algorithm selects an action to take based on the observations so far, (2) we transition the true state based on that action, and (3) based on the updated state, we produce an observation (which is always given to the algorithm) and a reward (which is given to the algorithm during training; it is only used for evaluation during testing). At the end of the horizon, we assess performance by summing the rewards accrued across steps. This simulation provides a rigorous held-out evaluation of our methodology under the assumption that our POMDP transitions, observations, and rewards accurately reflect the real task.


\paragraph{Baselines.}

We compare our adaptive POMDP sequencing algorithm against several baselines, including a binary-state HMM sequencing rule (i.e., the existing Bayesian knowledge tracing algorithm), a LSTM-based PPO policy, random sequencing, and fixed sequencing, as well as ablations of our algorithm that differ in how they initialize and update learner beliefs (no historical data and uniform over historical data). We describe these approaches in detail below.

\begin{algorithm}[H]
\caption{Simulation Evaluation of the HMM-Based Decision-Making Policy}
\label{alg:hmm}
\KwIn{Student ID, module IDs, ground truth $(c_1, c_2, \tau)$}
\KwOut{Sequences $\{a_{t}\}, \{o_{t}\}, \{r_{t}\}, \{p^{\text{learned}}_{t}\}$}

\textbf{Initialize:}\\
$p^{\text{learned}}_{0} \leftarrow 0$ \\
$a_{0} \leftarrow 0$ \\
\textbf{True State:} $SE^{\text{true}}_{0} \leftarrow c_1 + c_2$

\For{$t = 1,2,\ldots,T$}{

\textbf{Action Selection:}\\
\eIf{$p^{\text{learned}}_{t-1} \ge 0.95$ \textbf{and} $a_{t-1} < 3$}{
$a_t \leftarrow a_{t-1} + 1$
}{
$a_t \leftarrow a_{t-1}$
}

\textbf{Environment Step:}\\
Execute $a_t$; observe $o_t$, reward $r_t$ \\
$o^{\text{bin}}_t \leftarrow \mathbf{1}[o_t > 5]$

\textbf{True State Update:}\\
$\Delta^{\text{true}} \leftarrow \tau \cdot (1.2 + a_{t}\delta_{knowledge})$ \\
$SE^{\text{true}}_{t} \leftarrow (SE^{\text{true}}_{t-1} - c_2) \cdot \exp(-\Delta^{\text{true}}) + c_2$

\textbf{Belief Prediction:}\\
$P_{01} \leftarrow 1 - \exp(-\tau (1 + a_t))$ \\
$p^{\text{learned}}_{t\mid t-1}
\leftarrow
p^{\text{learned}}_{t-1}
+
(1 - p^{\text{learned}}_{t-1}) P_{01}$

\textbf{Belief Update:}\\
$\ell_L \leftarrow P(o^{\text{bin}}_t \mid \text{Learned})$ \\
$\ell_N \leftarrow P(o^{\text{bin}}_t \mid \text{Not Learned})$ \\
$p^{\text{learned}}_{t}
\leftarrow
\dfrac{
p^{\text{learned}}_{t\mid t-1} \ell_L
}{
p^{\text{learned}}_{t\mid t-1} \ell_L
+
(1 - p^{\text{learned}}_{t\mid t-1}) \ell_N
}$

\If{maximum steps reached or all modules are completed}{
\textbf{break}
}
}

\Return All sequences
\end{algorithm}

\begin{itemize}
\item \textbf{Hidden Markov Model (HMM)}: We first implement a Hidden Markov Model (HMM) baseline method for adaptive difficulty sequencing in the simulation. The method models student learning as a two-state Markov process and employs a threshold-based policy to dynamically adjust question difficulty based on probabilistic estimates of learning state. The HMM models student learning using a two-state space: (i) State N (Not Learned): The student has not yet mastered the material; performance is characterized by an initial latent effort variable $SE_0$ that evolves over time, and (ii) State L (Learned): The student has mastered the material; performance is characterized by $c_2$. The system maintains a belief state $p_{\text{learned}} \in [0,1]$, representing the probability that the student is in learned state.

The transition from ``Not Learned'' to ``Learned'' follows an exponential decay model 
$$P(N \rightarrow L \mid \tau, a) = 1 - \exp(-\tau (1 + a)).$$

Given the current belief $p^{\text{learned}}_{t}$ and action $a_{t}$, the predicted belief after the transition is 
$$p^{\text{learned}}_{t+1 \mid t} = p^{\text{learned}}_{t} + (1 - p^{\text{learned}}_{t}) \cdot P(N \rightarrow L \mid \tau, a_{t}).$$

For the observations, we convert the Poisson observation $o$ from \ref{sup:obs} into a binary variable using a cutoff threshold:
$$o^{\text{bin}} = 
\begin{cases} 
1 & \text{if } o > 5 \\
0 & \text{otherwise}
\end{cases}$$
The observation likelihoods are:
$$P(o^{\text{bin}} = 1 \mid \text{Learned}) = c_2
\qquad\qquad
P(o^{\text{bin}} = 1 \mid \text{Not Learned}) = SE_{t}$$
The belief state is updated based on $o_{\text{bin}}$ using Bayes' rule:
$$p^{\text{learned}}_{t+1} = \frac{p^{\text{learned}}_{t+1 \mid t} \cdot P(o^{\text{bin}} \mid \text{Learned})}{p^{\text{learned}}_{t+1 \mid t} \cdot P(o^{\text{bin}} \mid \text{Learned}) + (1 - p^{\text{learned}}_{t+1 \mid t}) \cdot P(o^{\text{bin}} \mid \text{Not Learned})}$$
Finally, the action selection policy uses a simple threshold rule:
$$a_{t+1} = \begin{cases}
a_{t}+1 & \text{if } p^{\text{learned}}_{t} \geq 0.95 \\
a_{t} & \text{otherwise}.
\end{cases}$$
We show the evaluation of the HMM algorithm in Algorithm~\ref{alg:hmm}.

\item \textbf{Proximal Policy Optimization (PPO)}: We consider using model-free reinforcement learning to train a policy in our simulator. Specifically, we trained a neural network policy using proximal policy optimization (PPO)~\cite{schulman2017proximal}; since PPO does not maintain an explicit belief state, we use a long short-term memory (LSTM) policy to learn latent state. Furthermore, we provide the policy with a richer observation vector that summarizes the interaction history---the agent receives a 6-dimensional observation vector at each step $t$: 
$$\mathbf{x}_t = \left(\frac{t}{T},\; \frac{a_{t-1}}{3},\; \frac{o_{t-1}}{25},\; \frac{\bar{o}_{1:t-1}}{25},\; \frac{\hat{SE}_{t-1}}{20},\; \frac{\hat{SE}_{t-1} - \hat{SE}_1}{20} + 0.5\right),$$
where $T$ is the horizon length, $a_{t-1}$ is the previous action (0 at the start), $o_{t-1}$ is the most recent Poisson observation, $\bar{o}_{1:t-1} = \frac{1}{t-1}\sum_{i=1}^{t-1} o_i$ is the cumulative mean of all past observations, $\widehat{\text{SE}}_{t-1} = o_{t-1} / ((1 + a_{t-1})(1 - \text{slip}))$ estimates the underlying success rate, and the trend in estimated success rate (values below 0.5 indicate declining performance, above 0.5 indicate improvement). All components are clipped to $[0, 1]$.

The policy network uses a single-layer LSTM with 128 hidden units, followed by separate 64-unit policy and value heads. The agent is trained using generalized advantage estimation (GAE) with $\gamma = 0.99$ and $\lambda = 0.95$, a linearly decaying learning rate from $3 \times 10^{-4}$ to $3 \times 10^{-5}$, and an entropy coefficient of 0.1 to encourage exploration. The policy is trained on one million timesteps using the training split of student parameters, and evaluated on the held-out test split. While PPO uses the same difficulty-matching reward and monotonicity penalty as the POMDP, we found that training with the identical reward function led the agent to learn a degenerate policy that always selects the hardest difficulty level. To address this issue, we replace the single first-step penalty with an extended early-step penalty over the first three steps:
$$r^{\text{early}}_t = -\max(a_t - t, 0) \cdot \left(1 - \frac{t}{3}\right) \qquad (\forall t \in \{0, 1, 2\}).$$
\item \textbf{No historical:} We assume no prior knowledge of $\theta$, and maintain a uniform distribution over 10 particles sampled uniformly at random from the parameter domain.
\item \textbf{Uniform historical:} At each module, we maintain a uniform distribution over the estimated parameters of all students from the previous modules as our belief state. For the first module, we use the estimated parameters of the first 10 students (excluding the current student) from the current module.
\item \textbf{Greedy:} We replace our policy with a greedy one-step lookahead policy. At each step, we select the action that maximizes the immediate expected reward.
\item \textbf{Random sequencing:} We uniformly randomly sample $a_{t}\sim\text{Uniform}(\{0,1,2,3\})$.
\item \textbf{Fixed sequencing:} We use a fixed sequence of actions across the 15 steps.
$$[0, 0, 0, 0, 0, 0, 0, 0, 0, 1, 1, 2, 2, 3, 3].$$
\item \textbf{Oracle:} We initialize the belief state with the student's ground-truth parameters $(c_1, c_2, \tau)$ for each module, concentrating all probability mass on the true parameter configuration. Since this approach uses the ground truth state, it serves as an upper bound on the performance achievable by any policy in our POMDP framework.
\end{itemize}

\paragraph{Metrics.}

Our main outcome is the cumulative reward; average over the horizon to normalize it. While this is a shaped reward, it is intended to capture the desired behavior of our POMDP policy. In addition, we report the average difficulty level of problems (represented as an integer in $\{0,1,2,3\}$) across the horizon. This metric allows us to assess whether each algorithm is correctly assigning harder problems to high-type students than to low-type students.







%

\subsection{Simulation Results}
\label{sup:simresult}

\begin{table}[!t]
\centering
\small
\begin{tabular}{lcccccc}
\toprule
\textbf{Method} 
& \makecell{\textbf{Avg Reward}\\(High $\tau$)}
& \makecell{\textbf{Avg Reward}\\(Low $\tau$)}
& \makecell{\textbf{Avg Reward}\\(All)}
& \makecell{\textbf{Avg Diff.}\\(High $\tau$)}
& \makecell{\textbf{Avg Diff.}\\(Low $\tau$)}
& \makecell{\textbf{Avg Diff.}\\(All)} \\
\midrule
ML (Ours)       & \textbf{0.975} & \textbf{0.903} & \textbf{0.942} & 2.760 & 2.684 & 2.725 \\
ML (no LLM)     & 0.963 & 0.901 & 0.935 & 2.714 & 2.687 & 2.702 \\
HMM             & 0.793 & 0.527 & 0.673 & 2.000 & 0.000 & 1.098 \\
PPO             & 0.904 & 0.890 & 0.898 & 2.512 & 2.574 & 2.540 \\
No hist           & 0.821 & 0.871 & 0.843 & 2.145 & 2.029 & 2.093 \\
Uni hist      & 0.951 & 0.896 & 0.926 & 2.671 & 2.575 & 2.628 \\
Greedy          & 0.950 & 0.907 & 0.931 & 2.667 & 2.607 & 2.640 \\
Random seq     & 0.492 & 0.508 & 0.499 & 1.544 & 1.578 & 1.559 \\
Fixed seq      & 0.492 & 0.682 & 0.578 & 0.800 & 0.800 & 0.800 \\
\midrule
Oracle   & 0.988 & 0.975 & 0.982 & 2.800 & 2.265 & 2.559 \\
\bottomrule
\end{tabular}
\caption{Comparison of methods by average reward and difficulty across $\tau$ groups (difficulty ranges from 0 to 3).}
\label{tab:method_comparison}
\end{table}

\paragraph{Quantitative results.}

Table~\ref{tab:method_comparison} shows a comprehensive comparison of all methods; we also break down performance based on the subset of students with $\tau=\tau_{\text{high}}$ vs. $\tau=\tau_{\text{low}}$. The ``Oracle" method serves as a performance upper bound, since it knows the true state. Our POMDP-based policy with ML-based initialization (``ML'') achieves the highest overall average reward (0.942) and performs consistently well across for both high $\tau$ (0.975) and low $\tau$ (0.903) students. The ML method without LLM features (``ML no LLM'') shows slightly lower performance (0.935 overall), suggesting that LLM-derived features contribute to improved adaptation. The ``Greedy'' method performs comparably (0.931 overall), demonstrating that while one-step lookahead can be effective, there is value to accounting for long-term rewards. The ``PPO" method achieves competitive performance (0.898 overall) and, notably, exhibits very balanced rewards across high- and low-$\tau$ groups (0.904 vs. 0.890). These results suggest that PPO learns a robust global strategy, but does not distinguish as effectively between fast and slow learners as our method.

The ``Uniform'' historical baseline (0.926) performs slightly worse, highlighting that adaptation based on the initial state has some value, though online adaptation can mitigate shortcomings of prediction. However, the ``No historical'' baseline (0.843) shows substantially lower performance, particularly for high $\tau$ students (0.821); thus, using historical data to initialize the belief state is critical. The ``HMM'' method shows
significantly worse performance, suggesting limited ability to adapt to individual students. Both ``Random'' and ``Fixed'' sequencing baselines perform poorly (0.499 and 0.578 respectively), which demonstrates the importance of adapting to individual students.

\paragraph{Qualitative examples.}

\begin{figure}[!t]
\centering
\includegraphics[width=1.0\linewidth]{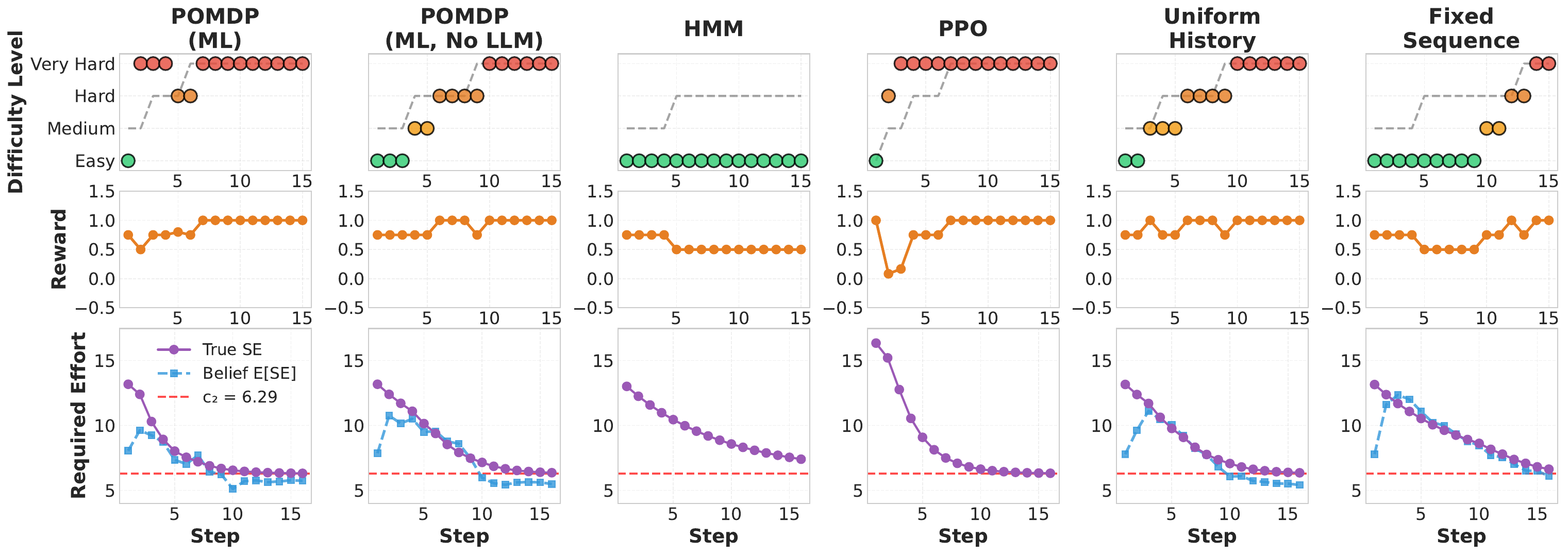}
\caption{Comparison of adaptive difficulty selection strategies for a student with $\tau_{\text{low}}$ and large initial gap between the true latent effort variable $\text{SE}$ and $c_2$, indicating limited initial familiarity with the topic. The figure presents three metrics across multiple methods: (top row) \textbf{Difficulty Level} showing selected difficulty levels (color-coded: Easy, Medium, Hard, Very Hard) with expected actions as a dashed reference line; (middle row) \textbf{Rewards} accumulated at each step; (bottom row) \textbf{Required Effort} displaying the true required effort (purple solid line), belief-expected required effort for POMDP methods (blue dashed line), and ground truth $c_2$ (red dashed line). Each column represents a different method.}
\label{fig:low_t_large_gap}
\end{figure}

\begin{figure}[!t]
\centering
\includegraphics[width=1.0\linewidth]{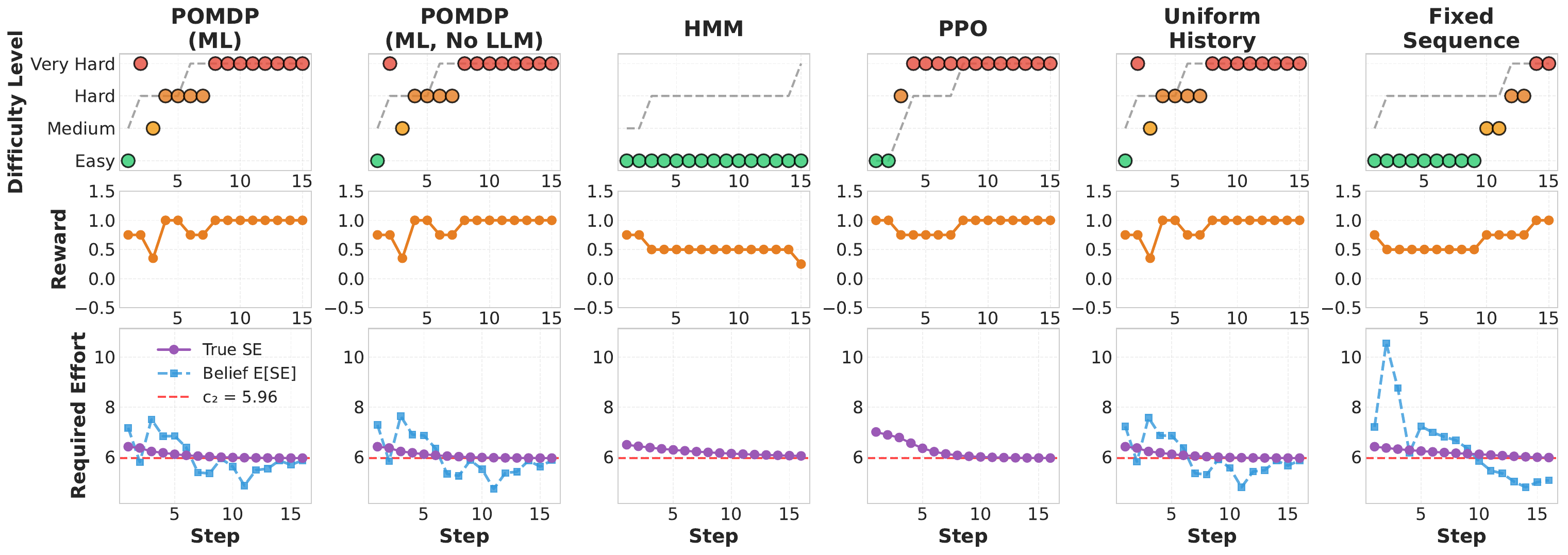}
\caption{Comparison of adaptive difficulty selection strategies for a student with $\tau_{\text{low}}$ and small initial gap between the true required effort $\text{SE}$ and $c_2$, indicating high initial familiarity with the topic. Same setup as Figure~\ref{fig:low_t_large_gap}.}
\label{fig:low_t_small_gap}
\end{figure}

\begin{figure}[!t]
\centering
\includegraphics[width=1.0\linewidth]{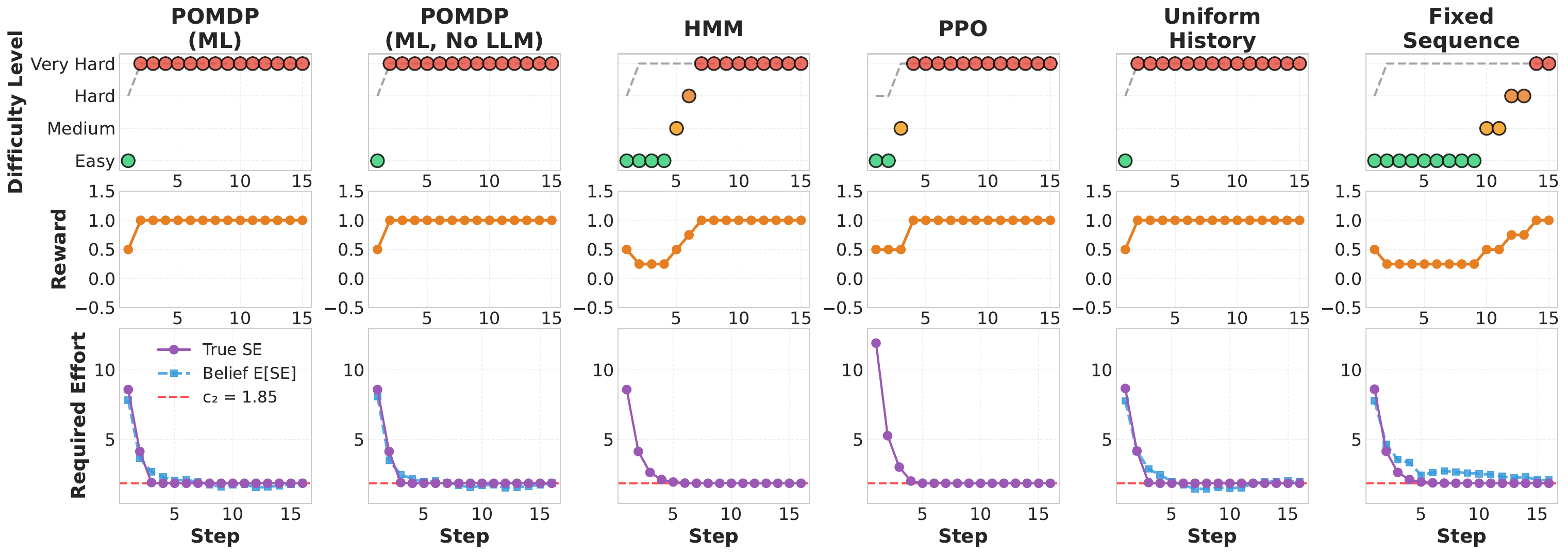}
\caption{Comparison of adaptive difficulty selection strategies for a student with $\tau_{\text{high}}$ and large initial gap between the true required effort $\text{SE}$ and $c_2$, indicating limited initial familiarity with the topic. Same setup as Figure~\ref{fig:low_t_large_gap}.}
\label{fig:high_t_large_gap}
\end{figure}

\begin{figure}[!t]
\centering
\includegraphics[width=1.0\linewidth]{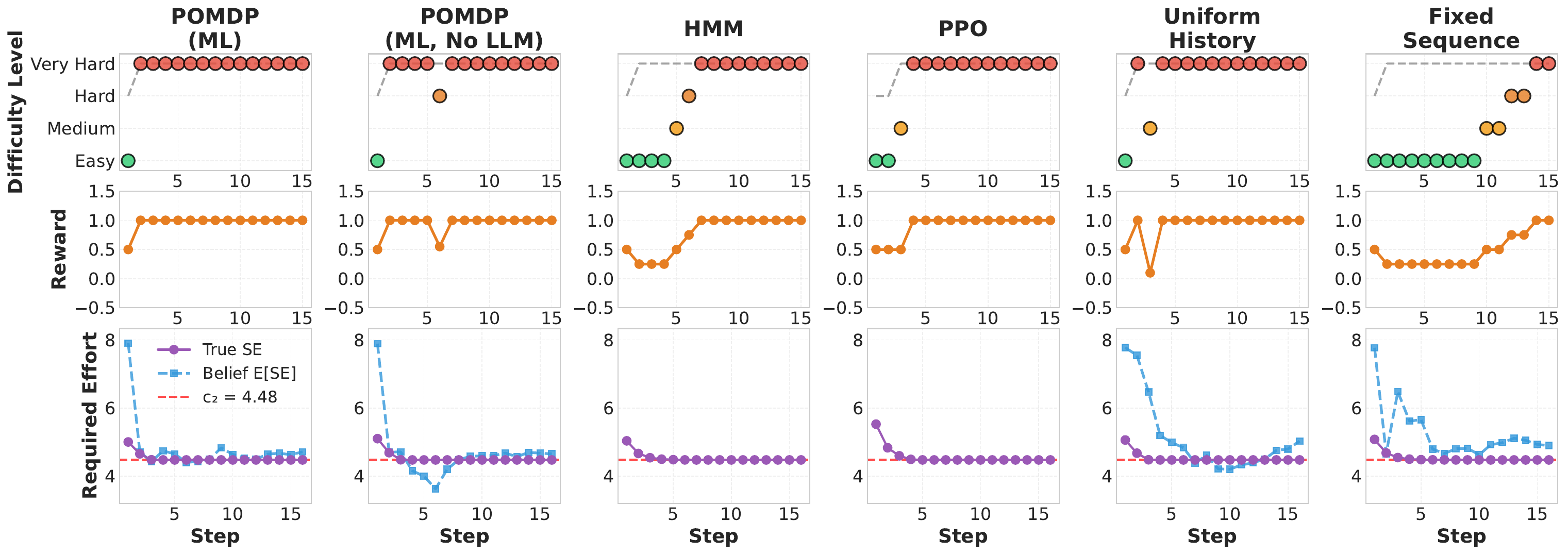}
\caption{Comparison of adaptive difficulty selection strategies for a student with $\tau_{\text{high}}$ and small initial gap between the true required effort $\text{SE}$ and $c_2$, indicating high initial familiarity with the topic. Same setup as Figure~\ref{fig:low_t_large_gap}.}
\label{fig:high_t_small_gap}
\end{figure}

To illustrate how different methods adapt to varying student types, we present qualitative examples across four representative cases that capture the key dimensions of student variability: learning speed ($\tau$) and initial topic familiarity (quantified by the initial gap between the true required effort $\text{SE}_0$ and the mastery ceiling $c_2$) in Fig.~\ref{fig:low_t_large_gap} (low $\tau$, large gap), Fig.~\ref{fig:low_t_small_gap} (low $\tau$, small gap), Fig.~\ref{fig:high_t_large_gap} (high $\tau$, large gap), and Fig.~\ref{fig:high_t_small_gap} (high $\tau$, small gap).

\paragraph{Prediction model evaluation.}

\begin{figure}[!t]
\centering
\begin{subfigure}{0.48\textwidth}
\centering
\includegraphics[width=\linewidth]{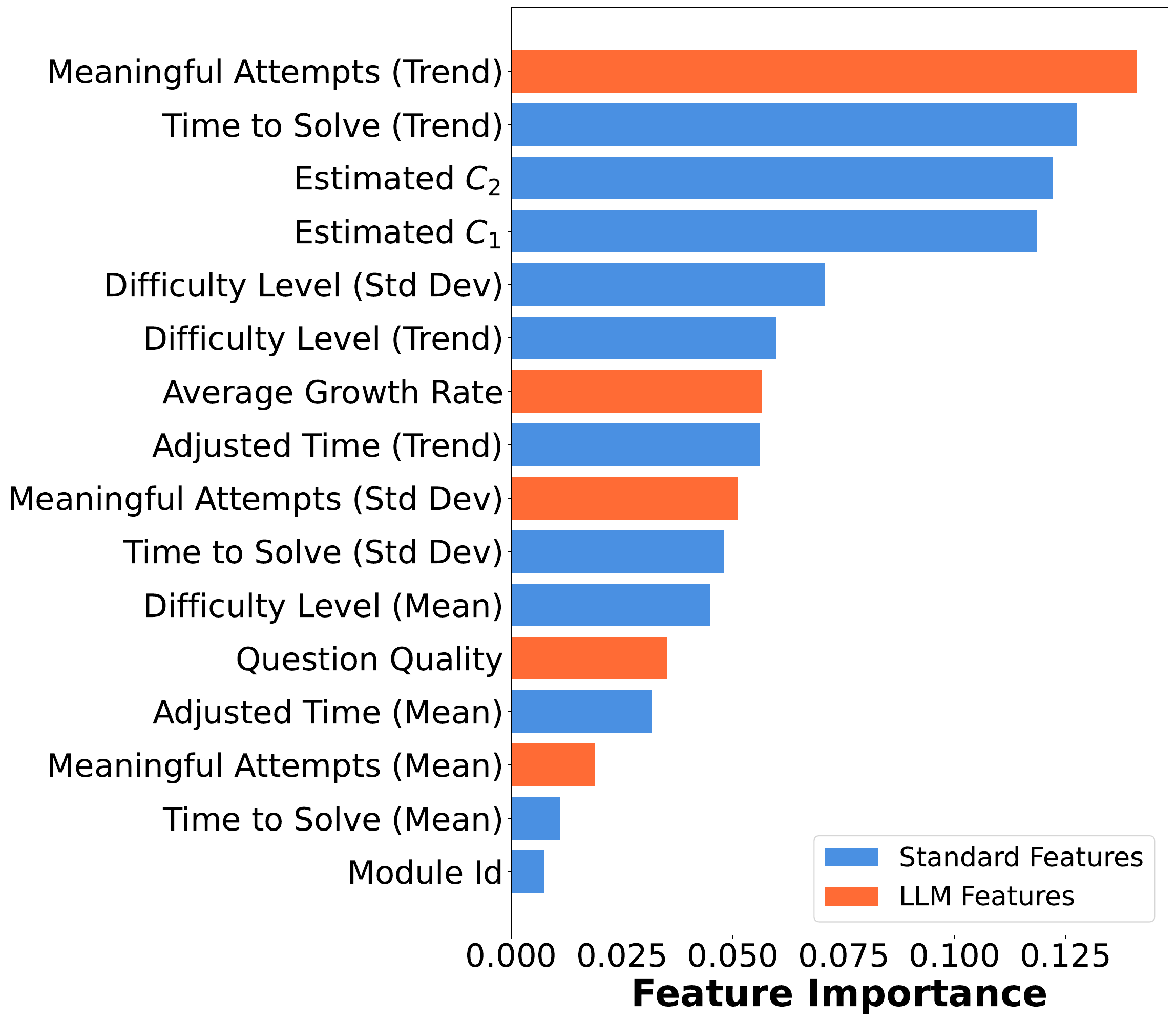}
\caption{Feature importance with LLM features}
\label{fig:feat_imp_a}
\end{subfigure}
\hfill
\begin{subfigure}{0.48\textwidth}
\centering
\includegraphics[width=\linewidth]{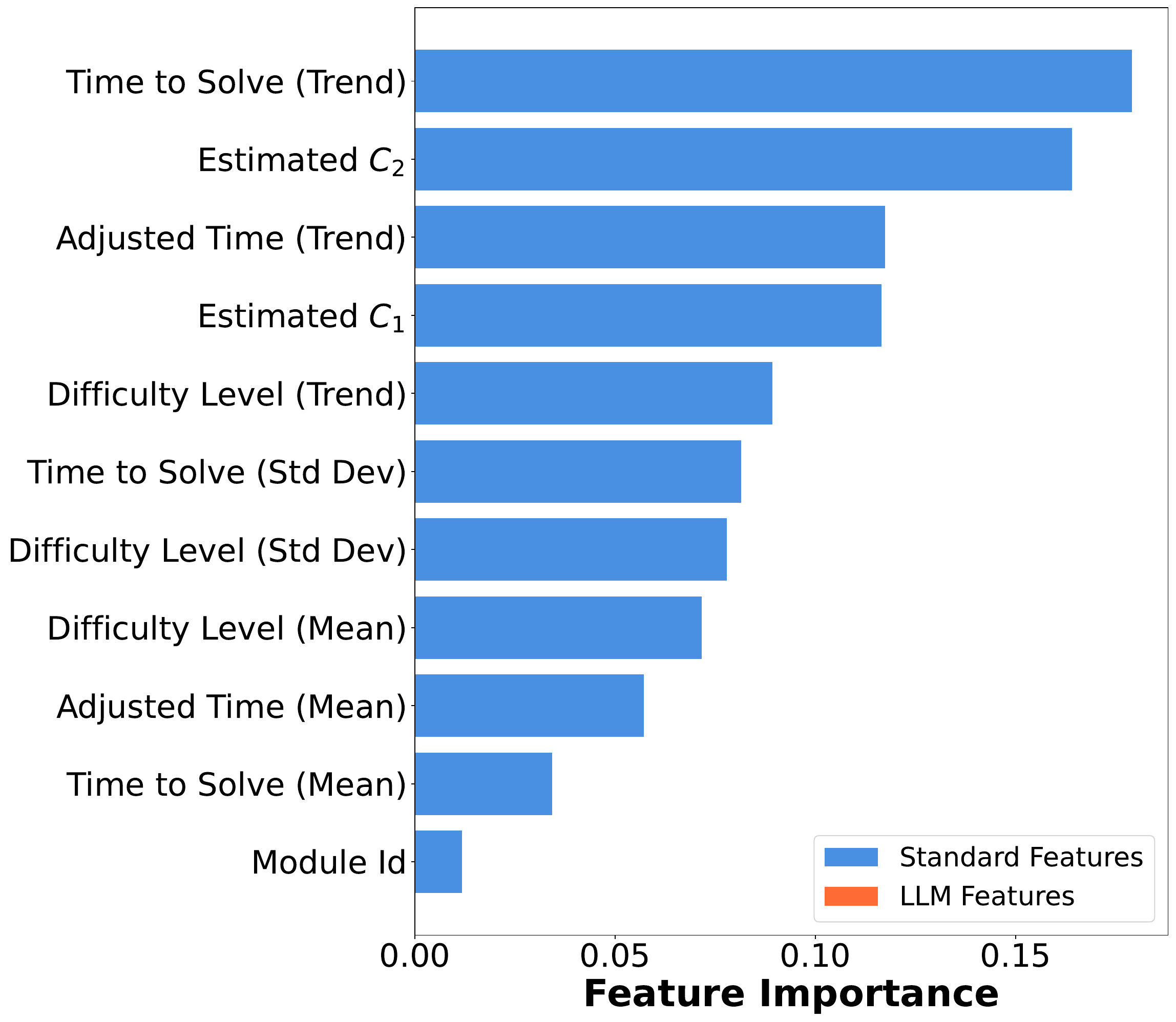}
\caption{Feature importance for without LLM features}
\label{fig:feat_imp_b}
\end{subfigure}
\caption{\textbf{Feature importance comparison for predicting $\tau$ with and without LLM features.}}
\label{fig:feature_importance}
\end{figure}

Finally, we also analyze the accuracy of our predictive model. First, Fig.~\ref{fig:feature_importance} summarizes which behavioral signals are most predictive of the learning speed $\tau$. Notably, several of our LLM-derived features achieve high importance. These features are also important according to several other metrics. Specifically, we found that including LLM features increases AUC from 63.66\% to 72.44\%; see Table~\ref{tab:predict_comparison}. These gains are consistent with the fact that meaningful edit attempts and chat-quality features capture engagement and help-seeking behavior that is difficult to measure in traditional tutoring systems in real time.

\begin{table}[!t]
\centering
\caption{\textbf{Classification performance of different models with and without LLM features.} Metrics are computed using a threshold of 0.5. AUC and Accuracy are reported as percentages.}
\begin{tabular}{lccccccc}
\hline
\textbf{Model Type} & \textbf{LLM Feats} 
& \textbf{AUC} 
& \textbf{Acc} 
& \textbf{TP} 
& \textbf{FP} 
& \textbf{TN} 
& \textbf{FN} \\
\hline
\multirow{2}{*}{Gradient Boosting}   & Yes  & \textbf{72.44} & \textbf{71.57} & 46 & 19 & 27 & 10 \\
& No & 63.66 & 69.61 & 47 & 22 & 24 & 9 \\
\midrule
\multirow{2}{*}{Logistic Regression} & Yes  & 61.88 & 57.84 & 31 & 18 & 28 & 25 \\
& No & 54.66 & 53.92 & 28 & 19 & 27 & 28 \\
\midrule
\multirow{2}{*}{Random Forest}       & Yes  & 70.30 & 69.61 & 41 & 16 & 30 & 15 \\
& No & 63.78 & 59.80 & 33 & 18 & 28 & 23 \\
\midrule
\multirow{2}{*}{XGBoost}             & Yes  & 69.57 & 64.71 & 42 & 22 & 24 & 14 \\
& No & 61.34 & 65.69 & 43 & 22 & 24 & 13 \\
\hline
\end{tabular}
\label{tab:predict_comparison}
\end{table}

\paragraph{Efficacy of LLM-based observations.}

A key feature of our approach is that our observations are based on an LLM judging the number of meaningful attempted solutions, rather than a conventional observation such as the number of raw attempted solutions. Intuitively, raw attempts are very noisy since students often make sequences of submissions with minor edits. To evaluate whether meaningful attempts (\textit{MeaningfulAttempt}) better reflects students' learning progression than the raw number of attempts (\textit{TotalAttempt}), we compare their explanatory power for student learning outcomes. Specifically, we first construct standardized versions of each predictor (namely, $z^{\text{MeaningfulAttempt}}$ and $z^{\text{TotalAttempt}}$), and then use a sequence of ordinary least squares (OLS) regression models to assess their explanatory power.

First, we consider two univariate specifications:
\begin{align}
\label{eq:attempt_only}
Y_i &= \alpha + \beta_1 z^{\text{TotalAttempt}}_i + \varepsilon_i, \\
\label{eq:mattempt_only}
Y_i &= \alpha + \beta_2 z^{\text{MeaningfulAttempt}}_i + \varepsilon_i,
\end{align}
as well as a joint specification:
\begin{equation}
\label{eq:full}
Y_i = \alpha + \beta_1 z^{\text{TotalAttempt}}_i + \beta_2 z^{\text{MeaningfulAttempt}}_i + \varepsilon_i.
\end{equation}
The raw-attempt model in Eq.~\eqref{eq:attempt_only} explains $R^2=0.08$ of the variation in student's learning outcome, whereas the meaningful-attempt model in Eq.~\eqref{eq:mattempt_only} explains $R^2=0.14$. Thus, meaningful attempts account for substantially more outcome variation than raw attempts. 

Next, we test whether each observation contributes incremental explanatory power beyond the other using nested-model comparisons. Adding $z^{\text{MeaningfulAttempt}}$ to Eq.~\eqref{eq:attempt_only} produces a large improvement in fit ($F=67.36$, $p<0.001$). In contrast, adding $z^{\text{TotalAttempt}}$ to Eq.~\eqref{eq:mattempt_only} yields a smaller (though still statistically significant) improvement ($F=15.66$, $p<0.001$). These results indicate that meaningful attempts contain substantial information about learning outcomes not captured by raw attempts, but less so vice versa.

Fianlly, because $z^{\text{TotalAttempt}}$ and $z^{\text{MeaningfulAttempt}}$ are correlated, coefficient magnitudes in the joint model can be difficult to interpret due to shared variance. Thus, we complement the nested tests with a variance decomposition based on the Lindeman-Merenda-Gold (LMG) metric~\cite{lindeman1980introduction}, which tells us the attribution of the model $R^2$ to regressors averaged over all possible orderings. Applying the LMG decomposition to Eq.~\eqref{eq:full} shows that $z^{\text{MeaningfulAttempt}}$ accounts for 0.69 of explained variance (69.06\%), whereas $z^{\text{TotalAttempt}}$ accounts for 0.31 (30.94\%). This result confirms that meaningful attempts contribute most of the explanatory power in the joint model.

\paragraph{Discussion on LLM-derived features and observations.}

Both our POMDP simulation results and our prediction results demonstrate the value of LLMs in extracting useful signals of student learning progression from student-platform interactions. First, our simulation results as well as the performance of our GBM prediction model on a held-out test set both demonstrate that LLM-derived signals from student-chatbot interactions benefit the prediction of the initial belief state. Second, our analysis of meaningful vs. total attempts demonstrate the value of LLM-derived observations from student code submissions on our platform. Together, these results provide strong support for our hypothesis that student-platform interactions are a valuable source of information that can be used to obtain more accurate estimates of student knowledge state.

\section{Pilot Study}
\label{sup:pilot}

Before launching our large-scale RCT, we conducted an IRB-approved pilot study to (i) validate the feasibility of the end-to-end AI learning system (question generation, platform logging, and tutoring workflow), (ii) collect early learner-trajectory data for training and calibrating our reinforcement learning (RL) algorithm, and (iii) incorporate user feedback to redesign and re-implement the platform and question bank.

\paragraph{Study context and sample.}
The pilot was run in the two month accelerated summer term at the National Taiwan University (NTU) in 2024 using the same course content as the main study here. A total of 52 students participated: 21 were high school students or first-year college students, 6 were graduate students, and participants represented multiple universities, with 22 students studying outside of Taipei. Throughout the course, students completed an in-person midterm and final exam covering the course concepts; all exams were graded by human grader to measure learning outcomes. All eligible university students and advanced high school students nationwide could register this class in the summer.

\paragraph{Pilot experimental design.}
The pilot used a $2\times2$ randomized controlled design that crossed (i) adaptive sequencing (RL vs. Random) with (ii) support modality (human tutor vs. GPT-4-based tutor). Participating students were uniformly randomly assigned to one of four arms:
\begin{enumerate}
\item \textbf{Control (No RL$+$Human Tutor).} Non-adaptive practice with a randomized difficulty sequence; access to standard human tutors.
\item \textbf{No RL$+$GPT-4.} Non-adaptive practice with a randomized difficulty sequence; access to a GPT-4-based tutor.
\item \textbf{RL$+$Human Tutor.} Adaptive practice intended to be personalized via RL; access to standard human tutors.
\item \textbf{RL$+$GPT-4.} Adaptive practice intended to be personalized via RL; access to a GPT-4-based tutor.
\end{enumerate}
To ensure perceived fairness, the pilot implemented a post-midterm switch (control $\leftrightarrow$ RL$+$GPT-4; no-RL$+$GPT-4 $\leftrightarrow$ RL$+$human), so that each student experienced multiple learning environments over the course.

\paragraph{Lessons for platform and algorithm.}
During the pilot, we found there were data infrastructure and integration issues that prevented the RL policy from being reliably updated online as intended. As a result, we treat the pilot primarily as a validation of our system and a data collection opportunity rather than a formal test of the fully functional adaptive algorithm. We used the pilot interaction logs---especially the control-arm trajectories with randomized difficulty---to (i) train and calibrate the RL components offline, and (ii) guide a comprehensive redesign, re-implementing the learning platform (logging pipeline, interface usability, and tutoring workflow) and question bank (generation, validation, and quality control). These changes were informed by both quantitative usage traces and structured feedback from students and teaching staff. Note that because the pilot included a mid-course arm switch, we restrict all pilot analyses used for scientific interpretation to data \emph{before} the switch, to avoid carryover effects across conditions.

\section{Experimental Protocol Details}
\label{sup:expprotocol}

\ref{sup:briefing} outlines teacher and student briefings; \ref{sup:programschedule} describes the Python certification program schedule and rules; \ref{sup:survey} lists the survey items; and \ref{sup:preregist} details our pre-registered analyses.

\subsection{Teacher and Student Briefing}
\label{sup:briefing}

Before the start of the program, our team held an in-person briefing with the high school teachers responsible for program implementation at participating schools, focused on introducing the study protocol and demonstrating the learning platform.\footnote{Note that instruction was delivered through pre-recorded videos on our learning platform, so these teachers did not need to prepare or teach the curriculum; their role was limited to supervising program implementation.} The briefing covered the overall course structure, the platform interface (lecture videos, practice mode, and the embedded AI tutor), and the rules governing student participation and academic integrity.

Following the briefing, teachers recruited students from their respective schools. All participants provided informed consent, but they were blinded to specific details regarding the random treatment assignment. Students who signed up consisted primarily of 10th-graders (59.5\%), followed by 11th graders (23.8\%) and 12th graders (16.7\%). Regarding prior Python learning experience, 28.4\% were first-time learners and 27.0\% identified as beginners. While 39.9\% reported intermediate experience, only a small fraction identified as proficient (3.8\%) or at a competition level (1.0\%). We provide more information about the student sample in our balance table (Table~\ref{tab:balance} in \ref{sup:databalanced}). 

Before the first module, teachers outlined the study’s general structure, explaining that the program involved a multi-week sequence of online learning modules culminating in a certification exam. Students were issued unique login credentials and invited to a brief synchronous online orientation. To assist them throughout the program, a platform demonstration video was made accessible for reference at any time. Furthermore, the course materials included a concise set of usage rules and tips for effective AI-tutor interaction (see Fig.~\ref{fig:aiusagerules}).

\begin{center} 
\begin{tcolorbox}[
    title=\textbf{AI Tutor Usage Tip and Rules for Students}, 
    width=\linewidth, 
    breakable,        
    enhanced,         
    colback=white,   
    colframe=black!75 
]
\begin{small}
\noindent\textbf{AI Tutor: Tips for Effective Use}

\begin{itemize}[leftmargin=1.2em, itemsep=2pt, parsep=0pt, topsep=3pt]
    \item Think carefully about what you are stuck on; the more specific your question, the more useful the response.
    \item State what you want explained: a concept, code logic, or the problem statement itself.
    \item The AI can be wrong. Always evaluate its answers critically---verify by asking follow-up questions, challenging assumptions, and checking the logic. Use this as practice for independent thinking.
    \item If the explanation is unclear, ask the AI to rephrase in simpler terms or to provide a concrete example.
    \item Use the AI to clarify any confusion. Do not worry if your question feels basic, and it is okay to ask similar questions again.
\end{itemize}

\vspace{6pt}
\noindent\textbf{AI Tutor: Usage Rules}

\begin{enumerate}[leftmargin=1.4em, itemsep=4pt, topsep=3pt, parsep=0pt]
    \item \textit{Rate limit.} To ensure fair access, usage is rate-limited: after every five questions, students must wait 10 seconds before submitting additional AI-tutor queries.
    \item \textit{Compliance with usage standards.} Students were required to follow the platform’s usage policy:
    \begin{itemize}[leftmargin=1.2em, itemsep=2pt, topsep=2pt, parsep=0pt]
        \item \textbf{No abuse.} Ask questions thoughtfully and read the response carefully before following up.
        \item \textbf{Respect privacy and rights.} Do not share or misuse personal information. Do not use AI-generated content to harass, discriminate, scam, or harm others.
        \item \textbf{Ethics and social responsibility.} Do not use the AI tutor for illegal activities or to manipulate or mislead others in harmful ways.
        \item \textbf{Safety and fairness.} Do not attempt to bypass safety mechanisms or conduct unauthorized testing. Report abnormal behavior or risks to the teaching assistants or instructor.
        \item \textbf{Law and academic integrity.} Do not use the AI tutor for cheating, plagiarism, or other dishonest academic conduct.
    \end{itemize}
\end{enumerate}

\vspace{4pt}
\noindent\textit{Students were instructed to use the AI tutor as a learning aid and to ensure all usage is ethical and compliant. Any violation resulted in revocation of platform access and removal from the program.}
\end{small}

\end{tcolorbox}

\captionof{figure}{\textbf{AI tutor usage tips and rules provided to student participants.}}
\label{fig:aiusagerules}

\end{center}

\subsection{Certification Program Schedule and Rules}
\label{sup:programschedule}

Table~\ref{tab:schedule} presents the course schedule, assignment deadlines, and module outlines for the certification program, along with the lengths of the instructional videos. The program concluded with an on-site paper-based assessment administered by participating schools (see \ref{sup:exam}).

\begin{longtable}
{|p{3cm}|p{5cm}|p{7cm}|}
\caption{\textbf{Program Schedule and Outline}} \\
\hline
\textbf{Date} & \textbf{Course Module \& Assignment} & \textbf{Learning Hours / Course Outline} \\
\hline
\endfirsthead
\multicolumn{3}{c}%
{{\bfseries \tablename\ \thetable{} -- continued from previous page}} \\
\hline
\textbf{Date} & \textbf{Course Module \& Assignment} & \textbf{Learning Hours / Course Outline} \\
\hline
\endhead

\hline \multicolumn{3}{|r|}{{Continued on next page}} \\ \hline
\endfoot

\hline
\endlastfoot

2/17 -- 2/23 \newline 
2/24 -- 3/2 & 
\textbf{Module 1-3: Basics} \newline 
Assignment 1 \newline 
\textit{*Deadline: 3/12 11:59pm} & 
This week introduces Python applications in modern business, as well as Python syntax and the programming development environment. Students will interact using basic commands and programs. \newline
\textbf{Video Length:} \newline
M1: 1:14:54 \newline
M2: 1:27:49 \newline
M3: 1:08:53 \\
\hline

3/3 -- 3/9 \newline 
3/10 -- 3/16 & 
\textbf{Module 4-5: Conditionals} \newline 
Assignment 2 \newline 
\textit{*Deadline: 3/23 11:59pm} & 
This week continues the introduction to different variable types and moves into conditional selection teaching, enabling programs to respond based on given conditions. \newline
\textbf{Video Length:} \newline
M4: 0:53:27 \newline
M5: 1:00:49 \\
\hline

3/17 -- 3/23 \newline 
3/24 -- 3/30 & 
\textbf{Module 6: Loops (Part 1)} \newline 
Assignment 3 \newline 
\textit{*Deadline: 3/30 11:59pm} & 
This week continues with logical operators and introduces loop concepts to build more complex conditional selections, enabling the program to repeat automated tasks. \newline
\textbf{Video Length:} \newline
M6: 0:49:11 \\
\hline

3/31 -- 4/6 \newline 
4/7 -- 4/13 & 
\textbf{Module 7: Loops (Part 2)} \newline 
Assignment 4 \newline 
\textit{*Deadline: 4/13 11:59pm} & 
Continuing with loops; learning the application of Lists in data management. \newline
\textbf{Video Length:} \newline
M7: 0:41:18 \\
\hline

4/14 -- 4/20 \newline 
4/21 -- 4/27 & 
\textbf{Module 8: Operations Applications} \newline 
Assignment 5 \newline 
\textit{*Deadline: 4/27 11:59pm} & 
This module explores program applications in business topics such as production scheduling and inventory control. \newline
\textbf{Video Length:} \newline
M8: 1:32:42 \\
\hline

4/28 -- 5/4 \newline 
5/5 -- 5/11 & 
\textbf{Module 9: Functions (Part 1)} \newline 
Assignment 6 \newline 
\textit{*Deadline: 5/11 11:59pm} & 
This week focuses on Functions. \newline
\textbf{Video Length:} \newline
M9: 1:06:57 \\
\hline

5/12 -- 5/18 \newline 
5/19 -- 5/25 & 
\textbf{Module 10: Functions (Part 2)} \newline 
Assignment 7 \newline 
\textit{*Deadline: 5/25 11:59pm} & 
Continuing with Functions. This week introduces the application of functions in program development. \newline
\textbf{Video Length:} \newline
M10: 0:56:19 \\
\hline

5/26 -- 6/6 & 
\textbf{On-site Paper-based Assessment} & 
Schools will select a date to administer the paper-based test to assess learning outcomes and determine eligibility for certificates. 
\label{tab:schedule}
\end{longtable}

\subsection{Survey Instrument}
\label{sup:survey}

We administered a baseline survey after the introductory module (when students had become familiar with the platform) to obtain statistics on student demographics, motivation, parental education, familiarity with the course material, etc. Table~\ref{tab:baseline_survey} lists the survey questions.

\subsection{Pre-Registration Details}
\label{sup:preregist}

Our pre-registration is available here: \href{https://aspredicted.org/6az9cd.pdf}{https://aspredicted.org/6az9cd.pdf}. The primary outcome was student performance on the certification exam. We pre-registered two primary analyses: (i) linear regression evaluating treatment effects on test performance with controls for baseline covariates and fixed effects, and (ii) evaluation of different pedagogical strategies in the system prompt. We report the first analysis in Fig.~\ref{fig:main}b and \ref{sup:main}; for the second analysis, as noted in \ref{sup:aitutor}, we find no meaningful variation across prompt conditions. We supplement our primary regression with unadjusted two-sided $t$-tests (Fig.~\ref{fig:main}a) and report effects on students in an accelerated track where the implementation of RL was severely delayed (Table~\ref{tab:rl_robust}; \ref{sup:main}).  

We pre-registered several secondary analyses examining heterogeneous treatment effects and student engagement patterns. We report heterogeneity by prior Python experience and school tier in Section~\ref{sec:hte} (with interaction-regression specifications in \ref{sup:hte}). Following our pre-registered plan to analyze engagement measures, we examine platform logs and AI chat conversations as potential mechanisms in Section~\ref{sec:mechanisms} and \ref{sup:mediation}.

\begin{table}[!t]
\centering
\caption{\textbf{Baseline survey questions.}}
\label{tab:baseline_survey}
\setlength{\tabcolsep}{4pt}
\renewcommand{\arraystretch}{1.12}
\footnotesize

\begin{tabularx}{\textwidth}{@{}p{0.95cm}X@{}}
\toprule
\textbf{Item} & \textbf{Survey question (response options)} \\
\midrule

Q1 &
\textbf{What is your gender?}
\emph{(Female; Male; Prefer not to say; Other)} \\

Q2 &
\textbf{What is the \textit{primary} motivation for you to take this course?}
\emph{(Parents required it; Classmates/friends invited me; School/graduation requirement; Enrich my resume; Learn relevant skills in the AI era)} \\

Q3 &
\textbf{What is the highest education level of your parents?}
\emph{(High school or below; Bachelor's degree; Master's degree; Doctoral degree)} \\

Q4 &
\textbf{Do you attend cram school (including private tutoring), or have you ever attended? If yes, which subjects have you attended for?}
\emph{(None; Chinese; English; Math; Science; Programming; Other subjects; Multiple selections allowed)} \\

Q5 &
\textbf{During the period you attended high school, how was your average academic performance?}
\emph{(Top 20; Top 50; Top 100; Top 150; Top 200; Top 300; Top 500 or $>$500 in class/school rank)} \\

Q6 &
\textbf{Before taking this course, have you taken any courses related to programming?
(Including self-study and online courses; any related learning of more than 20 hours counts. Multiple selections allowed.)}
\emph{(No, none or fewer than 20 hours; Yes, Python; Yes, C/C++; Yes, other programming languages; Multiple selections allowed)} \\

Q7 &
\textbf{What do you think your Python ability or level is?}
\emph{(First time learning; Little to no foundation; Average; Proficient; Very proficient and have participated in competitions)} \\

Q8 &
\textbf{What are your views on learning using generative AI?}
\emph{(AI can help me learn; AI may create dependence and hinder my learning)} \\


\bottomrule
\end{tabularx}

\vspace{2pt}
\begin{minipage}{\textwidth}
\small
\end{minipage}

\normalsize
\end{table}

Note that our pre-registration included two additional components that we do not emphasize in this paper. First, the pre-registration planned analyses for both university (NTU) and high school samples (conducted as separate deployments on different student populations). However, because the NTU term began earlier, unforeseen delays in integrating our platform with the government e-learning system prevented reliable delivery of our intervention during the NTU deployment. Thus, we only report on the high school student deployment. Second, we randomized the AI tutor prompt across three pedagogical frameworks (Constructivism, Behaviorism, and Social Learning). We find no meaningful or statistically significant differences across pedagogical conditions in our primary outcomes. Accordingly, we do not discuss these contrasts in this paper; however, we show that our main results remain robust when controlling for these features (see \ref{sup:main}).

\section{Data Processing, Attrition, and Covariate Balance}
\label{sup:databalanced}

\ref{sup:exclusion} details sample exclusion criteria; \ref{sup:balancedTable} shows covariate balance across arms; and \ref{sup:attrition} analyzes differential attrition.

\subsection{Exclusion Processing} 
\label{sup:exclusion}

A total of 1{,}047 students participated in the program. We construct the main analysis sample by excluding observations that do not follow our pre-specified rule in our pre-registered plan or are not comparable to the standard program (see Table~\ref{tab:attrition}).

Following our pre-registration, we exclude 94 students who dropped out of the program or were unable to attend the exam due to schedule conflicts (and therefore do not have an outcome measure) and 29 students who were flagged for cheating during the certification exam. The cheating incident only happened in a single classroom and arose from a misunderstanding of proctoring instructions. We additionally exclude 154 senior high school students due to complications and delays with the implementation of RL. Specifically, (1) the course for seniors was accelerated due to their graduation timeline, (2) the program launch for senior students was moved substantially earlier than initially planned, and (3) key platform integrations were delayed due to governmental constraints. As a consequence, our RL algorithm was only active one month after senior students started the course, which is about half of their course period due to their accelerated timeline. Thus, we would not expect RL to be effective among this student population. For transparency, we report our main analysis on the sample including the accelerated senior-student track (see \ref{sup:main}).

\begin{table}[!t]
\centering
\caption{\textbf{Sample construction and attrition.}}
\label{tab:attrition}
\setlength{\tabcolsep}{6pt}
\renewcommand{\arraystretch}{1.08}
\footnotesize
\begin{tabular}{lrrrr}
\toprule
\textbf{Step} & \textbf{Excluded} & \textbf{\% of total} & \textbf{Remaining $N$} \\
\midrule
Total enrolled in program & -- & -- & 1{,}047 \\
Exclude: Certification score = 0 (did not complete exam) & 94 & 8.98\% & 953 \\
Exclude: Flagged for cheating during certification exam & 29 & 2.77\% & 924 \\
Exclude: Accelerated senior-student track (non-comparable schedule/exam) & 154 & 14.71\% & 770 \\
\midrule
\textbf{Total excluded} & 277 & 26.46\% & - \\
\textbf{Final analysis sample} & - & - & \textbf{770} \\
\bottomrule
\end{tabular}
\vspace{2pt}

{\footnotesize \emph{Notes:} Percentages are relative to the total enrolled sample ($N=1{,}047$). }
\end{table}

\subsection{Balance Table} 
\label{sup:balancedTable}

After these exclusions, our final analysis sample contains 770 students. We assess baseline comparability between the treatment and control groups within this final analysis sample. Table~\ref{tab:balance} reports covariate balance and shows no evidence of systematic differences in observed pre-treatment characteristics across treatment conditions.

\begin{table}
\centering
\footnotesize
\begin{threeparttable}
\caption{\textbf{Baseline Covariate Balance Between Treatment and Control Groups (N=770)}}
\label{tab:balance}

\begin{tabular}{lcccc}
\toprule
\textbf{Variable} & \textbf{Control} & \textbf{Treated} & \textbf{p-value} & \textbf{p-adj (BH)} \\
\midrule
\multicolumn{5}{l}{\textit{Continuous Variables, Mean (SD)}} \\
\midrule
Parent education (ordinal)         & 2.37 (0.88)   & 2.31 (0.78)   & 0.727 & 0.844 \\
Has tutoring                & 0.93 (0.26)   & 0.91 (0.28)   & 0.399 & 0.776 \\
Academic performance in school (ordinal)     & 4.84 (1.68)   & 4.79 (1.62)   & 0.726 & 0.844 \\
Has programming experience               & 0.71 (0.46)   & 0.72 (0.45)   & 0.959 & 0.959 \\
Python ability (ordinal)     & 1.17 (0.96)   & 1.16 (0.93)   & 0.938 & 0.959 \\
AI positive view           & 0.97 (0.16)   & 0.98 (0.15)   & 0.632 & 0.830 \\
\midrule
\midrule
\multicolumn{5}{l}{\textit{Categorical Variables, n (\%)}} \\
\midrule
\textbf{Gender}              & & & 0.271 & 0.715 \\
\quad Female                 & 121 (31.5\%) & 131 (33.9\%) & & \\
\quad Male                   & 212 (55.2\%) & 206 (53.4\%) & & \\
\quad Prefer not to say      & 19 (4.9\%)   & 23 (6.0\%)   & & \\
\quad Other                  & 5 (1.3\%)    & 1 (0.3\%)    & & \\
\quad Missing                & 27 (7.0\%)   & 25 (6.5\%)   & & \\[4pt]

\textbf{Motivation}          & & & 0.013 & 0.130 \\
\quad Parents                & 7 (1.8\%)    & 5 (1.3\%)    & & \\
\quad Classmates or friends         & 33 (8.6\%)   & 18 (4.7\%)   & & \\
\quad School or graduation           & 125 (32.6\%) & 110 (28.5\%) & & \\
\quad Enrich CV               & 68 (17.7\%)  & 60 (15.5\%)  & & \\
\quad Learn skill in AI era & 123 (32.0\%) & 168 (43.5\%) & & \\
\quad Missing                & 28 (7.3\%)   & 26 (6.7\%)   & & \\[4pt]

\textbf{Parent Education}    & & & 0.012 & 0.130 \\
\quad High school \& below          & 59 (15.4\%)  & 56 (14.5\%)  & & \\
\quad University                   & 142 (37.0\%) & 152 (39.4\%) & & \\
\quad Master                   & 119 (31.0\%) & 136 (35.2\%) & & \\
\quad PhD                   & 36 (9.4\%)   & 16 (4.1\%)   & & \\
\quad Missing                & 28 (7.3\%)   & 26 (6.7\%)   & & \\[4pt]

\textbf{Python Ability}      & & & 0.658 & 0.830 \\
\quad First time learner          & 111 (28.9\%) & 113 (29.3\%) & & \\
\quad Beginner           & 97 (25.3\%)  & 90 (23.3\%)  & & \\
\quad Intermediate               & 129 (33.6\%) & 143 (37.0\%) & & \\
\quad Proficient                & 14 (3.6\%)   & 13 (3.4\%)   & & \\
\quad Competition level & 5 (1.3\%)   & 1 (0.3\%)    & & \\
\quad Missing                & 28 (7.3\%)   & 26 (6.7\%)   & & \\
\midrule
\bottomrule
\end{tabular}

\begin{tablenotes}
\item \emph{Notes}: p-values computed from t-test (continuous) or chi-square test (categorical). BH-adjusted p-values control the false discovery rate across all covariates.
\end{tablenotes}

\end{threeparttable}
\end{table}

\subsection{Differential attrition} 
\label{sup:attrition}
To assess differential attrition by treatment status, we define an indicator $A_i=1$ if student $i$ is excluded from the main analysis sample (i.e., not present in the post-exclusion dataset) and $A_i=0$ otherwise. As shown in Table~\ref{tab:attrition_balance}, exclusion rates are nearly identical across arms: $140/524=26.7\%$ in control and $137/523=26.2\%$ in treatment. We fail to reject equality of attrition rates using both a two-sample proportion test ($p=0.85$) and a chi-squared test of equal rates ($p=0.90$).

\begin{table}[!htbp]
\centering
\caption{\textbf{Attrition balance by treatment condition.}}
\label{tab:attrition_balance}
\setlength{\tabcolsep}{3.5pt}
\renewcommand{\arraystretch}{1.05}
\footnotesize
\begin{threeparttable}
\begin{tabular}{lcccc}
\toprule
 & \textbf{Control} & \textbf{Treatment} & \textbf{Diff.\ (T--C)} & \textbf{$p$-value} \\
\midrule
$N$ (full) & 524 & 523 & -- & -- \\
Excluded $N$ & 140 & 137 & -- & -- \\
Attrition rate & 0.267 & 0.262 & $-0.005$ & -- \\
\midrule
Two-sample proportion test & \multicolumn{3}{c}{--} & 0.848 \\
Chi-squared test of equal rates & \multicolumn{3}{c}{--} & 0.903 \\
\bottomrule
\end{tabular}

\vspace{2pt}
\begin{tablenotes}
\footnotesize
\item \textit{Notes:} ``Attrition'' indicates exclusion from the main analysis sample. Diff.\ is Treatment minus Control.
\end{tablenotes}
\end{threeparttable}
\end{table}

Furthermore, we test for differential attrition within schools by estimating a linear probability and a logit model of $A_i$ on treatment assignment with school fixed effects:
\begin{equation}
A_i = \alpha + \beta\text{Treatment}_i + \sum_{s}\gamma_s\mathbbm{1}\{\text{School}_i=s\} + \varepsilon_i,
\label{eq:attrition_lpm}
\end{equation}
where $\text{Treatment}_i$ indicates assignment to the RL-personalized sequence and the school indicators absorb school-level differences in participation. 
\begin{equation}
\Pr(A_i=1) = \Lambda\!\left(\alpha + \beta\text{Treatment}_i + \sum_{s}\gamma_s\mathbbm{1}\{\text{School}_i=s\}\right),
\label{eq:attrition_logit}
\end{equation}
where $\Lambda(\cdot)$ denotes the logistic CDF. 

Table~\ref{tab:attrition_reg} shows that the estimated treatment effect on attrition is small and statistically indistinguishable from zero in both specifications (Logit: $\hat\beta=-0.057$, $p=0.73$; LPM: $\hat\beta=-0.78$ percentage points, $p=0.74$). Overall, these results suggest there is negligible differential attrition between treatment and control, alleviating concerns that sample exclusions bias our estimated treatment effects.

\begin{table}
\centering
\caption{\textbf{Regression tests for differential attrition.}}
\label{tab:attrition_reg}
\setlength{\tabcolsep}{6pt}
\renewcommand{\arraystretch}{1.15}
\footnotesize
\begin{tabular}{lcc}
\toprule
 & \textbf{Logit} & \textbf{LPM} \\
\midrule
Treatment indicator & $-0.057$ & $-0.0078$ \\
 & $(0.169)$ & $(0.0232)$ \\
$p$-value & 0.733 & 0.736 \\
\midrule
School fixed effects & Yes & Yes \\
$N$ & 1{,}047 & 1{,}047 \\
\bottomrule
\end{tabular}

\vspace{3pt}
\begin{minipage}{0.95\linewidth}
\footnotesize
\textit{Notes:} The dependent variable equals 1 if a student is excluded from the main analysis sample and 0 otherwise.
The Logit specification estimates Equation~\eqref{eq:attrition_logit} and reports coefficients on the log-odds scale. The LPM (linear probability model) estimates Equation~\eqref{eq:attrition_lpm}, where coefficients are interpreted as percentage-point changes in the probability of exclusion. Both specifications include school fixed effects. Standard errors are shown in parentheses.
\end{minipage}
\end{table}

\section{Evaluation}
\label{sup:eval}

\ref{sup:main} details the regression specifications for our main analysis and several robustness checks; \ref{sup:ruleout} provides evidence ruling out an alternative explanation that increased problem difficulty drives our findings; \ref{sup:hte} presents heterogeneity analyses; \ref{sup:mediation} reports mediation analyses; and \ref{sup:studentfeedback} summarizes student feedback on the AI tutoring program.

\subsection{Main Analysis and Robustness Checks}
\label{sup:main}

We analyze the impact of RL-based personalization on exam performance using ordinary least squares (OLS) regressions. For student $i$, our baseline specification is
\begin{equation}
    Y_i = \alpha + \beta \textit{RL}_i + \mathbf{X}_i' \boldsymbol{\gamma}
          + \delta_{\text{school}(i)} + \delta_{\text{grade}(i)}
          + \varepsilon_i,
    \label{eq:main-reg}
\end{equation}
where $Y_i$ is either the raw exam score or the standardized exam score (computed as a z-score using the mean and standard deviation of the full sample); $\textit{RL}_i$ is an indicator for treatment assignment to the RL personalization condition,
$\mathbf{X}_i$ is a vector of baseline covariates (including prior programming experience, parental education, self-reported prior Python ability, gender, baseline academic performance, and self-reported motivation), and $\delta_{\text{school}(i)}$ and $\delta_{\text{grade}(i)}$ are school and grade-level fixed effects, respectively.

\begin{table}
\centering
\footnotesize
\begin{threeparttable}
\caption{\textbf{Effect of RL Personalization on Exam Performance}}
\label{tab:rl_main}
\begin{tabular}{lcccc}
\toprule
 & \multicolumn{4}{c}{\textit{Dependent variable}} \\
\cmidrule(lr){2-5}
 & (1) & (2) & (3) & (4) \\
 & Standardized score & Raw score & Standardized score & Raw score \\
\midrule
(Intercept)               
 & -0.078  & $31.820^{***}$ & $-1.273^{***}$  & 3.957 \\
 & (0.051)       & (1.189)        & (0.229)       & (5.339) \\[2pt]

RL                         
 & $0.156^{*}$ & $3.636^{*}$  & $0.150^{*}$   & $3.503^{*}$ \\
 & (0.072)           & (1.679)      & (0.067)       & (1.572) \\[4pt]

Has programming experience 
 &                   &             & $0.246^{**}$  & $5.745^{**}$ \\
 &                   &             & (0.085)       & (1.982) \\[2pt]

Parent education (ordinal) 
 &                   &             & 0.018         & 0.425 \\
 &                   &             & (0.041)       & (0.959) \\[2pt]

Python ability (ordinal)  
 &                   &             & $0.125^{**}$ & $2.919^{**}$ \\
 &                   &             & (0.042)       & (0.985) \\[2pt]

Male                      
 &                   &             & $0.224^{**}$   & $5.231^{**}$ \\
 &                   &             & (0.078)       & (1.813) \\[2pt]

Academic performance (ordinal) 
 &                   &             & $0.079^{***}$ & $1.843^{***}$ \\
 &                   &             & (0.021)       & (0.499) \\[4pt]
\midrule
School fixed effects      
 & NO & NO & YES & YES \\
Grade level fixed effects 
 & NO & NO & YES & YES \\
Motivation controls       
 & NO & NO & YES & YES \\
\midrule
$R^{2}$                  
 & 0.006 & 0.006 & 0.253 & 0.253 \\
Adjusted $R^{2}$        
 & 0.005 & 0.005 & 0.226 & 0.226 \\
Observations            
 & 770   & 770   & 716   & 716   \\
F-statistic             
 & 4.69 (1, 770) & 4.69 (1, 770) & 9.34 (25, 716) & 9.34 (25, 716) \\
\midrule
\bottomrule
\end{tabular}
\begin{tablenotes}
\item \emph{Notes: Estimates use listwise deletion for missing baseline survey covariates in columns (3) and (4).$^{*} p<0.05$, $^{**} p<0.01$, $^{***} p<0.001$}.
\end{tablenotes}
\end{threeparttable}
\end{table}

Table~\ref{tab:rl_main} reports the effect of the RL-based problem sequencing on students' raw and standardized performance on the final exam. Columns (1) and (2) present OLS estimates without controls; columns (3) and (4) add baseline covariates and school and grade-level fixed effects.

As a robustness check, we also re-estimate the main specification using alternative standard error calculations to account for potential heteroskedasticity and within-school correlation of residuals. Results in Table~\ref{tab:rl_robustse} show that the positive effect of RL personalization remains statistically significant across all three specifications.

\begin{table}
\centering
\footnotesize
\begin{threeparttable}
\caption{\textbf{Effect of RL Personalization on Exam Performance with Alternative Standard Errors}}
\label{tab:rl_robustse}
\begin{tabular}{lccc}
\toprule
 & \multicolumn{3}{c}{\textit{Dependent variable: Standardized score}} \\
\cmidrule(lr){2-4}
 & (1) & (2) & (3) \\
 & OLS & HC1 robust & School-clustered \\
\midrule
(Intercept)               
 & $-1.273^{**}$ & $-1.273^{***}$ & $-1.273^{*}$ \\
 & (0.316) & (0.271) & (0.455) \\[2pt]
RL                         
 & $0.150^{*}$ & $0.150^{*}$ & $0.150^{*}$ \\
 & (0.067) & (0.067) & (0.069) \\[4pt]
Python ability (ordinal)  
 & $0.125^{**}$ & $0.125^{**}$ & 0.125 \\
 & (0.042) & (0.043) & (0.088) \\[2pt]
Male                      
 & $0.224^{**}$ & $0.224^{**}$ & 0.224 \\
 & (0.078) & (0.075) & (0.208) \\[2pt]
Has programming experience 
 & $0.246^{**}$ & $0.246^{**}$ & 0.246 \\
 & (0.085) & (0.081) & (0.164) \\[2pt]
Parent education (ordinal) 
 & 0.018 & 0.018 & 0.018 \\
 & (0.041) & (0.040) & (0.029) \\[2pt]
Academic performance (ordinal) 
 & $0.079^{***}$ & $0.079^{***}$ & $0.079^{***}$ \\
 & (0.021) & (0.022) & (0.020) \\[4pt]
\midrule
School fixed effects      
 & YES & YES & YES \\
Grade level fixed effects      
 & YES & YES & YES \\
Motivation controls       
 & YES & YES & YES \\
\midrule
$R^{2}$                  
 & 0.253 & 0.253 & 0.253 \\
Adjusted $R^{2}$        
 & 0.226 & 0.226 & 0.226 \\
Observations            
 & 716 & 716 & 716 \\
\midrule
\bottomrule
\end{tabular}
\begin{tablenotes}
\item \emph{Notes: Column (1) reports OLS standard errors. Column (2) reports heteroskedasticity-robust (HC1) standard errors. Column (3) reports school-clustered standard errors with CR2 small-sample correction. $^{*} p<0.05$, $^{**} p<0.01$, $^{***} p<0.001$.}
\end{tablenotes}
\end{threeparttable}
\end{table}

Recall that our main analysis omitted senior students due to complications with the implementation: (1) the course for seniors was accelerated due to their graduation timeline, (2) the program launch for senior students was moved substantially earlier than initially planned, and (3) key platform integrations were delayed due to governmental constraints. As a consequence, our RL algorithm was only active one month after senior students started the course, which is about half of their course period due to their accelerated timeline. Thus, we would not expect to observe significant benefits for these students. Nevertheless, for transparency, we report results for an expanded sample that includes senior students (Table~\ref{tab:rl_robust}). First, column (1) presents the OLS estimate without controls, and column (2) adds baseline covariates as well as school and grade-level fixed effects; these results are directionally the same as our main analysis, but as expected, they are not statistically significant. To disentangle the effects on senior students from those on non-seniors, columns (3)–(4) include a senior indicator and an RL$\times$Senior interaction term (with and without controls and fixed effects, respectively). In both specifications, the main RL coefficient remains positive, whereas the interaction term is negative and statistically insignificant, which is consistent with the hypothesis that our system was effective for non-seniors but not for seniors.

Finally, we examine our primary treatment effect estimates controlling for the randomly-assigned pedagogical approach used in the prompt (these were also randomized at the student-level, see discussion in \ref{sup:preregist}). Specifically, the AI tutor prompt was randomized across three pedagogical frameworks (Constructivism, Behaviorism, and Social Learning) so that the tutor followed a consistent instructional style throughout a student’s interactions.\footnote{Constructivism uses a student-centered, discovery-based approach that prompts students to articulate the problem in their own words and explain their reasoning; Behaviorism emphasizes structured, repetitive practice with immediate feedback; Social Learning positions the tutor in a peer-like ``learning buddy'' role rather than a traditional instructor.} Table~\ref{tab:main_pedagogy} reports OLS estimates of Equation~\eqref{eq:main-reg} but additionally including indicators for the type of randomized AI-tutor pedagogical prompts. Consistent with our main result, assignment to RL-based personalization improves standardized exam scores by $0.145$ standard deviations ($p=0.026$). The estimated coefficients on the pedagogical prompt indicators are noisy and statistically insignificant, and including them leaves the estimated RL effect essentially unchanged, suggesting that the pedagogical style of the prompt does not materially affect our primary treatment estimate.

\begin{table}
\centering
\footnotesize
\begin{threeparttable}
\caption{\textbf{Results Including Delayed Implementation with Seniors}}
\label{tab:rl_robust}
\begin{tabular}{lcccc}
\toprule
 & \multicolumn{4}{c}{\textit{Dependent variable: Standardized score}} \\
\cmidrule(lr){2-5}
 & (1) & (2) & (3) & (4) \\
\midrule
(Intercept)
 & $0.088^{*}$ & $-1.016^{***}$ & $-0.078$ & $-1.238^{***}$ \\
 & (0.044) & (0.288) & (0.051) & (0.207) \\[2pt]

RL
 & 0.116$^{\dagger}$ & 0.114$^{\dagger}$ & $0.156^{*}$ & $0.144^{*}$ \\
 & (0.063) & (0.058) & (0.072) & (0.067) \\[2pt]

Senior
 &  &  & 0.110 & 0.237$^{\dagger}$ \\
 &  &  & (0.125) & (0.140) \\[2pt]

RL $\times$ Senior
 &  &  & $-0.219$ & $-0.185$ \\
 &  &  & (0.176) & (0.174) \\[4pt]

Has programming experience
 &  & $0.221^{**}$ &  & $0.229^{**}$ \\
 &  & (0.074) &  & (0.078) \\[2pt]

Parent education (ordinal)
 &  & 0.015 &  & 0.019 \\
 &  & (0.035) &  & (0.038) \\[2pt]

Python ability (ordinal)
 &  & $0.115^{**}$ &  & $0.124^{**}$ \\
 &  & (0.037) &  & (0.039) \\[2pt]

Male
 &  & $0.234^{***}$ &  & $0.258^{***}$ \\
 &  & (0.067) &  & (0.072) \\[2pt]

Academic performance (ordinal)
 &  & $0.070^{***}$ &  & $0.075^{***}$ \\
 &  & (0.018) &  & (0.019) \\[4pt]

\midrule
School fixed effects
 & NO & YES & NO & YES \\
Grade level fixed effects
 & NO & YES & NO & YES \\
Motivation controls
 & NO & YES & NO & YES \\
\midrule
$R^{2}$
 & 0.004 & 0.279 & 0.005 & 0.235 \\
Adjusted $R^{2}$
 & 0.003 & 0.255 & 0.002 & 0.209 \\
Observations
 & 924 & 839 & 924 & 839 \\
F-statistic
 & 3.41 (1, 922) & 12.05 (26, 812) & 1.61 (3, 920) & 9.21 (27, 811) \\
\bottomrule
\end{tabular}

\begin{tablenotes}
\item \emph{Notes:} Columns (1)--(2) uses the main regression specification, while Columns (3)--(4) additionally includes the senior interaction term (RL $\times$ Senior). $^{\dagger}p<0.1$, $^{*}p<0.05$, $^{**}p<0.01$, $^{***}p<0.001$.
\end{tablenotes}
\end{threeparttable}
\end{table}

\begin{table}[!t]
\centering
\footnotesize
\begin{threeparttable}
\caption{\textbf{Result of Main Analysis with Pedagogy Controls}}
\label{tab:main_pedagogy}
\begin{tabular}{lc}
\toprule
& \textit{Dependent variable: Standardized score} \\
\cmidrule(lr){2-2}
& (1) \\
\midrule
RL
& 0.151$^{*}$ \\
& (0.067) \\[2pt]

Prior Python ability (ordinal)
& 0.125$^{**}$ \\
& (0.042) \\[2pt]

Prior programming experience
& 0.246$^{**}$ \\
& (0.085) \\[2pt]

Parent education (ordinal)
& 0.022 \\
& (0.041) \\[2pt]

Baseline academic performance (ordinal)
& 0.078$^{***}$ \\
& (0.021) \\[2pt]

Male
& 0.232$^{**}$ \\
& (0.078) \\[2pt]

AI Pedagogy: Constructivism
& -0.101 \\
& (0.083) \\[2pt]

AI Pedagogy: Behaviorism
& -0.022 \\
& (0.081) \\[2pt]
\midrule
Grade-level fixed effects
& YES \\
School fixed effects
& YES \\
Motivation controls
& YES \\
\midrule
$R^2$
& 0.255 \\
Adjusted $R^2$
& 0.225 \\
Observations
& 716 \\
F-statistic
& 8.703 (27, 688) \\
\midrule
\bottomrule
\end{tabular}

\begin{tablenotes}[flushleft]
\footnotesize
\item \emph{Notes:} OLS estimates with standard errors in parentheses. $^{*}p<0.05$, $^{**}p<0.01$, $^{***}p<0.001$.
\end{tablenotes}
\end{threeparttable}
\end{table}

\subsection{Ruling Out ``Harder is Better''}
\label{sup:ruleout}

Recall that we found that the treatment group received somewhat higher difficulty questions on average than the control group (see Table~\ref{tab:pilot_difficulty}). Thus, one alternative explanation for our results is that the performance gains were driven simply by exposure to more difficult questions. We are unable to rule out this possibility using our RCT data because difficulty assignment was endogenous in both conditions of our study: in the treatment group, difficulty was dynamically determined by the RL policy, whereas in the control group, it followed a fixed, increasing sequence, which means that exposure to harder problems was confounded with student persistence. However, our pilot data---performed on the same course material and on a comparable student population (see~\ref{sup:pilot})---does not have this limitation, since in the pilot setting, difficulty was assigned exogenously: in the control arm, difficulty is randomized rather than adaptively chosen. Therefore, we estimate regressions within the pilot control group relating assigned difficulty to learning outcomes (e.g., exam or quiz performance), controlling for baseline covariates and AI tutor assignment. Because difficulty is randomized in this arm, this analysis isolates the causal effect of receiving harder questions holding other factors constant.  Specifically, letting student $i$'s average practice-question difficulty be denoted by \textit{AvgDif}, we estimate:
\begin{equation}
Y_{i}
= \alpha
+ \beta\textit{AvgDif}_{i}
+ \eta\textit{AI}_{i}
+ \theta\textit{PriorPython}_{i}
+ \mathbf{X}_{i}' \boldsymbol{\gamma}+
\delta_{\text{school}(i)} + \delta_{\text{grade}(i)}
+ \varepsilon_{i},
\label{eq:pilot_dif_fe_main}
\end{equation}
where $Y_i$ denotes standardized exam score, $\textit{PriorPython}_{i}$ is a dummy indicator of whether students have prior Python coding background, $\mathbf{X}_i$ includes baseline demographic controls, and $\delta_{\text{school}(i)},~\delta_{\text{grade}(i)}$ are school and grade-level fixed effects.

We also look at heterogeneous treatment effects with respect to students' prior Python coding background (\textit{PriorPython}). To this end, we estimate
\begin{equation}
\begin{aligned}
Y_{i} =& \alpha
+ \beta_1\textit{AvgDif}_{i}
+ \beta_2\textit{PriorPython}_{i}
+ \beta_3\big(\textit{AvgDif}_{i}\times \textit{PriorPython}_{i}\big) \\
&+ \eta\textit{AI}_{i}
+ \mathbf{X}_{i}\boldsymbol{\gamma}+
\delta_{\text{school}(i)} + \delta_{\text{grade}(i)}+
+ \varepsilon_{i}.
\end{aligned}
\label{eq:pilot_dif_fe_interaction}
\end{equation}
Results reported in Table~\ref{tab:pilot_difficulty} show that increased difficulty alone does not account for improved performance and could potentially harm beginners, supporting the interpretation that the gains in our main study are driven by adaptive sequencing rather than by mechanical difficulty inflation.

\begin{table}[!t]
\centering
\footnotesize
\begin{threeparttable}
\caption{\textbf{Pilot study: Practice-question difficulty exposure and assessment performance}}
\label{tab:pilot_difficulty}
\begin{tabular}{lcc}
\toprule
& \multicolumn{2}{c}{\textit{Dependent variable: Standardized score}} \\
\cmidrule(lr){2-3}
& (1) & (2) \\
\midrule
AvgDif
& 0.417 & 0.271 \\
& (0.358) & (0.252) \\[2pt]

PriorPython
& 0.909$^{*}$ & -3.739 \\
& (0.375) & (1.598) \\[2pt]

AvgDif $\times$ PriorPython
&  & 2.542$^{*}$ \\
&  & (0.863) \\[2pt]

AI
& 0.685$^{*}$ & 0.357 \\
& (0.285) & (0.226) \\[2pt]

Parent Education
& 0.542$^{*}$ & 0.710$^{**}$ \\
& (0.213) & (0.158) \\[2pt]

ProgrammingBackground
& 0.296 & 0.676 \\
& (0.370) & (0.286) \\[2pt]

Gender
& -0.110 & -0.028 \\
& (0.306) & (0.213) \\[2pt]

(Intercept)
& -1.414 & -0.751 \\
& (0.697) & (0.531) \\
\midrule
School fixed effects
& YES & YES \\
Study Major control
& YES & YES \\
\midrule
Observations
& 28 & 28 \\
Adjusted $R^{2}$ & 0.800 & 0.905 \\
\bottomrule
\end{tabular}

\begin{tablenotes}[flushleft]
\footnotesize
\item \emph{Notes:} OLS estimates with standard errors in parentheses. AvgDif denotes average assigned practice-question difficulty. 3 observations were deleted due to missingness in survey response (listwise deletion). $^{*}p<0.05$, $^{**}p<0.01$, $^{***}p<0.001$.
\end{tablenotes}
\end{threeparttable}
\end{table}

\subsection{Heterogeneity Analysis}
\label{sup:hte}

In addition to using subgroup regression to assess the heterogeneous effect of RL-based personalization, we also extend the main specification in Equation~\eqref{eq:main-reg} by interacting the treatment indicator with subgroup indicators. First, we define $\textit{PythonFamiliar}_i$ as an indicator equal to 1 for students who reported \emph{some} prior Python exposure (e.g., basic familiarity but not competition-level proficiency) before entering the program, and 0 otherwise (Python beginners).\footnote{We exclude 6 students (1\%) who reported national competition-level Python skill, since they had already
mastered the curriculum prior to the course and would reflect ceiling effects.} Then, we estimate:
\begin{equation}
Y_i = \alpha + \beta_1\textit{RL}_i
+ \beta_2\textit{PythonFamiliar}_i
+ \beta_3\big(\textit{RL}_i \times \textit{PythonFamiliar}_i\big)
+ \mathbf{X}_i' \boldsymbol{\gamma}
+ \delta_{\text{school}(i)} + \delta_{\text{grade}(i)}
+ \varepsilon_i,
\label{eq:hetero-reg-python}
\end{equation}
where $Y_i$ is the standardized outcome and $\mathbf{X}_i$ contains the same baseline controls as in the main analysis, along with school and grade-level fixed effects. In this specification, $\beta_1$ is the treatment effect for Python beginners, and $\beta_3$ captures the \emph{differential} effect for Python-familiar students relative to beginners; thus, the treatment effect for Python-familiar students is $\beta_1+\beta_3$. Next, we examine heterogeneity with respect to school selectivity by replacing $\textit{PythonFamiliar}_i$ with an indicator $\textit{SchoolTier}_i$, defined using entrance exam scores as described in the main paper. It is equal to 1 for higher-tier school otherwise 0:
\begin{equation}
Y_i = \alpha + \beta_1\textit{RL}_i
+ \beta_2\textit{SchoolTier}_i
+ \beta_3\big(\textit{RL}_i \times \textit{SchoolTier}_i\big)
+ \mathbf{X}_i' \boldsymbol{\gamma}
+ \delta_{\text{school}(i)} + \delta_{\text{grade}(i)}
+ \varepsilon_i.
\label{eq:hetero-reg-tier}
\end{equation}

Table~\ref{tab:hteall} reports the estimates from both types of heterogeneous treatment effects. Consistent with Fig.~\ref{fig:hteforest} in the main paper, the point estimates indicate that the gains from RL personalization are largest among Python beginners. In the Python-skill interaction specification (Table~\ref{tab:hteall}, cols.~(1)--(2)), the RL effect is positive and statistically significant for Python beginners (the reference group). In contrast, the incremental effect for Python-familiar students is negative (RL$\times$PythonFamiliar $=-0.249$ SD, $p=0.066$), implying that the treatment effect for Python-familiar students ($\beta_1+\beta_3$) is much smaller and not statistically distinguishable from zero. In the school-tier interaction specification (Table~\ref{tab:hteall}, cols.~(3)--(4)), we find no evidence that RL disproportionately benefits students in higher-tier schools: the interaction term is statistically insignificant (RL$\times$SchoolTier $=-0.145$ SD, $p=0.328$), indicating that performance gains do not systematically favor already advantaged students.

\begin{table}[!t]
\centering
\caption{\textbf{Heterogeneous effects of RL personalization by prior Python skill and school tier.}
This table reports interaction regressions that extend Eq.~\eqref{eq:main-reg} by interacting the RL assignment indicator with (i) prior Python familiarity and (ii) school tier. Columns (1)--(2) use the Python-skill interaction; columns (3)--(4) use the school-tier interaction.}
\label{tab:hteall}
\footnotesize
\setlength{\tabcolsep}{6pt}
\renewcommand{\arraystretch}{1.15}

\begin{tabular}{lcccc}
\toprule
& \multicolumn{2}{c}{Prior Python skill interaction} & \multicolumn{2}{c}{School tier interaction} \\
\cmidrule(lr){2-3}\cmidrule(lr){4-5}
& (1) Standardized score & (2) Raw score & (3) Standardized score & (4) Raw score \\
\midrule

RL
& 0.235$^{**}$ & 5.498$^{**}$ & 0.191$^{*}$ & 4.459$^{*}$ \\
& (0.088) & (2.045) & (0.079) & (1.850) \\

RL $\times$ Subgroup
& $-0.249^{\dagger}$ & $-5.816^{\dagger}$ & $-0.145$ & $-3.393$ \\
& (0.135) & (3.155) & (0.148) & (3.464) \\

Subgroup indicator
& 0.414$^{***}$ & 9.674$^{***}$ & 0.809$^{**}$ & 18.882$^{**}$ \\
& (0.103) & (2.400) & (0.263) & (6.133) \\

\addlinespace[4pt]
Prior Python ability (ordinal)
&  &  & 0.126$^{**}$ & 2.936$^{**}$ \\
&  &  & (0.042) & (0.985) \\

Has programming experience
& 0.242$^{**}$ & 5.659$^{**}$ & 0.243$^{**}$ & 5.672$^{**}$ \\
& (0.083) & (1.940) & (0.085) & (1.984) \\

Male
& 0.218$^{**}$ & 5.093$^{**}$ & 0.225$^{**}$ & 5.251$^{**}$ \\
& (0.077) & (1.803) & (0.078) & (1.813) \\

Parent education (ordinal)
& 0.022 & 0.511 & 0.017 & 0.398 \\
& (0.041) & (0.963) & (0.041) & (0.959) \\

Academic performance (ordinal)
& 0.078$^{***}$ & 1.811$^{***}$ & 0.079$^{***}$ & 1.849$^{***}$ \\
& (0.021) & (0.500) & (0.021) & (0.499) \\
\midrule
\addlinespace[4pt]
School fixed effects & YES & YES & YES & YES \\
Grade-level fixed effects  & YES & YES & YES & YES \\
Motivation controls  & YES & YES & YES & YES \\

\midrule
Observations & 710 & 710 & 716 & 716 \\
$R^2$        & 0.261 & 0.261 & 0.254 & 0.254 \\
Adj.\ $R^2$  & 0.233 & 0.233 & 0.226 & 0.226 \\
\midrule
\bottomrule

\multicolumn{5}{l}{\footnotesize Notes: Standard errors in parentheses. $^{***}p<0.001$, $^{**}p<0.01$, $^{*}p<0.05$, $^{\dagger}p<0.10$.} \\
\end{tabular}
\end{table}

\subsection{Mediation Analysis}
\label{sup:mediation}

We conduct a mediation analysis in the potential-outcomes framework following \cite{imai2010general}. We examine two measures of student engagement as mediators: (i) total time spent on the platform (\textit{totalTime}), and (ii) total number of attempts (\textit{totalAttempt}), showing that the RL policy increased performance by driving students to spend more time and effort.

For each mediator $M_i \in \{\textit{totalTime}, \textit{totalAttempt}\}$, we estimate a mediator model and an outcome model, using the same baseline covariates and fixed effects as in the main analysis:
\begin{align}
M_i &= \alpha_M + \tau\textit{RL}_i + \mathbf{X}_i'\boldsymbol{\gamma}_M
+ \delta_{\text{school}(i)} + \delta_{\text{grade}(i)} + u_i, 
\label{eq:med-m} \\
Y_i &= \alpha_Y + \beta\textit{RL}_i + \lambda M_i + \mathbf{X}_i'\boldsymbol{\gamma}_Y
+ \delta_{\text{school}(i)} + \delta_{\text{grade}(i)} + \varepsilon_i,
\label{eq:med-y}
\end{align}
where $Y_i$ is the standardized score outcome and $\mathbf{X}_i$ includes the same controls as in the main specification. We then use the \texttt{mediation} package in R to estimate the average mediation effect (ACME; indirect effect) and the average direct effect (ADE), with nonparametric bootstrap percentile confidence intervals (1{,}000 simulations), as recommended by \cite{imai2010general}. 

Table~\ref{tab:med_results} summarizes the results. Across both mediators, the estimated ACME is positive and precisely estimated, while the ADE is statistically indistinguishable from zero. For \textit{totalTime}, the ACME is $0.193$ SD ($p<0.001$) and the ADE is $-0.031$ SD ($p=0.682$); for \textit{totalAttempt}, the ACME is $0.152$ SD ($p<0.001$) and the ADE is $0.009$ SD ($p=0.854$). These estimates imply that the overall treatment effect ($\approx 0.15$ SD) operates primarily through increased engagement rather than through a residual direct channel.\footnote{The estimated proportion mediated can exceed 1 when the estimated direct and indirect components have opposite signs and/or due to sampling variability, which is common in finite samples. We interpret these estimates as evidence that mediation through engagement accounts for almost all of the total effect.}

\begin{figure}[!t]
\centering
\label{fig:med_sensitivity}
\begin{subfigure}[t]{0.49\linewidth}
  \centering
  \includegraphics[width=\linewidth]{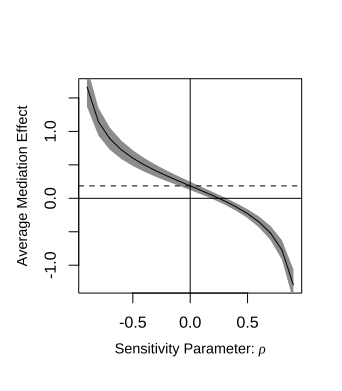} 
  \caption{Mediator: total time spent.}
  \label{fig:med_sens_time}
\end{subfigure}\hfill
\begin{subfigure}[t]{0.49\linewidth}
  \centering
  \includegraphics[width=\linewidth]{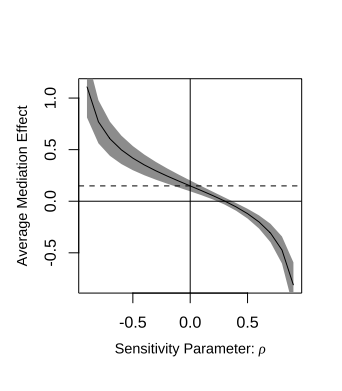} 
  \caption{Mediator: total number of attempts.}
  \label{fig:med_sens_attempt}
\end{subfigure}
\caption{\textbf{Sensitivity analysis for mediation through engagement.}
Sensitivity plots for the estimated average causal mediation effect (ACME) under potential violations of sequential ignorability. Each panel shows ACME as a function of the sensitivity parameter $\rho$, the correlation between unobserved determinants of the mediator and the outcome. The solid curve gives the point estimate and the shaded band indicates the 95\% confidence interval. The vertical line marks $\rho=0$ (no unobserved confounding), and the horizontal line marks zero mediated effect.}
\end{figure}

Finally, we assess robustness to violations of sequential ignorability using the sensitivity analysis in \cite{imai2010general}, parameterized by $\rho$, the correlation between the unobserved components of the mediator and outcome models. The ACME would be attenuated to zero only if $\rho$ were approximately $0.3$ for \textit{totalTime} and $0.3$ for \textit{totalAttempt} (see Fig.~\ref{fig:med_sens_time}--\ref{fig:med_sens_attempt}), indicating that the mediated effect is robust to moderate levels of unobserved confounding.

\begin{table}[!t]
\centering
\caption{\textbf{Mediation analysis: engagement as a mechanism.}
ACME denotes the average causal mediation effect (indirect effect) and ADE denotes the average direct effect. Confidence intervals are nonparametric bootstrap percentile intervals with 1{,}000 simulations.}
\label{tab:med_results}
\small
\setlength{\tabcolsep}{7pt}
\renewcommand{\arraystretch}{1.15}
\begin{tabular}{lcc}
\toprule
& \multicolumn{2}{c}{\textit{Mediators}} \\
\cmidrule(lr){2-3}
& \textbf{Total Time Spent} & \textbf{Total Attempts} \\
\midrule
ACME (indirect effect) 
& 0.193 [0.137,0.252]$^{***}$ 
& 0.153 [0.108,0.200]$^{***}$ \\
ADE (direct effect)  
& -0.031 [-0.170,0.109] 
& 0.009 [-0.109,0.139] \\
Total effect          
& 0.162 [0.034,0.298]$^{*}$ 
& 0.162 [0.037,0.297]$^{*}$ \\
Proportion mediated   
& 1.189 [0.571,5.113]$^{*}$ 
& 0.943 [0.499,3.148]$^{*}$ \\
\addlinespace[3pt]
$\rho$ at which ACME $=0$ (sensitivity) 
& 0.3 & 0.3 \\
\midrule
Sample size & 712 & 712 \\
Bootstrap simulations & 1{,}000 & 1{,}000 \\
\bottomrule
\multicolumn{3}{p{0.92\linewidth}}{\footnotesize \textit{Notes:} Entries report point estimates with 95\% confidence intervals in brackets. $^{***}p<0.001$, $^{**}p<0.01$, $^{*}p<0.05$. Sensitivity parameter $\rho$ captures the correlation between unobserved determinants of the mediator and the outcome. The sample size (N=712) is reduced from the main analysis with controls (N=716) due to missing total attempt and total time data for four participants. These participants showed minimal engagement (no homework completion) despite remaining enrolled through the final certification exam.}
\end{tabular}
\end{table}

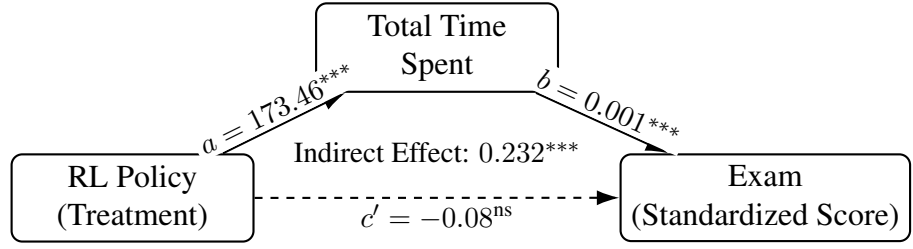
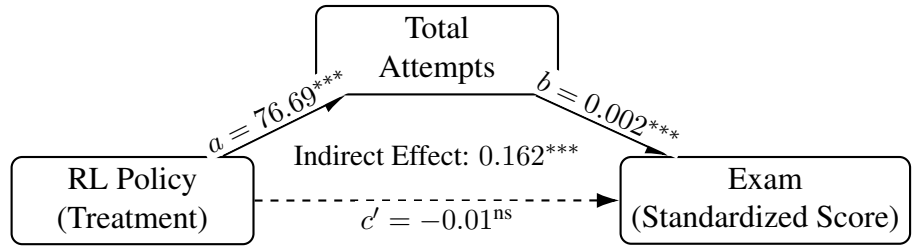
\begin{figure}[!t]
\centering

\tikzset{
node_style/.style={draw, rectangle, rounded corners, minimum width=3.2cm, minimum height=1cm, align=center, fill=white, thick},
arrow_style/.style={->, >={Latex[length=3mm, width=2mm]}, thick},
label_style/.style={midway, fill=white, font=\small, inner sep=2pt}
}

\begin{subfigure}[b]{0.8\textwidth}
\centering
\begin{tikzpicture}[node distance=1.5cm and 2.5cm]
\node[node_style] (RL) {RL Policy\\(Treatment)};
\node[node_style, above right=0.8cm and 0.8cm of RL] (Med) {Total Time \\ Spent};
\node[node_style, below right=0.8cm and 0.8cm of Med] (Score) {Exam\\(Standardized Score)};

\draw[arrow_style] (RL) -- (Med) node[label_style, sloped, above] {$a=173.46^{***}$};

\draw[arrow_style] (Med) -- (Score) node[label_style, sloped, above] {$b=0.001^{***}$};

\draw[arrow_style, dashed] (RL) -- (Score) node[label_style, below] {$c'=-0.08^{\text{ns}}$};

\node[font=\small, align=center, below=0.5cm of Med] {Indirect Effect: $0.232^{***}$};
\end{tikzpicture}
\caption{Mediator: Total Time Spent per Practice Problem (sec)}
\label{fig:path_a}
\end{subfigure}

\vspace{1cm} 

\begin{subfigure}[b]{0.8\textwidth}
\centering
\begin{tikzpicture}[node distance=1.5cm and 2.5cm]
\node[node_style] (RL) {RL Policy\\(Treatment)};
\node[node_style, above right=0.8cm and 0.8cm of RL] (Med) {Total \\ Attempts};
\node[node_style, below right=0.8cm and 0.8cm of Med] (Score) {Exam\\(Standardized Score)};

\draw[arrow_style] (RL) -- (Med) node[label_style, sloped, above] {$a=76.69^{***}$};

\draw[arrow_style] (Med) -- (Score) node[label_style, sloped, above] {$b=0.002^{***}$};

\draw[arrow_style, dashed] (RL) -- (Score) node[label_style, below] {$c'=-0.01^{\text{ns}}$};

\node[font=\small, align=center, below=0.5cm of Med] {Indirect Effect: $0.162^{***}$};
\end{tikzpicture}
\caption{Mediator: Total Number of Attempts}
\label{fig:path_b}
\end{subfigure}

\caption{\textbf{Path Analysis Results.} Structural equation models testing the mediation of the treatment effect via (a) Total Time Spent and (b) Total Attempts. Path $a$ represents the effect of RL on the mediator; Path $b$ represents the effect of the mediator on the exam score; Path $c'$ represents the direct effect of RL on the exam score after controlling for the mediator. \textit{Note: Values represent coefficients. ***$p<0.001$; ns = not significant.}}
\label{fig:path_analysis}
\end{figure}

As a robustness analysis, we conducted path analyses using the \emph{lavaan} package in R to examine whether the effect of the RL policy on exam performance is mediated by increased engagement, measured by two metrics: \emph{total time spent to solve the question} and \emph{total number of attempts}. We use structural equation modeling (SEM) with 1,000 bootstrap resamples to ensure robust inference. 
The results, summarized in Fig.~\ref{fig:path_analysis}, show a consistent mechanism across both measures. First, the total effect of the RL intervention on exam scores is positive and significant ($\beta_{\text{total}} = 0.158, p=0.023$); furthermore, this gain is driven entirely by the indirect path through engagement. 

For time taken, RL significantly increased total time taken ($a=173.46, p<0.001$), which in turn strongly predicted higher exam performance ($b=0.001, p<0.001$), yielding a significant positive indirect effect ($\beta_{\text{indirect}} = 0.232, p<0.001$). Similarly, for practice intensity, the RL policy significantly increased the number of attempts ($a=76.69, p<0.001$), which also predicted higher scores ($b=0.002, p<0.001$), resulting in a significant indirect effect ($\beta_{\text{indirect}} = 0.162, p<0.001$). 

In both models, the direct effect of the RL policy on performance—after accounting for the mediator—was small and statistically indistinguishable from zero ($\beta_{\text{direct (time)}} = -0.081, p=0.235$; $\beta_{\text{direct (attempts)}} = -0.011, p=0.869$). This pattern of full mediation across both time spent and practice intensity suggests that the primary driver of the RL system's efficacy is its ability to sustain student effort and extend productive practice. 

\subsection{Student Feedback}
\label{sup:studentfeedback}

Halfway through the study (before starting Module 7), we asked students one open-ended question ``What are your thoughts and suggestions about using this learning platform?'' Overall, the students' responses reflect the successful implementation of our learning platform. We give a summary and example quotes below:

\paragraph{Overall Educational Experience.}
Despite the complexities of introducing a new system, students responded positively to the core learning experience. Students found the platform effective, specifically noting the clarity of instruction and the diversity of practice materials. The consensus was that the system successfully facilitated a meaningful learning process.

\begin{quote}
\itshape
``It is very easy to use, and the explanations are very clear.'' \\
``The practice questions are very diverse; I like them a lot.'' \\
``The instructor's explanations are very clear...this time I truly felt that I learned.''
\end{quote}

\paragraph{Addressing System Stability.} During the first two weeks, we encountered stability issues—specifically lag and session timeouts—stemming from traffic congestion on the government's digital learning system (SSO) linkage. However, our team moved quickly to mitigate the impact by fixing the technical glitches and extending the deadline for the first required homework. These early issues were resolved quickly and did not persist through the remainder of the curriculum.

\begin{quote}
\itshape
``The work session often expires; there are webpage errors that require refreshing.'' \\
``Sometimes it crashes and I have to reload repeatedly.''
\end{quote}

\paragraph{AI Tutor.} Students found it to be a convenient companion for identifying mistakes, though we acknowledge there is still room for improvement regarding the precision of its answers.

\begin{quote}
\itshape
``Having an AI on the side to ask questions is great---it is very convenient!'' \\
``The AI is useful, but its effectiveness is limited---there is still a lot of room for improvement.''
\end{quote}

Overall, our work is a proof of the concept that tightly integrating a carefully-designed GenAI chatbot with a reinforcement learning algorithm for sequencing practice problems can be an effective way of improving student learning outcomes. With further improvements in the design of the tutoring platform, chatbot, and delivery, we may anticipate even larger gains.

\section{Summary of Related Work on Adaptive Learning Policies}
\label{sup:literature}

\begin{table}[!t]
\tiny
\caption{\textbf{Summary of related work on adaptive learning}}
\label{tab:lit}
\begin{tabular}{lllrllll}
\toprule
\textbf{Paper} & \textbf{Domain} & \textbf{Population} & \textbf{$N$} & \textbf{Adaptation Policy} & \textbf{Baseline} & \textbf{Outcome} & \textbf{Significance} \\
\midrule
%
%
%
Rafferty et al. (2016)~\cite{rafferty2016faster} & Alpha. Arith. & Online (self-recruit) & 40 & POMDP & Random & Time taken & Sig. (Conditional) \\
\addlinespace
Sen et al. (2018)~\cite{sen2018machine} & Chemistry & Online (AMT) & 325 & ML-generated seq. & Random/Human & Test score & Sig. \\
\addlinespace
Clement et al. (2015)~\cite{clement2015multiarmed} & Arithmetic & Ages 7-8 & 400 & MAB Intel. System & Human Expert & Test score & Mixed \\
\addlinespace
David et al. (2016)~\cite{david2016sequencing} & Mathematics & K-12 & 73 & BKT & Ascending Algo. & Test score & Not Sig. \\
\addlinespace
Schatten (2017)~\cite{schatten2017sequencing} & Mathematics & K-12 & 98 & Vygotsky Policy (VPS) & Rule-based & Test score & Not Sig. \\
\addlinespace
Doroudi et al. (2017)~\cite{doroudi2019wheres} & Mathematics & Grades 4-5 & 69 & BKT-based adaptive & Spiral difficulty & Test score & Not Sig. \\
\addlinespace
Segal et al. (2018)~\cite{segal2018combining} & Mathematics & Grade 7 & 92 & Multi-Armed Bandit & BKT \& Ascending & Test score & Not Sig. \\
\addlinespace
Zhou et al. (2017)~\cite{zhou2017towards} & Probability & Undergrad & 153 & RL & Random & Test score & Sig. \\
\addlinespace
Shen \& Chi (2016)~\cite{shen2016reinforcement} & Logic & Undergrad & 106 & RL & Random & Test score & Not Sig. \\
\addlinespace
Shen et al. (2018)~\cite{shen2018improving} & Logic & Undergrad & 77 & Constrained POMDP & Random & Test score & Sig. (Low knowledge only) \\
\addlinespace
Shen et al. (2018) S1~\cite{shen2018empirically} & Logic & Undergrad & 105 & POMDP & Random & Test score & Not Sig. \\
\addlinespace
Shen et al. (2018) S2~\cite{shen2018empirically} & Logic & Undergrad & 181 & POMDP (w/ features) & Random & Test score & Sig. \\
\addlinespace
Rowe \& Lester (2015)~\cite{rowe2015improving} & Microbiology & Grade 8 & 75 & Modular RL & Random & Test score & Not Sig. \\
\addlinespace
Beck et al. (2000)~\cite{beck2000advisor} & Arithmetic & Grade 6 & 97 & RL & Human Expert & Time taken & Sig. \\
\addlinespace
Koedinger et al. (1997)~\cite{koedinger1997intelligent} & Algebra & Grade 9 & 470 & PUPM + Algebra Tutor & Trad. Curriculum & Test score & Sig. \\
\addlinespace
Pane et al. (2014)~\cite{pane2014effectiveness} & Algebra & Grades 9-12 & 18,700 & CTAI & Existing Curr. & Test score & Mixed \\
\addlinespace
Pane et al. (2014)~\cite{pane2014effectiveness} & Algebra & Grades 6-8 & 6,800 & CTAI & Existing Curr. & Test score & Not Sig. \\
\addlinespace
Belfer et al. (2022)~\cite{belfer2022raising} & Data Science & Online (self-recruit) & 44 & Contextual Bandit & Rule-based & Completion rate & Sig. \\
\addlinespace
Ritter et al. (2007)~\cite{ritter2007evidence} & Algebra & Grade 9 & 426 & Cognitive Tutor (+Train) & Existing Curr. & Test score & Sig. \\
\addlinespace
Jaciw et al. (2007)~\cite{jaciw2007comparative} & Algebra & Grades 8-13 & 541 & Cognitive Tutor (+Train) & Existing Curr. & Test score & Not Sig. \\
\addlinespace
Pane et al. (2010)~\cite{pane2010experiment} & Geometry & Grades 9-12 & 1,640 & Cognitive Tutor (+Train) & Existing Curr. & Test score & Not Sig. \\
\addlinespace
Roschelle et al. (2016)~\cite{roschelle2016online} & Mathematics & Grade 7 & 2,850 & ASSISTments (+PD) & Existing Practice & Test score & Sig. \\
\addlinespace
Clément et al. (2024)~\cite{clement2024improved} & Mathematics & Grade 2 & 265 & Multi-Armed Bandit & Human Expert & Test score & Sig. \\
\addlinespace
Zhou et al. (2022)~\cite{zhou2022leveraging} & Discrete Math & Undergrad & 180 & HRL & Random & Test score & Sig. \\
\addlinespace
Bassen et al. (2020)~\cite{bassen2020reinforcement} & Lin. Algebra & Amazon Employees & 1,987 & RL (activities) & Self-directed & Test score & Sig. \\
\addlinespace
Mandel et al. (2014)~\cite{mandel2014offline} & Fractions & Online (Game recruit) & 2,000 & POMDP/HMM & Random/Expert & Concept count & Sig. \\
\addlinespace
Schmucker et al. (2025)~\cite{schmucker2023learning} & Biology & Online Class & 27,268 & Bandit (select hint) & Random & Accuracy & Sig. \\
\addlinespace
Prihar et al. (2023)~\cite{prihar2022automatic} & Mathematics & Middle School & 3,602 & Contextual MAB & Random & Accuracy & Not Sig. \\
\addlinespace
Harrison et al. (2024)~\cite{harrison2024study} & Mathematics & UK Secondary School & 2,901 &
PVAE + Bayesian Optimization
& Existing Curr. & Test score & Sig. \\
\addlinespace
Alam et al. (2024)~\cite{alam2024much} & Mathematics & Undergrad & 212 & Deep RL & Ascending difficulty & Test score & Not Sig. \\
\addlinespace
Naslund-Hadleyetal et al. (2025)~\cite{Naslund-Hadleyetal2025} & Mathematics and reading & Age 10-12 & 1900 & rule-based personalization & Existing Curr. & Test score & Not Sig. (Math); Sig (English)\\
\bottomrule
\end{tabular}
\begin{minipage}{\linewidth}
\tiny
\textit{Notes:} AMT denotes Amazon Mechanical Turk; BKT denotes Bayesian Knowledge Tracing; MAB denotes Multi-Armed Bandit; PVAE denotes Partial Variational Autoencoder; HRL denotes Hierarchical Reinforcement Learning; CTAI denotes Cognitive Tutor Algebra I.
\end{minipage}
\end{table}

To identify real-world evaluations of adaptive learning algorithms related to our work, we began with the empirical literature reviewed by Doroudi's work~\cite{doroudi2019wheres}, including only those conducted in real-world educational settings (e.g., classrooms or deployed online platforms) and focused on STEM domains such as mathematics, science, or computer science. We expanded the search using the deep research capabilities of frontier large language models (GPT-4.5, Claude Opus 4.5, and Gemini Pro), each prompted with a structured query designed to return studies meeting strict inclusion criteria. Specifically, studies had to involve adaptive learning algorithms such as Bayesian Knowledge Tracing (BKT), multi-armed bandits, or reinforcement learning, and apply them to adaptively sequence or select learning content. In addition, we only included experimental studies that reported learning outcomes that were comparable between treatment and control groups; we excluded studies focused solely on personalized pedagogy, such as those adapting instructional style rather than content or difficulty order, as well as those limited to simulated settings or offline-only evaluations. This process yielded a preliminary list of candidate papers, which we then reviewed manually to verify that they met all criteria and to extract and summarize key information from each study (see Table~\ref{tab:lit}).

While many studies on BKT and Intelligent Tutoring Systems (ITS) have shown gains over traditional instruction~\cite{vsaric2024twenty, ma2014intelligent}, evidence is mixed when it comes to adaptive learning systems that aim to improve existing instructional sequencing in STEM education. Studies showing significant effects often include or are confounded with additional supports (e.g., teacher training with Cognitive Tutor or ASSISTments), compare against weak baselines (e.g., random sequencing), or measure outcomes not directly tied to learning (e.g., time or completion rate). Many are short-term or conducted entirely online with crowd-recruited participants, lacking institutional context and the academic stakes that are inherent to a formal education system. The two large-scale field RCTs closely aligned with our work are the evaluation of the Eedi platform~\cite{harrison2024study} and Doodle personalized learning platform~\cite{Naslund-Hadleyetal2025}. However, Eedi focuses on demonstrating the efficacy of their overall platform rather than studying personalized problem sequencing in isolation; Doodle Learning in Belize reported no significant improvement in math scores.

To the best of our knowledge, our study is the first to show a statistically significant educational impact of mastery-based adaptive learning over expert-informed fixed-sequence practice problems in a large-scale, long-term, institutionally grounded deployment.

\section{Materials}
\label{sup:material}

\ref{sup:practiceQ} provides example practice problems; \ref{sup:exam} provides the in-person certification exam.

\subsection{Practice Problem Examples}
\label{sup:practiceQ}

Each module contains 40 AI-generated questions (Section~\ref{sec:rag}) for homework practice, with 10 at each difficulty level (easy, medium, hard, very hard). We prepared 10 additional backup questions per module to address potential issues, totaling approximately 500 questions in our question bank. For generated questions that passed our automated RAG validation procedure, six independent reviewers (three teaching assistants from National Taiwan University and three from our research team) manually verified each question for accuracy and language clarity. Example questions for each difficulty level are shown below:

\begin{exampleq}{Easy (True/False)}
\textbf{Statement.} The \texttt{len()} function can be used to calculate the length of both lists and strings.\\
\textbf{Task.} Indicate whether the statement is \texttt{true} or \texttt{false}.
\end{exampleq}

\begin{exampleq}{Medium (Debugging)}
\textbf{Task.} Write a program that reads a single line of space-separated integers representing students' ages.
The program should swap the first and last elements of the list and then output the modified list.
The provided code contains bugs that cause the swap to be performed incorrectly.
Fix and provide corrected program. (The error will not be in any \texttt{print()} statement.)

\vspace{4pt}
\textbf{Buggy code:}
\begin{lstlisting}[language=Python]
ages_str = input()             
ages = ages_str.split()        
for i in range(len(ages)):     
    ages[i] = int(ages[i])     
ages[0] = ages[-1]             
ages[-1] = ages[0]             
print(ages)                    
\end{lstlisting}

\vspace{4pt}

\textbf{Example test cases:}

\begin{lstlisting}
Input:
18 20 22 24

Output:
[24, 20, 22, 18]
5
\end{lstlisting}
\end{exampleq}

\begin{exampleq}{Hard (Coding)}
\textbf{Task.} Write a function that calculates tax based on the given income (integer) and returns the tax (float).
Rules: if income is less than 10,000 dollars, the tax rate is 3\%;
if income is between 10,000 and 20,000 dollars (inclusive), the tax rate is 4\%;
if income is above 20,000 dollars, the tax rate is 5\%.

\vspace{4pt}
\textbf{Example test cases.}
\begin{lstlisting}[language=Python]
assert calculate_tax(25000) == 1250.0
assert calculate_tax(10000) == 400.0
assert calculate_tax(20000) == 800.0
assert calculate_tax(1000000) == 50000.0
\end{lstlisting}
\end{exampleq}

\begin{exampleq}{Very Hard (Longer coding)}
\textbf{Task.} Assume you are a manufacturing manager responsible for scheduling jobs to minimize the maximum completion time (makespan). Given job processing times and the number of machines, compute the minimized makespan under the following strategy:
(1) sort job times in descending order, then
(2) assign each job (in that order) to the machine with the smallest current load.

\textbf{Input.} Two lines:
(1) space-separated job processing times (integers),
(2) number of machines (integer).

\textbf{Output.} Print the maximum machine load after assignment (the makespan).

\vspace{4pt}

\textbf{Example test cases:}
\begin{lstlisting}
Input:
3 3 3 4 4 5 5 6 7 8
3

Output:
18
\end{lstlisting}
\end{exampleq}

\subsection{Certification Exam}
\label{sup:exam}
This section provides the bilingual (English/Chinese) certification exam used in our study. We manually designed the exam (i.e., it was not AI-generated) based on the instructor’s course materials and evaluation data from previous terms of the same course. The exam was administered in person as a handwritten, paper-based assessment with no digital devices allowed, and lasted 120 minutes.

\setlength{\fboxrule}{0.6pt}

\includepdf[
  pages=1,
  pagecommand={\thispagestyle{plain}},
  trim=1.0cm 1.2cm 1.0cm 1.2cm,
  clip,
  scale=0.8,
  frame
]{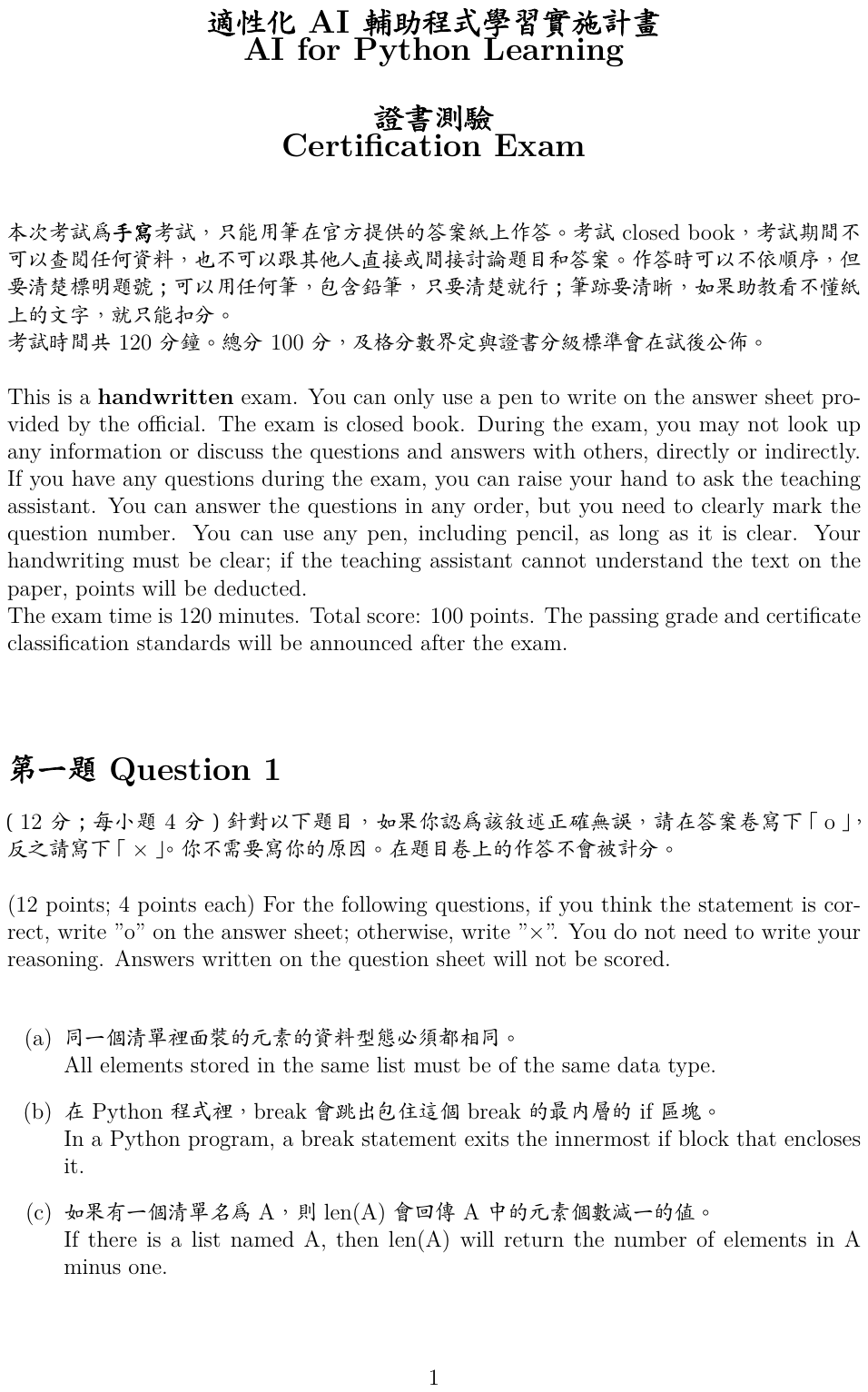}

\includepdf[
  pages=2-,
  pagecommand={\thispagestyle{plain}},
  trim=1.0cm 1.2cm 1.0cm 1.2cm,
  clip,
  scale=0.78,
  frame
]{TaiwanDoE_Exam-13.pdf}

\end{document}